\documentclass[options,onecolumn]{mn2e}
\usepackage{psfig}
\newif\ifAMStwofonts
\newcommand{\mt}[1]{\mbox{$\mathbfss{#1}$}}

\newcommand{\VEV}[1]{\langle#1\rangle}

\title[Local power spectrum estimates of the \emph{WMAP} data]
{Testing the cosmological principle of isotropy: local power spectrum estimates of the \emph{WMAP} data} 
  
\author[Frode K. Hansen, A. J. Banday and Krzysztof M. G\'orski]
  {{F. K. Hansen$^1$, \thanks{E-mail: frodekh@roma2.infn.it}}, {A. J. Banday$^2$, \thanks{E-mail: banday@MPA-Garching.MPG.DE}}, {K. M. G\'orski$^{3}$, \thanks{E-mail: Krzysztof.M.Gorski@jpl.nasa.gov}}\\
$1$ Dipartimento di Fisica, Universit\`a di Roma `Tor Vergata', Via della Ricerca Scientifica 1, I-00133 Roma, Italy\\
$^2$ Max Planck Institut f\"ur Astrophysik, Karl-Scharzschild-Strasse 1, Postfach 1317, D-85741 Garching bei M\"unchen, Germany\\
$^3$ JPL, M/S 169/327, 4800 Oak Grove Drive, Pasadena CA 91109 \\}
\pagerange{\pageref{firstpage}--\pageref{lastpage}}
\pubyear{2002}

\begin{document}

\label{firstpage}

\maketitle

\begin{abstract}
We apply the Gabor transform methodology proposed in (Hansen et al. 2002, 2003)
to the \emph{WMAP} data in order to test the statistical properties of the
CMB fluctuation field and specifically to evaluate the fundamental
assumption of cosmological isotropy.
In particular, we apply the transform with several apodisation scales,
thus allowing the determination of the positional dependence of
the angular power spectrum with either high spatial localisation
or high angular resolution (ie. narrow bins in multipole space).
Practically, this implies that
we estimate the angular power spectrum locally in discs of various sizes 
positioned in different directions: 
small discs allow the greatest sensitivity to positional dependence,
whereas larger discs allow greater sensitivity to variations 
over different angular scales. 
In addition, we determine whether the spatial position of
a few outliers in the angular power spectrum could suggest the presence 
of residual foregrounds or systematic effects. 
For multipoles close to the first peak, the most deviant local estimates
from the best fit \emph{WMAP} model are associated with
a few particular areas close to the Galactic plane. Such deviations
also include the ``dent'' in the spectrum just shortward of the first peak which 
was remarked upon by the \emph{WMAP} team. Estimating the angular power spectrum
excluding these areas gives a slightly higher first Doppler peak amplitude. 

Finally, we probe the isotropy of the largest angular scales by
estimating the power spectrum on hemispheres and reconfirm
strong indications of a north-south asymmetry previously
reported by other authors. Indeed, there is a remarkable lack of power 
in a region associated with the north ecliptic pole.
With the greater fidelity in $\ell$-space allowed by this larger sky coverage, 
we find tentative evidence for residual foregrounds in the range $\ell=2-4$, which 
could be associated with the low measured quadrupole amplitudes
and other anomalies on these angular scales (eg. planarity and alignment).
However, over the range $\ell=5-40$ the observed asymmetry is much
harder to explain in terms of residual foregrounds and known systematic effects. 

By reorienting the coordinate axes, we partition the sky into
different hemispheres and search for the reference frame which
maximises the asymmetric distribution of power.
The north pole for this coordinate frame is found to intersect the sphere at
$(80^\circ,57^\circ)$ in Galactic co-latitude and longitude over almost
the entire multipole range $\ell=5-40$. Furthermore, the strong negative  
outlier at $\ell=21$ and the strong positive outlier at $\ell=39$ 
as determined from the global power spectrum by the \emph{WMAP} team, 
are found to be associated with the northern and southern hemispheres
respectively (in this frame of maximum asymmetry). Thus, these two 
outliers follow the general tendency of the multipoles 
$\ell=5-40$ to be of systematically lower amplitude in the north 
and higher in the south. Such asymmetric distributions of power on the sky provide a serious test for the cosmological principle of isotropy.

\end{abstract}

\section{introduction}

One of the fundamental assumptions of cosmology is that the universe
is isotropic. Such an  assumption necessarily implies therefore
that the statistical properties of the cosmic microwave 
background (CMB) should be the same in all directions on the sky. 
The FIRAS experiment on the COBE satellite \cite{cobe1,cobe2} demonstrated
that the mean temperature of the CMB is isotropic to high precision,
but it is only now with the superior sensitivity of the 
\emph{WMAP}\footnote{\emph{Wilkinson Microwave Anisotropy Probe}} \cite{WMAP}
data that the isotropic nature of the angular fluctuations
in the CMB can be tested. 

In this paper we will consider the variation in the angular power
spectrum of the CMB determined from patches sampling different
locations on the sky. In an isotropic universe, there should be no preferred 
direction and these locally determined angular power spectra, after allowing for some
sample variance, should demonstrate the same expectation value over
all positions. Here, 
we adopt the Gabor transform method proposed in \cite{hansen1,hansen2} 
and estimate the power spectrum in discs with 9.5 and 19 degree radius
and then on hemispheres, thus allowing either high spatial localisation
or high angular resolution, both of which are used to 
probe the directional dependence of the anisotropy.
The discs on which we estimate the power spectrum are positioned such that
they cover as much area as possible without overlapping so much as to
introduce strong correlations between the locally estimated
spectra. 
With the smaller discs, we can test the isotropy of the spectrum at a high 
spatial resolution, but small patches limit the resolution in 
multipole space. By increasing the size of the discs we test the 
isotropy of the spectrum at a refined resolution in multipole space 
but at the cost of a low spatial resolution.
We apply different tests of consistency between the spectra
in different multipole bins, comparing with a Monte Carlo ensemble
generated with the best fit model spectrum obtained by the \emph{WMAP} team
and using the noise and beam properties of \emph{WMAP}. We have chosen
to focus on the co-added W+V map as these are the channels in which
the foreground contamination is expected to be lowest.
In all cases, only data outside of the \emph{WMAP} Kp2 sky cut is included
in the analysis. 

With this approach we can 
also test for the presence of contaminating residual 
foregrounds or systematic effects. For example, 
incorrectly subtracted Galactic foregrounds would most likely manifest
asymmetry in the Galactic frame of reference, with an increasingly
likely level of residual contamination towards the plane. Similarly,
systematic effects would plausibly align with the ecliptic frame of
reference reflecting the scanning strategy of the satellite.
In addition, contamination of the data due to solar system emission would
be confined to the ecliptic plane. For these reasons, 
we compare the statistical distribution of the power spectrum
estimates in both the Galactic and ecliptic northern and
southern hemispheres, as well as the pole and equatorial regions for
these reference frames.

We will in particular check the positional dependence of some
features in the power spectrum observed by \emph{WMAP} \cite{hinshaw}: 
two at large scales corresponding to bins centered around
$\ell=21$ and $\ell=39$, one close to the first Doppler peak and one
close to the first trough (the latter was not seen in the analysis by \emph{WMAP}). Moreover we investigate the recent claims of an asymmetry of the large
scale structure between the northern and southern Galactic hemispheres
\cite{park,eriksen1,vielva,eriksen2,copi,curvat}. As the asymmetry is
reported to be seen at the largest scales, we use the results of the
power spectra estimated on hemispheres where the multipole resolution
is large enough to resolve different features between $\ell=2-63$. 

The outline of the paper is as follows.
In section \ref{sect:gabtrans} we outline the method described in 
\cite{hansen1,hansen2} to estimate the power spectrum on a patch on 
the sky using the so called Gabor transforms. 
The details of the data set used in the paper is outlined in section \ref{sect:data}. 
In section \ref{sect:small} we show the results of the local \emph{WMAP} 
spectra estimated on discs of radius $9.5^\circ$. We check for 
consistency between the locally estimated spectra in different 
multipole bins and different directions. To check the results at a 
higher multipole resolution, the analysis is repeated in section 
\ref{sect:medium} with discs of radius $19^\circ$. In section 
\ref{sect:hemis} we check for an asymmetry in the large scale 
structure and report results of the \emph{WMAP} spectra estimated on 
hemispheres. In particular, we test the robustness of the results
using other frequency channels and a variety of sky cuts.
Finally, section \ref{sect:concl} summarise our results.

\section{The Gabor Transformation on the sphere}
\label{sect:gabtrans}

In \cite{hansen1} (from now on HGH) we introduced the method
of Gabor transforms on the sphere and showed how its application
could be used to analyse the CMB power spectrum, using the 
pseudo-spectral information from a given patch on the sky. The
pseudo spectrum is used as input data to a likelihood analysis 
for an axi-symmetric patch on the sky. For such a sky partition,
the correlation properties of the sky signal can be
calculated analytically. In this paper, we extend the method to an arbitrary
asymmetric patch on the sky, by using Monte-Carlo pre-calculations of
the correlation matrix. In this section we will outline the method as
described in HGH. The details of the extension to arbitrary patches are
deferred to the appendix, here we will focus on the results.

We define the pseudo power spectrum $\tilde C_\ell$ as the power spectrum 
obtained using the spherical harmonic transform on a CMB temperature 
field $T({\mathbf{\hat n}})$ multiplied with a window function 
$G({\mathbf {\hat n}})$. Letting $a_{\ell m}$ be the spherical 
harmonic coefficients of the CMB field and $C_\ell$ the corresponding 
power spectrum, the corresponding quantity on the windowed sky is 
indicated by a tilde and can be written as

\begin{equation}
\tilde C_\ell=\sum_m \frac{\tilde a_{\ell m}^*\tilde a_{\ell m}}{2\ell+1},
\end{equation}
where
\begin{equation}
\label{eq:psalm}
\tilde a_{\ell m}=\int d{\mathbf{\hat n}} T({\mathbf{\hat n}})G({\mathbf{\hat n}})Y^*_{\ell m}({\mathbf{\hat n}}).
\end{equation}

This pseudo spectrum can be written in terms of the full sky spectrum by means of the following formulae:
\begin{equation}
\label{eq:pcl}
\VEV{\tilde C_\ell}=\frac{1}{2\ell+1}\sum_{\ell'}C_{\ell'}K(\ell,\ell')
\end{equation}
where the kernel $K(\ell,\ell')$ can be calculated analytically by the 
formulae given in HGH, or by Monte-Carlo methods as described here in the appendix.

The method proposed in HGH was to sample a disk-like region of the observed sky 
$T({\mathbf{\hat n}})$ using a top-hat or a Gaussian window 
$G({\mathbf{\hat n}})$. The pseudo power spectrum of this disc 
was then used as the input vector to a likelihood analysis, assuming 
the distribution of the $\tilde C_\ell$ to be Gaussian. It was demonstrated
that this is a good approximation when the multipole $\ell$ is 
sufficiently high, which for a Gaussian window with $15^\circ$ FWHM 
corresponded to $\ell\sim 100$. HGH employed the following likelihood ansatz:
\begin{equation}
\label{eq:multilik}
\mathcal{L}=\frac{\mathrm{e}^{-\frac{1}{2}{\mathbf{ d}}^\mathrm{T} {\mt{M}}^{-1}{\mathbf{
d}}}}{\sqrt{2\pi \det{\mt{M}}}},
\end{equation}
where the elements of the data vector are
\begin{equation}
d_i=\tilde{C}_{\ell_i}-\VEV{\tilde{C}^S_{\ell_i}}-\VEV{\tilde{C}^N_{\ell_i}}.
\end{equation}
Here the first term is the observed pseudo power spectrum including noise 
and the second term is the expected signal pseudo spectrum obtained 
from equation (\ref{eq:pcl}). The last term contains the expected 
noise pseudo spectrum for the patch. Similarly, the matrix expressing 
the correlation between the pseudo power spectrum coefficients can be written as 
 \begin{equation}
M_{ij}=<\tilde C_{\ell_i}\tilde C_{\ell_j}>-<\tilde C_{\ell_i}><\tilde C_{\ell_j}>=M_{ij}^\mathrm{S}+M_{ij}^\mathrm{N}+M_{ij}^\mathrm{X},
\end{equation}
where there is one contribution from the signal (S), one from the
noise (N) and one from signal-noise mixing (X). The form of these
matrices and how they can be computed are discussed in HGH for an
axi-symmetric region of the sky. Here, in the appendix, we show how to
extend the calculation of these matrices to an arbitrarily shaped
patch using a Monte-Carlo approach. Note that this likelihood
expression is a function of the underlying full sky power spectrum
$C_\ell$ which we want to estimate. The best estimate is the power
spectrum which maximises this likelihood.\\

Note that one cannot use all $\tilde C_\ell$ in the analysis, as the
correlation matrix will then be overdetermined. Following HGH, the
optimal spacing in $\ell$-space is given by the width of the kernel
and can be expressed approximately as $\Delta\ell=115/\theta$ where
$\theta$ is the radius of the window in degrees. We will call the
number of samples ($\tilde C_\ell$) to be used, $N_\textrm{sam}$. Also, one cannot
estimate all the full sky power spectrum coefficients $C_\ell$ as
there is not enough information in a disc to obtain an estimate of
every single multipole. For this reason we use a limited number
$N_\textrm{bin}$ of flat power spectrum bins. As discussed in HGH this number
should be smaller than $N_\textrm{sam}$ to allow for at least 3-4
observations ($\tilde C_\ell$) per bin. A more complete discussion on the 
number of input multipoles and full sky bins to use is given in HGH.

In HGH it was also demonstrated that one can simultaneously include the pseudo power spectra 
from several discs in order to analyse a full sky observation. In 
this case, the log-likelihood can be written as
\begin{equation}
\label{eq:jointlik}
L=\sum_{I=1}^{N_\mathrm{disc}}{\mathbf{ d}}_I^\mathrm{T}{\mt{M}}_I^{-1}{\mathbf{
d}}_I+\sum_{I=1}^{N_\mathrm{disc}}\ln\det {\mt{M}}_I,
\end{equation}
where $d_I$ is the full data vector of disc number $I$, $M_I$ is 
similarly the correlation matrix of disc $I$ and $N_\mathrm{disc}$ 
is the number of discs considered. This form of the likelihood 
assumes the $\tilde C_\ell$ between different discs to be uncorrelated. 
For minimally overlapping discs this can remain a good approximation.
It will be demonstrated later that for the set of 
discs used here, only a tiny correlation between the estimated power spectra 
of overlapping discs is introduced.

The purpose of this paper is to use these locally estimated spectra 
to investigate the isotropy of the CMB in the \emph{WMAP} data. We will 
assume the best fit \emph{WMAP} running index spectrum, make a Monte-Carlo 
ensemble of estimated power spectra on discs of different sizes and 
compare these to the spectra estimated on the \emph{WMAP} data set.

\section{Preparing the data}

\label{sect:data}

In this paper we analyse the publicly available \emph{WMAP} 
data\footnote{These can be obtained at the LAMBDA website:
\emph{http://lambda.gsfc.nasa.gov/}} after correction for Galactic foregrounds.
We consider data from the V and W frequency bands only
since these are less likely to be contaminated by residual foregrounds. 
The maps from the different channels are coadded using inverse noise-weighting;
\[
T=\frac{\sum_cT_c/\sigma_c^2}{\sum_c1/\sigma_c^2},
\]
where $c$ runs over the frequency channels V1,V2,W1,W2,W3,W4. 
In the simulations we have been generating maps smoothing with a 
beam and adding noise equal to the effective beam and noise of 
the co-added W+V channel (also available on the LAMBDA website). 
The input power spectrum in the simulation was the best fit \emph{WMAP} 
running-index spectrum. Most of the analysis in this paper has
been performed using the \emph{WMAP} Kp2 sky cut, including the point source mask. In 
those cases where, in order to test the robustness of the results, 
we have used either an alternate power spectrum, different 
frequency channels or different masks, the differences 
have been clearly stated.

It should be noted that the \emph{WMAP} team have utilised the cross-spectra between 
different channels for their estimation of the power spectrum, which 
is therefore less dependent on the exact noise properties 
of the experiment than in our case. Here we have adopted the 
approximate white noise model provided by the \emph{WMAP} team, and 
uncertainties in this model could result in erroneous estimates on
smaller angular scales where the multipoles are noise dominated. 
In fact, it has been determined that the noise level estimated 
using the tail of the pseudo power spectrum on a given 
disc is slightly higher than that obtained using the 
white noise model. In principle, this could be
accounted for using the 110 noise realizations per channel,
which include all of the effects of the full data processing pipeline, 
that have been made available by the \emph{WMAP} team on the LAMBDA website.
In practise, this number is not sufficient to converge to an accurate estimate of the 
statistical properties of the noise for individual discs.

\section{Results of the spectra estimated on small discs}

\label{sect:small}

In this section, we test the isotropy of the CMB with high 
spatial sensitivity by estimating the power spectrum on discs of 
radius $9.5^\circ$. As a consequence of the small disc size, the 
multipoles are strongly coupled and wide power spectrum bins must
be employed. Using a set of simulated maps we have created a Monte-Carlo ensemble 
of estimates of the power spectrum on individual small discs from 
the disc set, and additionally
estimates of the power spectrum corresponding to larger sample regions
by combining several discs. These are then compared to the estimates
derived from the \emph{WMAP} data.

\subsection{The analysis}
\begin{figure}
\begin{center}
\leavevmode
\psfig {file=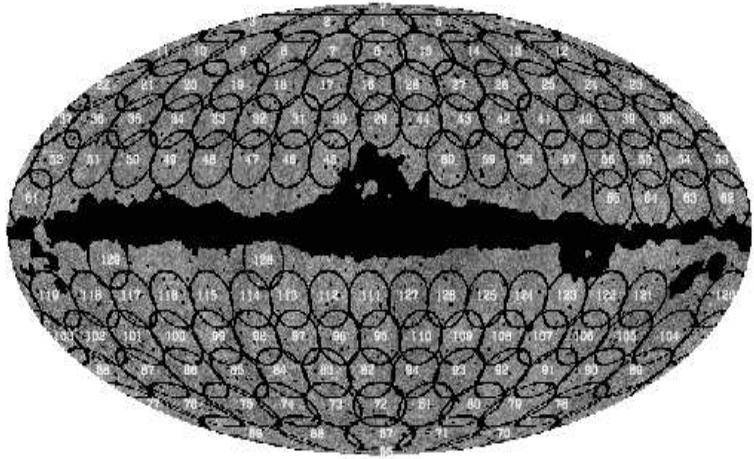,width=10cm,height=7cm}
\caption{The position and numbering of the $9.5^\circ$ discs on which the local power spectra have been calculated.}
\label{fig:discnumbers}
\end{center}
\end{figure}

With the method described above, we have estimated the angular power spectrum for
164 discs with an angular radius of $9.5^\circ$ uniformly distributed on the
part of the sphere outside of the \emph{WMAP} Kp2 sky cut. The disc position
and disc size was chosen in order to cover as much as possible of the
observed sky with as many samples (discs) as possible without losing
too much multipole resolution (i.e. to maintain enough power spectrum
bins to resolve the shape of the power spectrum). The 
overlap of the discs was also kept to a minimum in order to reduce
correlations between the locally estimated spectra. We sampled the
pseudo spectrum in $N_\textrm{sam}=84$ points which we used as input
data to the likelihood estimation procedure. We then estimated
$N_\textrm{bin}=12$ power spectrum bins in the multipole range
$\ell=2-850$. As the correlation matrix in the likelihood can become
singular if the observed area is too small, we have excluded
all discs where less than $90\%$ of the disc area remains after
application of the  Kp2 mask. In figure \ref{fig:discnumbers} we show
the remaining 130 discs with the numbering which will be used in the
following treatment. The distance in the latitudinal $\theta$ direction between the disc centres going from
the Galactic poles to the equator is roughly $14^\circ$ and in the
longitudinal $\phi$ direction the centres are all at an equal distance for a given
Galactic latitude. The 12 power spectrum bins $b$ are summarised in table
(\ref{tab:smallbins}). As described in HGH, the $C_\ell$ coupling
kernel has a similar width at all scales, giving a
similar resolution over all multipoles. Therefore we have adopted
bins of a fixed size $\Delta\ell$ in the whole
multipole range.

\begin{center}
\begin{table}
\caption{The power spectrum bins for the $9.5^\circ$ disc estimates.\label{tab:smallbins}}

%\begin{ruledtabular}
\begin{tabular}{|l|c|c|c|c|c|c|c|}
\hline
bin & 1 & 2 & 3 & 4 & 5 & 6 & 7 \\
\hline
range & $\ell=2-63$&$\ell=64-113$&$\ell=114-163$&$\ell=164-213$&$\ell=214-263$&$\ell=264-313$&$\ell=314-363$\\
centre & $\ell=33$&$\ell=89$&$\ell=139$&$\ell=189$&$\ell=239$&$\ell=289$&$\ell=339$\\
\hline
bin & 8 & 9 & 10 & 11 & 12 & &  \\
\hline
range & $\ell=364-413$&$\ell=414-463$&$\ell=464-513$&$\ell=514-563$&$\ell=564-613$\\
centre & $\ell=389$&$\ell=439$&$\ell=489$&$\ell=539$&$\ell=589$&\\
\hline
\end{tabular}
%\end{ruledtabular}
\end{table}
\end{center}

\begin{figure}
\begin{center}
\leavevmode
\psfig {file=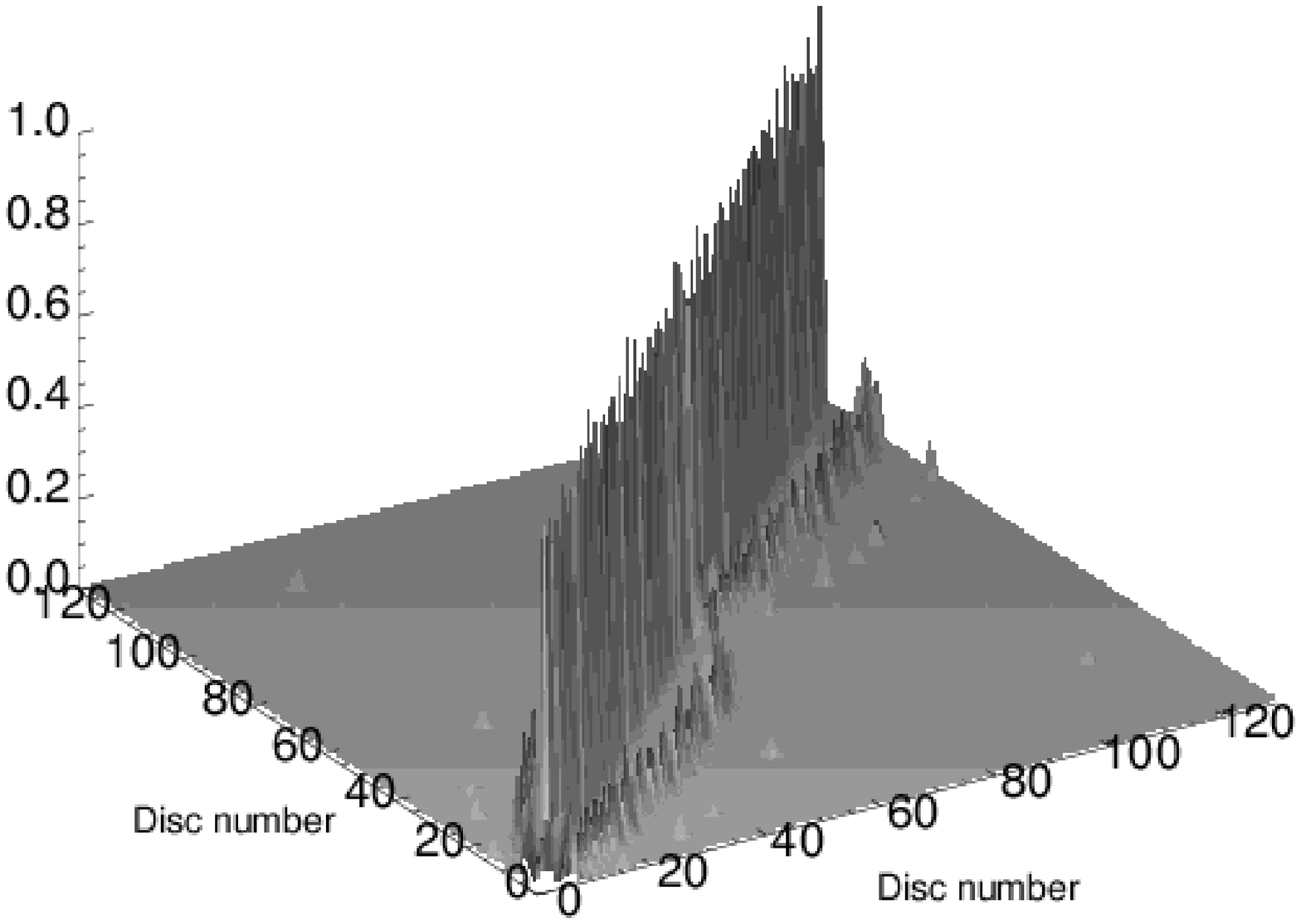,width=8cm,height=8cm}
\psfig {file=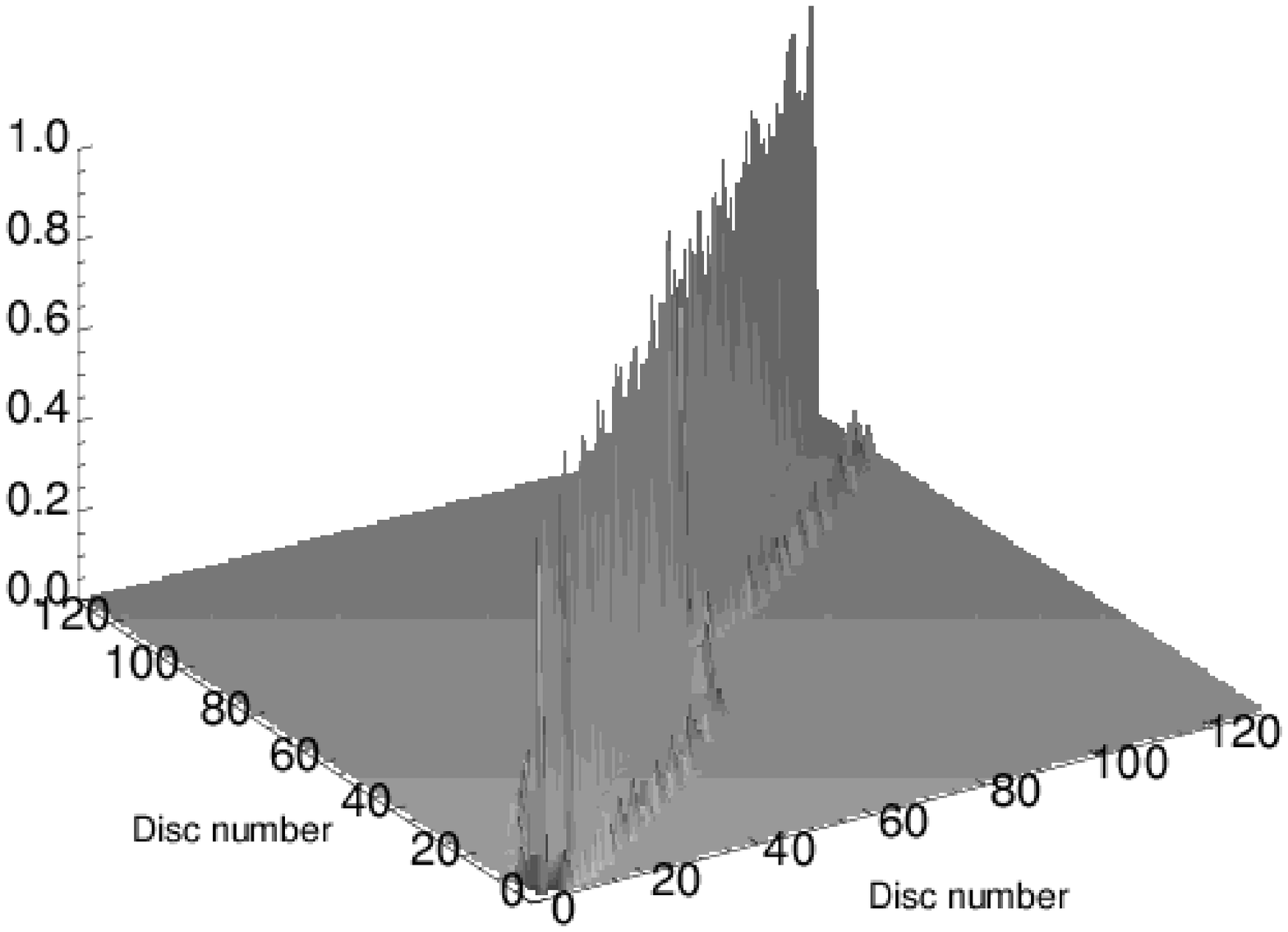,width=8cm,height=8cm}
\psfig {file=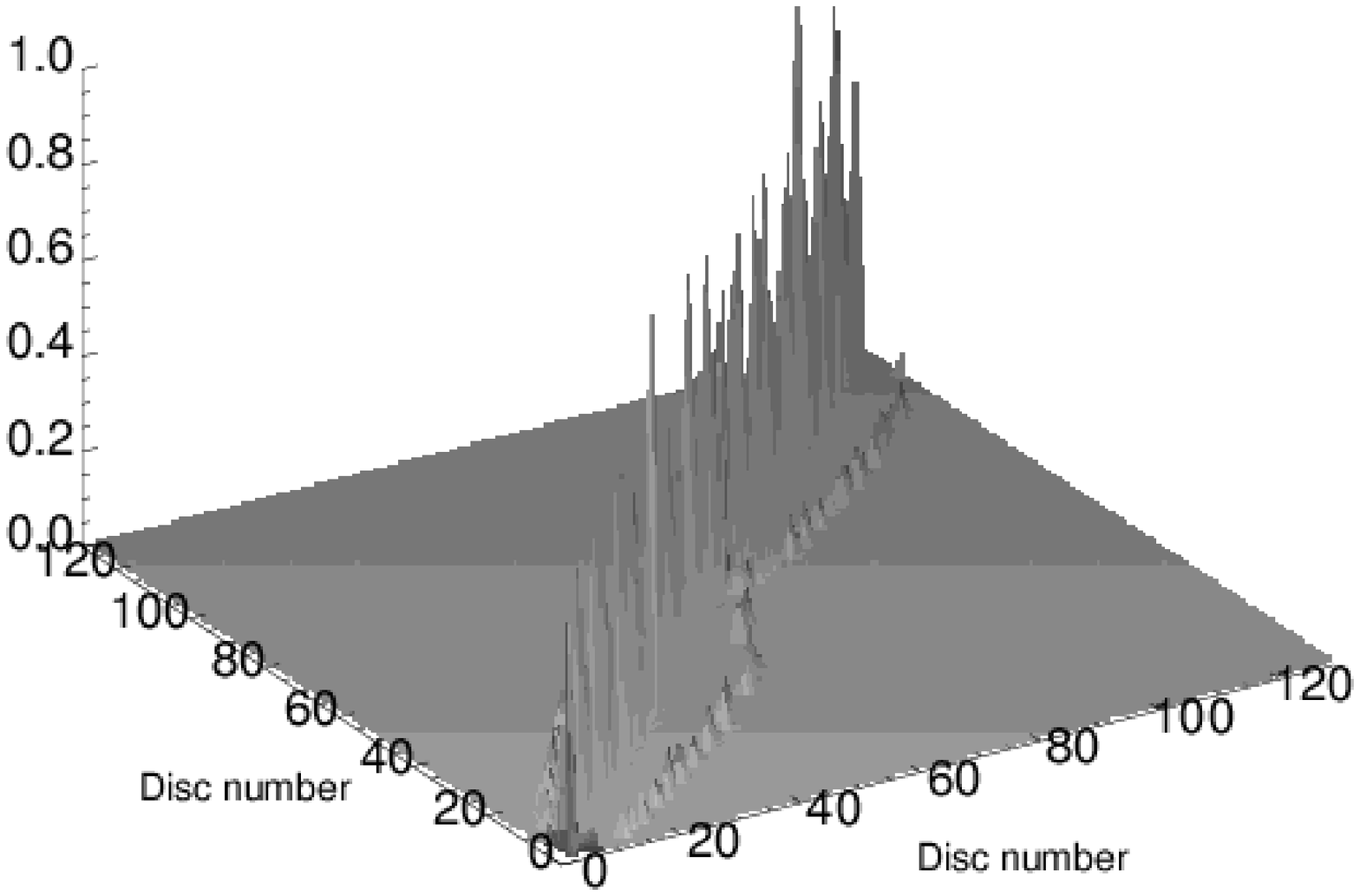,width=8cm,height=8cm}
\caption{The disc-disc correlation matrix for the multipole bins 0, 3 and 6 centred at  $\ell=33$, $\ell=189$ and $\ell=339$ respectively.}
\label{fig:corplot}
\end{center}
\end{figure}

\begin{figure}
\begin{center}
\leavevmode
\psfig {file=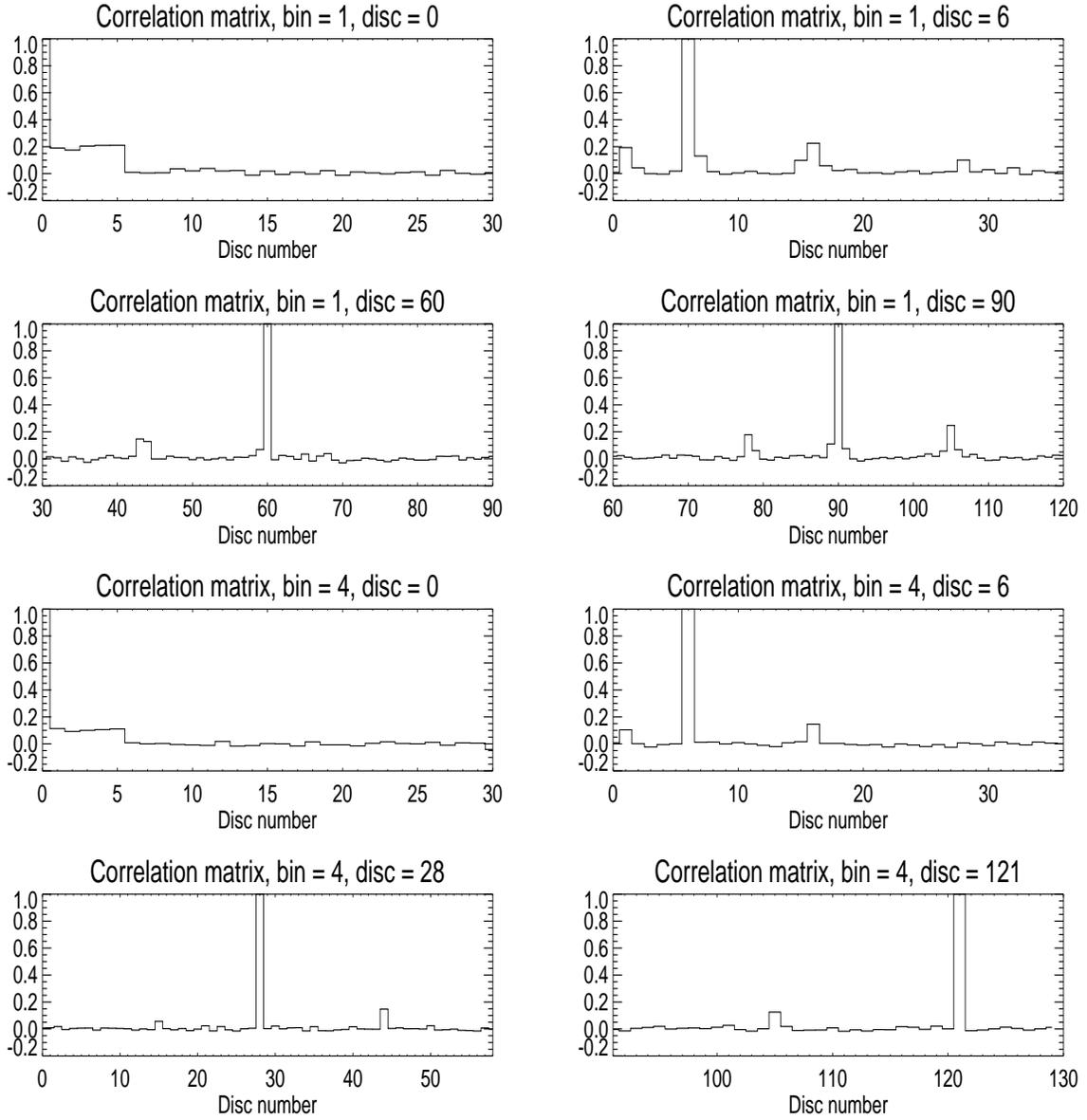,width=16cm,height=16cm}
\caption{Slice of the disc correlation matrix for some selected discs at the lowest multipole bin $\ell=33$ and a bin centred close to the first power spectrum peak $\ell=189$. We have chosen the discs 0,6 and 60 in order to study the correlations for discs which have many (disc 0) and few (disc 6 and 60) overlapping neighbours. The other discs have been chosen as particular bin values of these discs have been found in the \emph{WMAP} data. We have chosen particular discs close to the ecliptic poles as some particular features are found in this area (see the text and figure \ref{fig:plotdisc4_b034})}.
\label{fig:corcut}
\end{center}
\end{figure}

\begin{figure}
\begin{center}
\leavevmode
\psfig {file=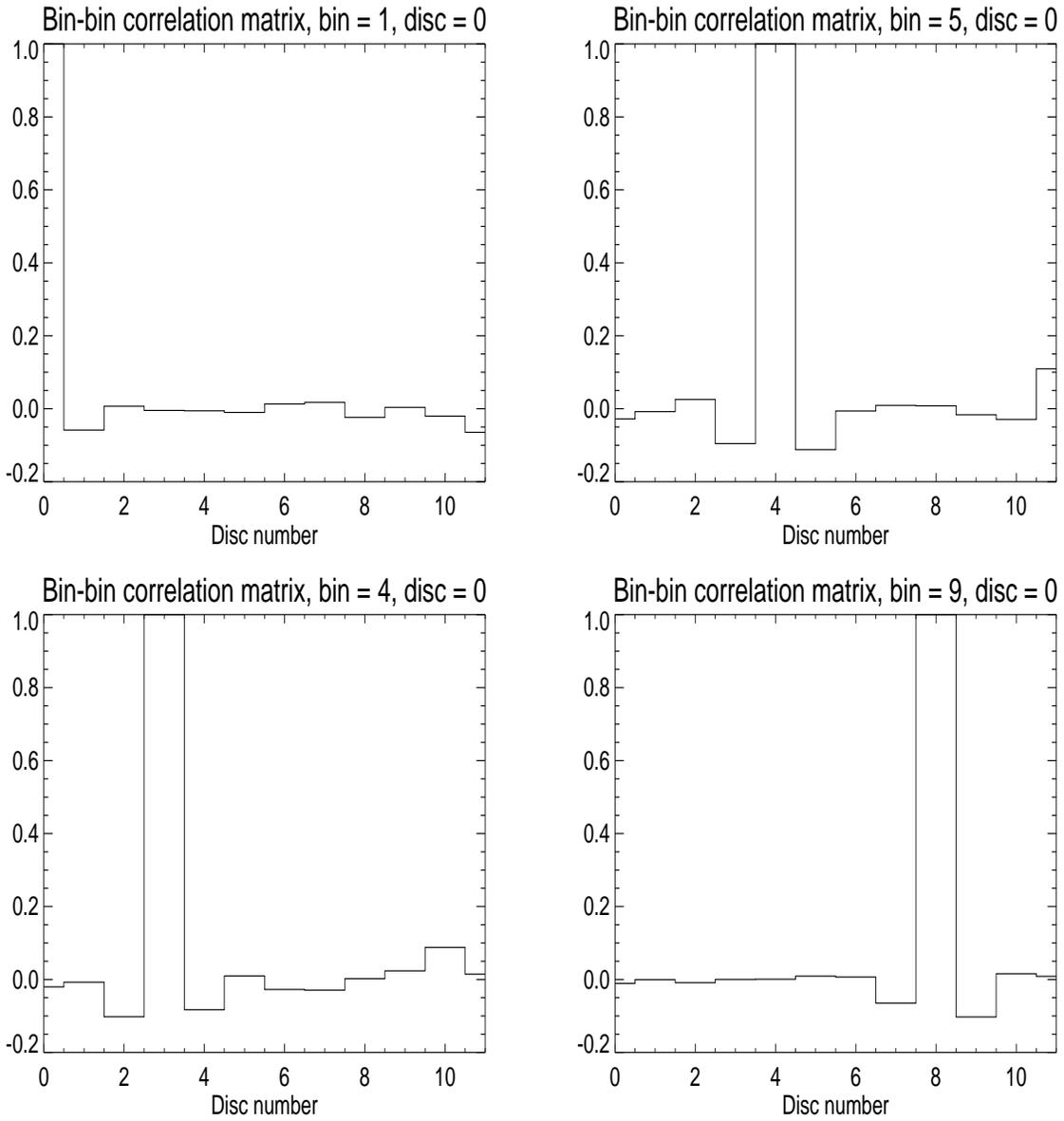,width=16cm,height=16cm}
\caption{Slice of the bin to bin correlation matrix for disc 0 taken at bins 1,4,5 and 9.}
\label{fig:binbincor}
\end{center}
\end{figure}

%We will first check for possible correlations between the estimated 
%spectra in different discs. 
Figure (\ref{fig:corplot}) shows the disc-disc correlation matrix for the bins 1,4 and 7 
(centred at $\ell=33$, $\ell=189$ and $\ell=339$). In figure 
(\ref{fig:corcut}) some slices of the correlation matrix are shown 
for disc 0 -- which has significant overlap with several other discs (1-5), 
for disc 6 -- which has overlap with two other discs (1 and 16) 
and for disc 60 -- which shows little intersection with other regions.
The other discs plotted were selected because of the properties
of certain of their multipole bins, as will become evident later in
this section. For the lowest 
bin one can see some correlation between several discs, whereas 
for the higher bins the correlations are limited to the neighbouring 
discs. Taking into account the extension of the structures at the 
largest scales, a stronger correlation could be expected and in the 
figure one can even see correlation between non-overlapping discs 
for the lowest multipole bin. Note that for higher multipole bins, 
the off-diagonal elements of the disc-disc correlation matrix are 
at a maximum $10-20\%$ of the diagonal. Moreover the disc-disc 
correlations for these multipoles are only significant between strongly 
overlapping discs. Finally figure (\ref{fig:binbincor})
shows the bin-to-bin correlation matrix for a single disc. One 
can see that, apart from a weak anti-correlation between neighbouring 
bins, the disc estimates are essentially uncorrelated.\\

\subsection{The results of the analysis applied to the \emph{WMAP} data on individual discs}

\begin{figure}
\begin{center}
\leavevmode
\psfig {file=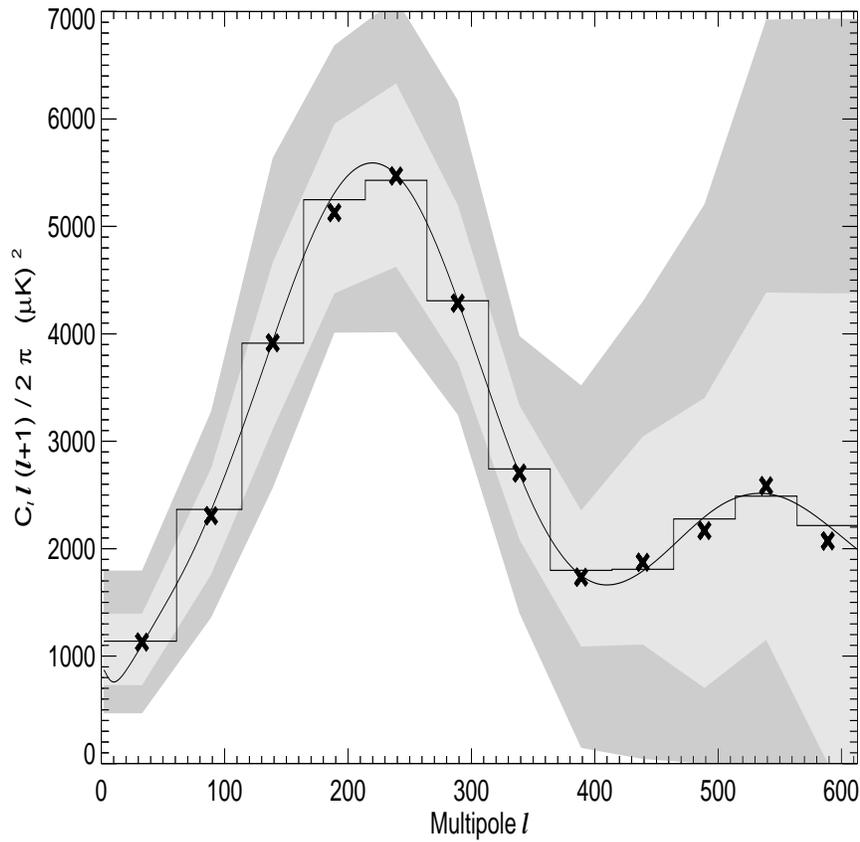,width=12cm,height=12cm}
\caption{The best fit \emph{WMAP} running index power spectrum (solid line). The histogram shows the spectrum binned using the same bins as for the disc estimates. The crosses show the binned full sky \emph{WMAP} spectrum obtained by the \emph{WMAP} team and the shaded areas show the spread of the \emph{WMAP} spectrum over 130 discs of radius $9.5^\circ$, found from the Gabor analysis. The two shaded areas indicate where 67 and 95 percent of the 130 \emph{WMAP} disc estimates are contained. Note that the shaded areas are NOT error bars for the \emph{WMAP} estimates, but the spread found by estimating the spectrum in different positions.}
\label{fig:fullspectrum}
\end{center}
\end{figure}

As a first test of isotropy, we will look at the distribution of the 
power spectrum estimates over the sphere. In figure
(\ref{fig:fullspectrum}) 
we have plotted the best fit \emph{WMAP} running index spectrum (solid line), 
the same spectrum binned as in our estimation procedure 
(histogram) and the power spectrum results of the \emph{WMAP} team with 
this binning (crosses). The shaded areas indicate the spread of 
our power spectrum estimates on the \emph{WMAP} data over the different 
discs. The bands represent the areas containing 67 and 95 percent 
of the disc estimates for a given bin. Note that the mean of the 
disc estimates close to the first peak (bin 4)
lies below the best-fit prediction. This corresponds
to the 'dip' in the power spectrum pointed out by the 
\emph{WMAP} team. Moreover, the distribution of estimates for the
spectrum does not contain strong outliers due to some
particular location on the sky.

\begin{figure}
\begin{center}
\leavevmode
\psfig {file=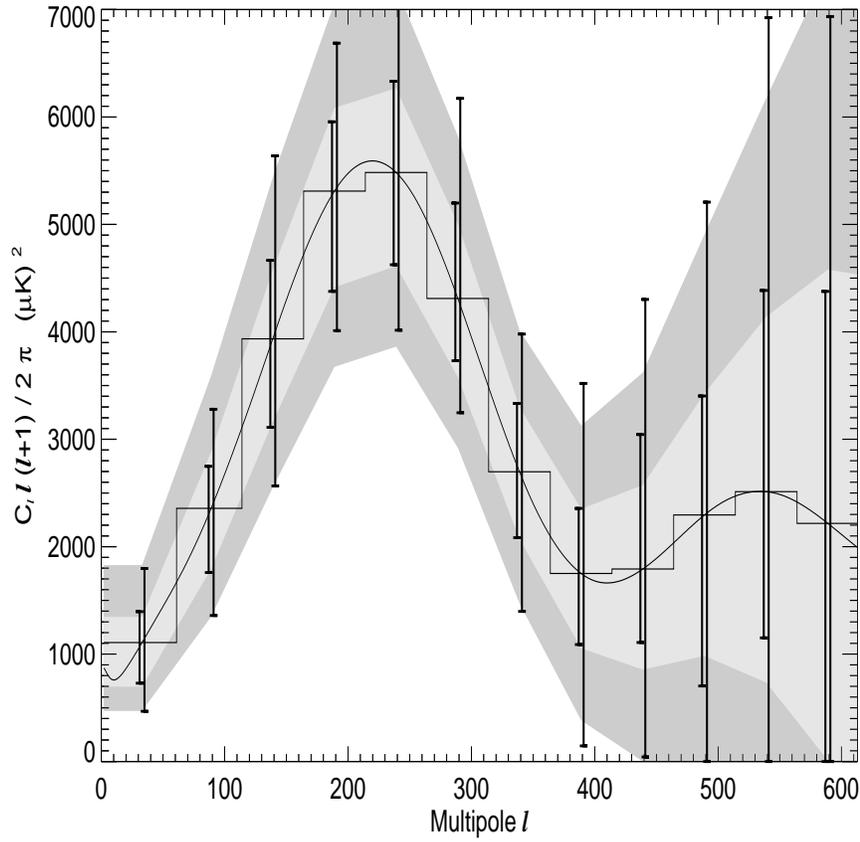,width=12cm,height=12cm}
\caption{The best fit \emph{WMAP} running index power spectrum (solid
line). The histogram shows the spectrum binned in the same way as the
disc estimates. The two shaded areas indicate where 67 and 95 percent
of the disc estimates in the simulations are contained. The errorbars
at each bin show the same distribution for the \emph{WMAP} data, the left
being the $1\sigma$ level and the right being the $2\sigma$ level.}
\label{fig:fulldist}
\end{center}
\end{figure}

The aim of the disc estimation procedure is to check the isotropy of
the CMB fluctuation field. But as explained in the introduction, if 
residual foregrounds or systematic effects are present in the data set, 
these could show up as asymmetries in the Galactic or ecliptic frames
of reference. Therefore, we evaluate the distribution of the
power spectrum estimates in the Galactic and ecliptic northern and
southern hemisphere as well as the pole and equatorial regions in
these reference frames. As a first check, we looked at the
distribution of all the 130 disc estimates compared to the same 
distribution in the 6144 simulations. In figure (\ref{fig:fulldist}) 
the errorbars at each power spectrum bin show where 67 (left bar) 
and 95 (right bar) percent of the disc estimates of \emph{WMAP} data fall 
(same as the coloured zones in figure (\ref{fig:fullspectrum})). 
The shaded zones show the same distribution taken from the 
simulations. We see that the width of the distributions for 
\emph{WMAP} is in good agreement with the simulations.

\begin{figure}
\begin{center}
\leavevmode
\psfig {file=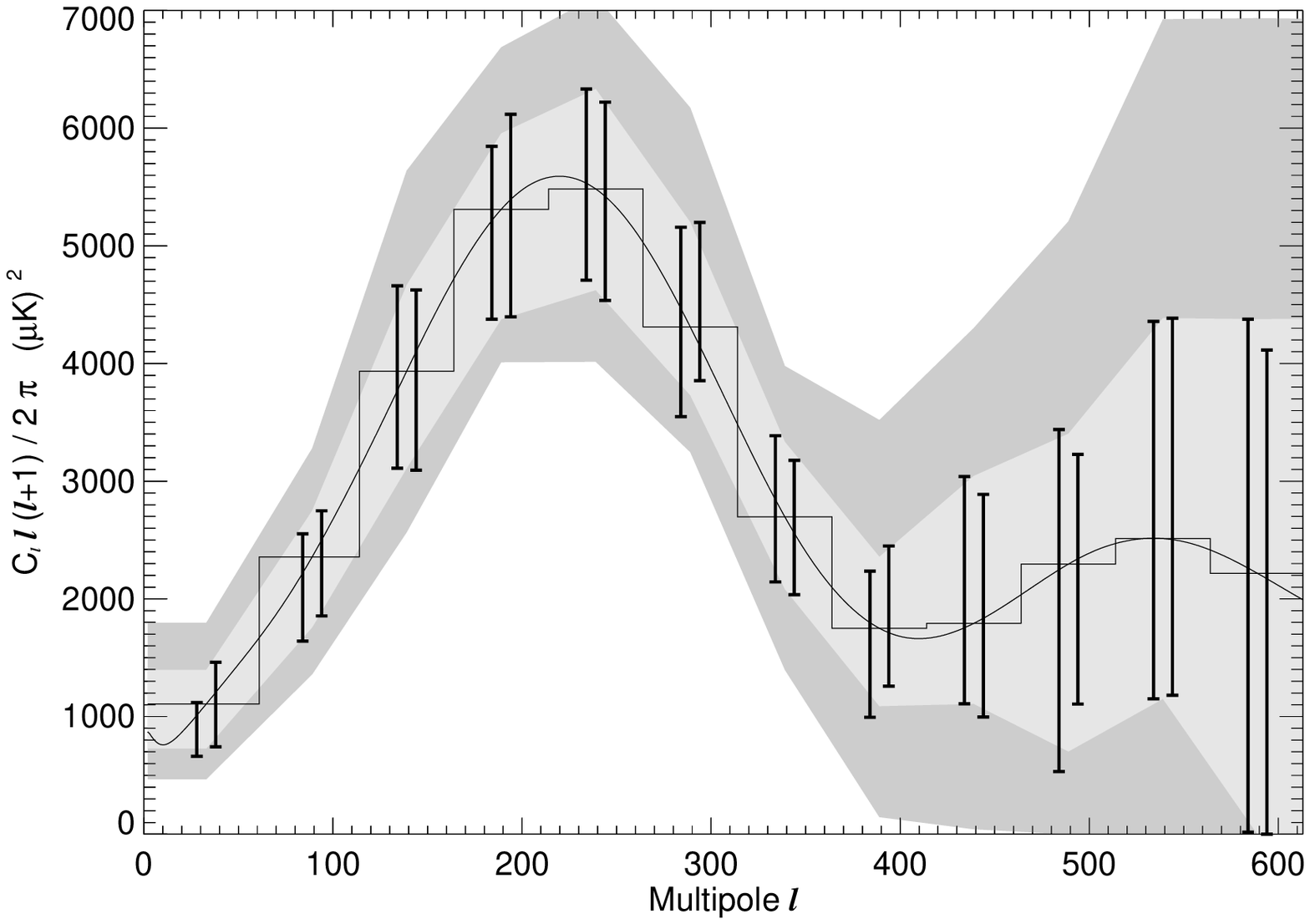,width=8cm,height=8cm}
\psfig {file=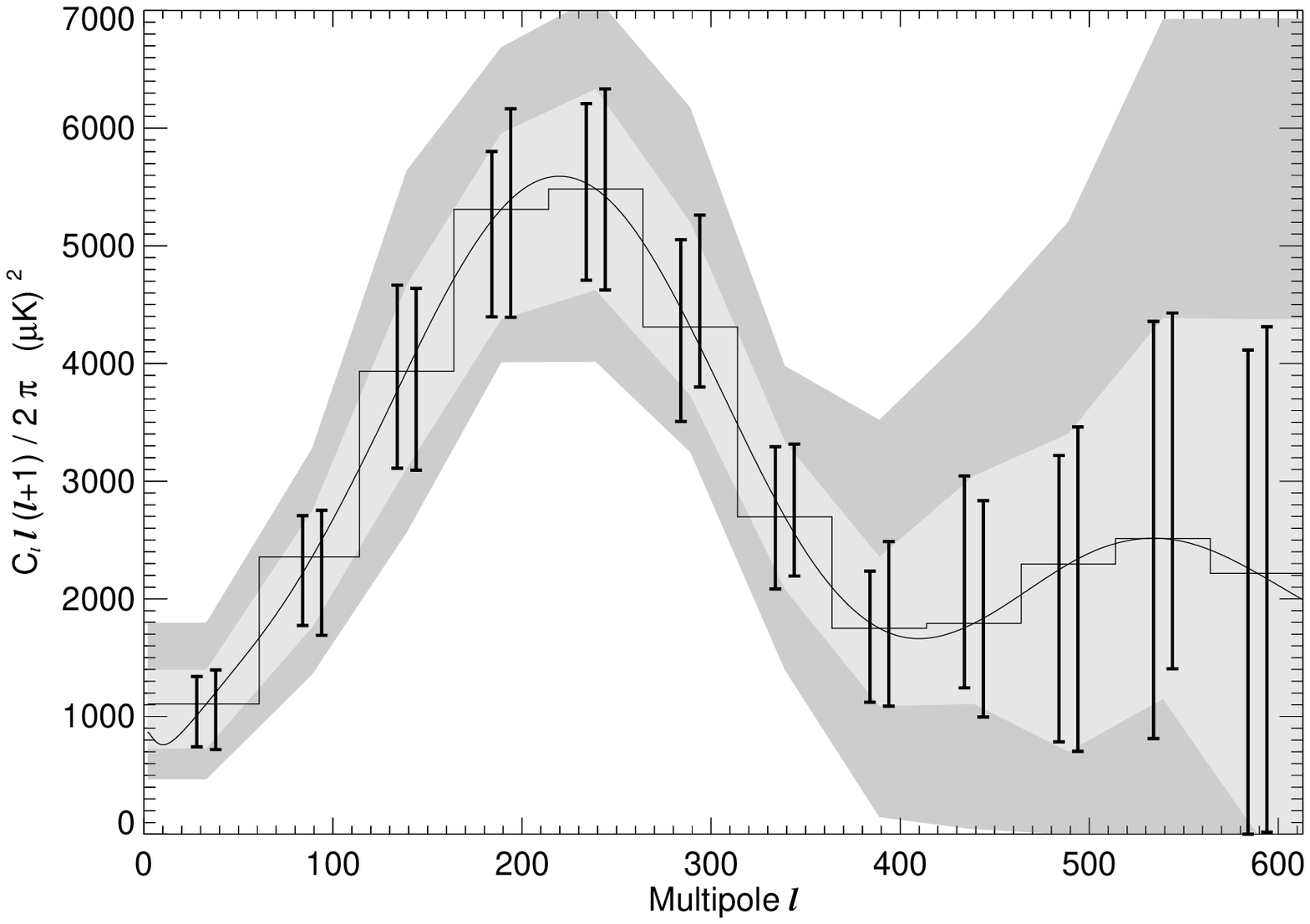,width=8cm,height=8cm}
\psfig {file=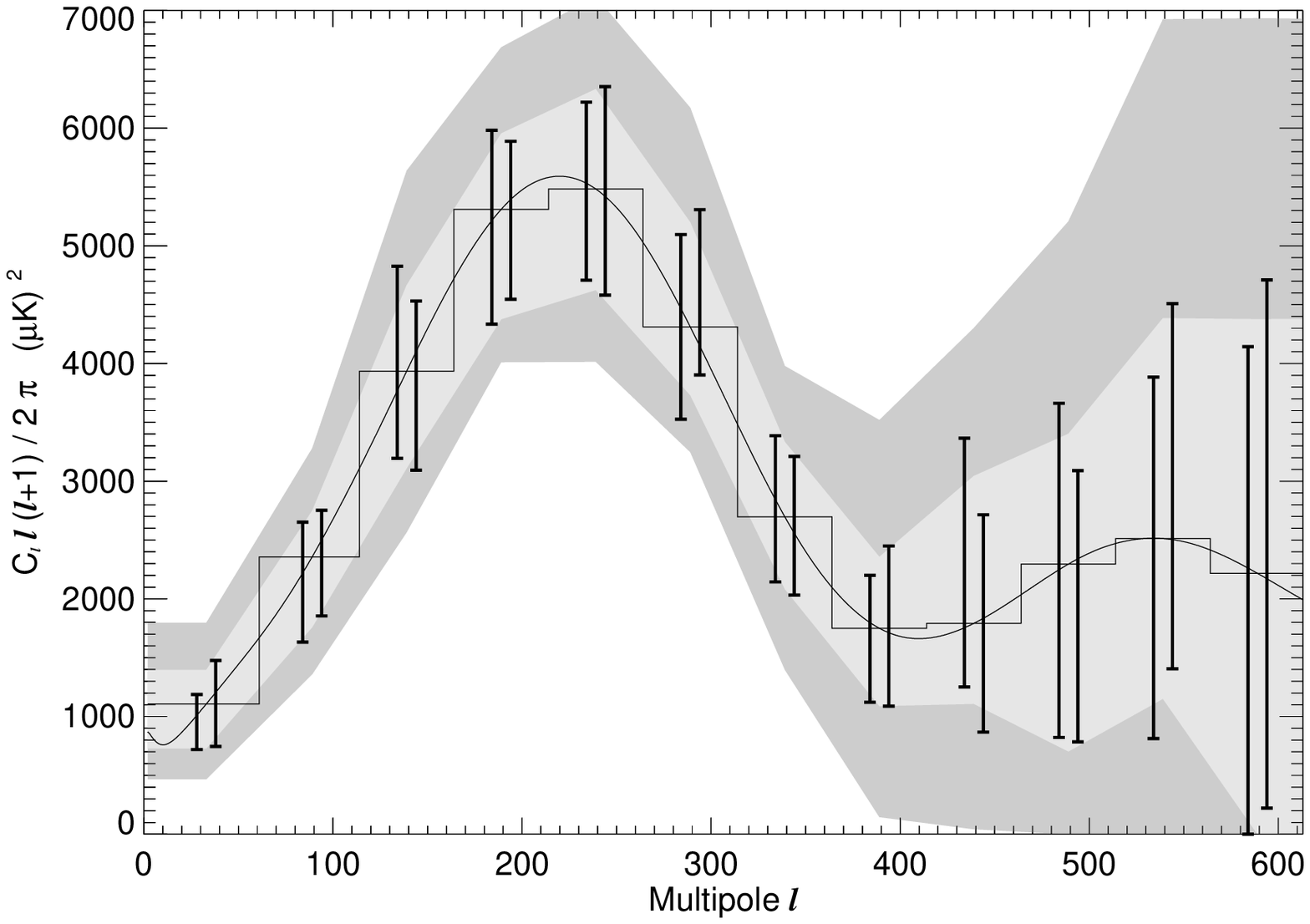,width=8cm,height=8cm}
\psfig {file=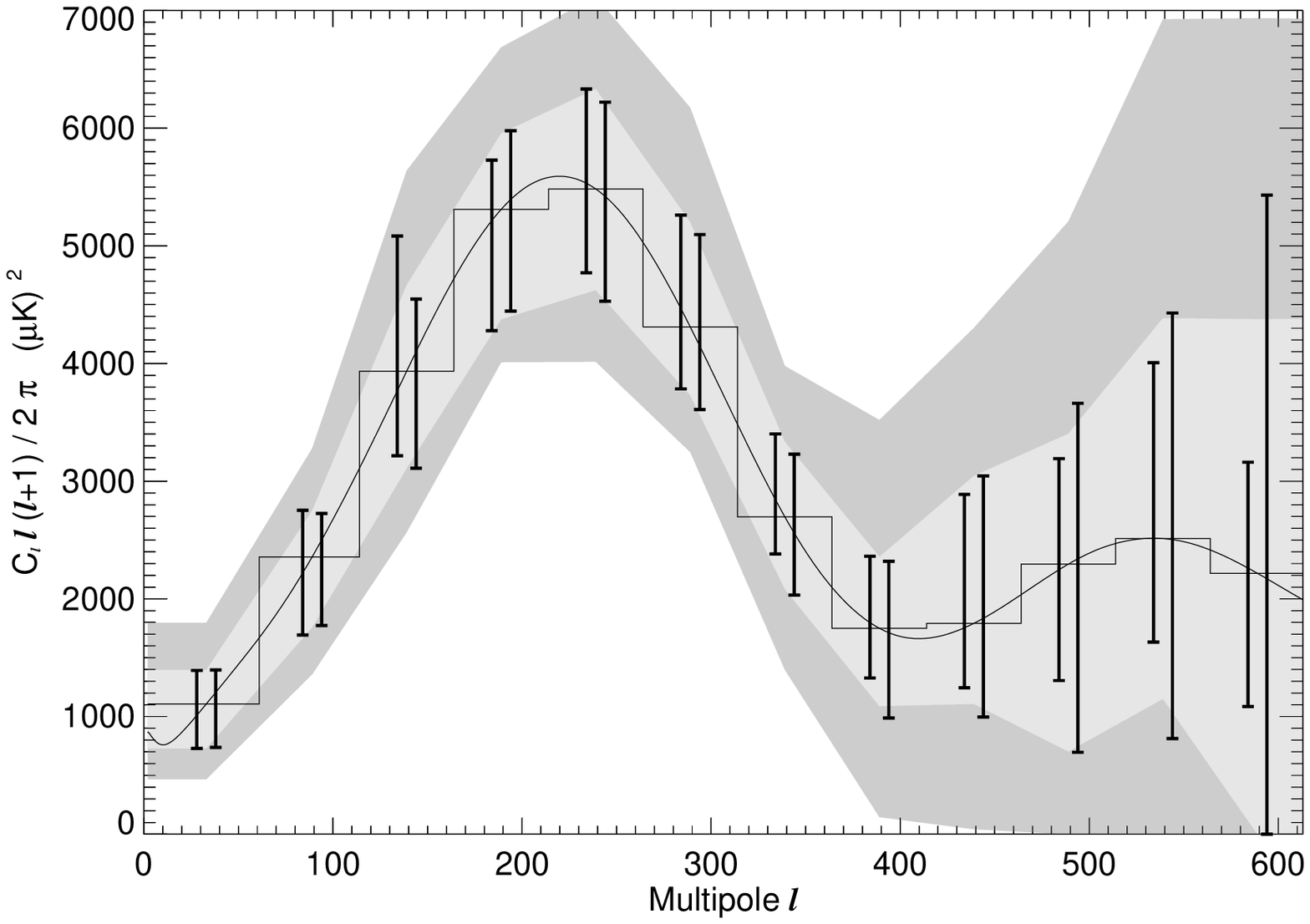,width=8cm,height=8cm}
\caption{The best fit \emph{WMAP} running index power spectrum (solid
line). The histogram shows the spectrum binned in the same way as the
estimates on $9.5^\circ$ discs. The two shaded areas indicate where 67
and 95 percent of the disc estimates in the \emph{WMAP} map are
contained. The errorbars at each bin show the same distribution (67
percent level) for a part of the \emph{WMAP} data. {\bf Upper left:} north Galactic
hemisphere (left bar) / south Galactic hemisphere (right bar), {\bf upper right:} Galactic
polar region (left bar), Galactic equatorial region (right bar),
{\bf lower left:} north ecliptic hemisphere (left bar) / south ecliptic
hemisphere (right bar), {\bf lower right:} ecliptic polar region (left)
/ ecliptic equatorial region (right).}
\label{fig:partdist}
\end{center}
\end{figure}

In figure (\ref{fig:partdist}) we show the disc distributions of 
the power spectrum on the northern and southern Galactic hemispheres 
(NGH and SGH upper left plot), Galactic polar and equatorial region 
(GPR and GER upper right plot), the northern and southern ecliptic 
hemispheres (NEH and SEH lower left plot) and the ecliptic polar 
and equatorial regions (EPR and EER lower right plot). The error 
bars indicate the $1\sigma$ spread of the power spectrum over the 
discs in the given regions for \emph{WMAP}, whereas the shaded zones 
show the corresponding 1 and 2 $\sigma$ spreads for the simulated 
maps. We find the distributions of the \emph{WMAP} spectra consistent with 
the simulations, the only exception being bin 1 for the north/south 
(Galactic and ecliptic) hemispheres shown in the plots in the left 
column. For the northern hemisphere, this bin value is below 
the average value for all discs within the $1\sigma$ band (indicated 
by the left error bar). For the southern hemisphere however (the right error bar), the distribution seems consistent with simulations. 

\begin{figure}
\begin{center}
\leavevmode
\psfig {file=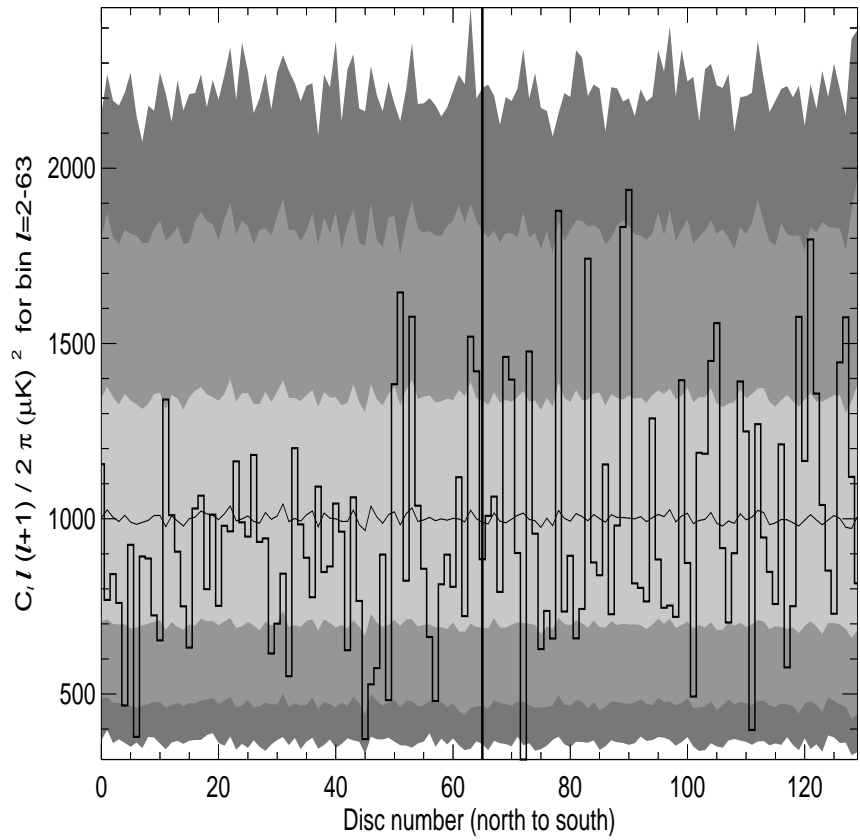,width=12cm,height=12cm}
\caption{The disc estimates of the bin $\ell=2-63$. The disc numbers correspond to the disc numbers in figure (\ref{fig:discnumbers}). The shaded zones indicate the 1, 2 and 3 sigma spread of the estimated bin for the given disc as found from Monte Carlo simulations.}
\label{fig:plotdiscs2bin0}
\end{center}
\end{figure}

In figure (\ref{fig:plotdiscs2bin0}) we investigate further the difference 
between the distribution of the power in bin 1 between the northern and 
southern hemisphere. The figure shows the power in this bin as a function 
of disc number (given in figure \ref{fig:discnumbers}). The discs 
on the left side (1-65) belong to the northern Galactic hemisphere 
and the discs on the right side (66-129) to the southern hemisphere. 
The figure confirms the observation that the large scale power 
estimated in the discs on the northern hemisphere is systematically 
lower than in the southern. Note that for the northern hemisphere 
only one fourth of the discs have values above the average whereas 
for the southern hemisphere the values are distributed equally above 
and below the average. Below the lower $1\sigma$ band there are only 
half as many discs in the southern hemisphere as in the northern. The 
significance of this north-south asymmetry will be studied further in 
section \ref{sect:hemis} using hemispheres for which the multipole 
resolution at large scales is higher.

\begin{figure}
\begin{center}
\leavevmode
\psfig {file=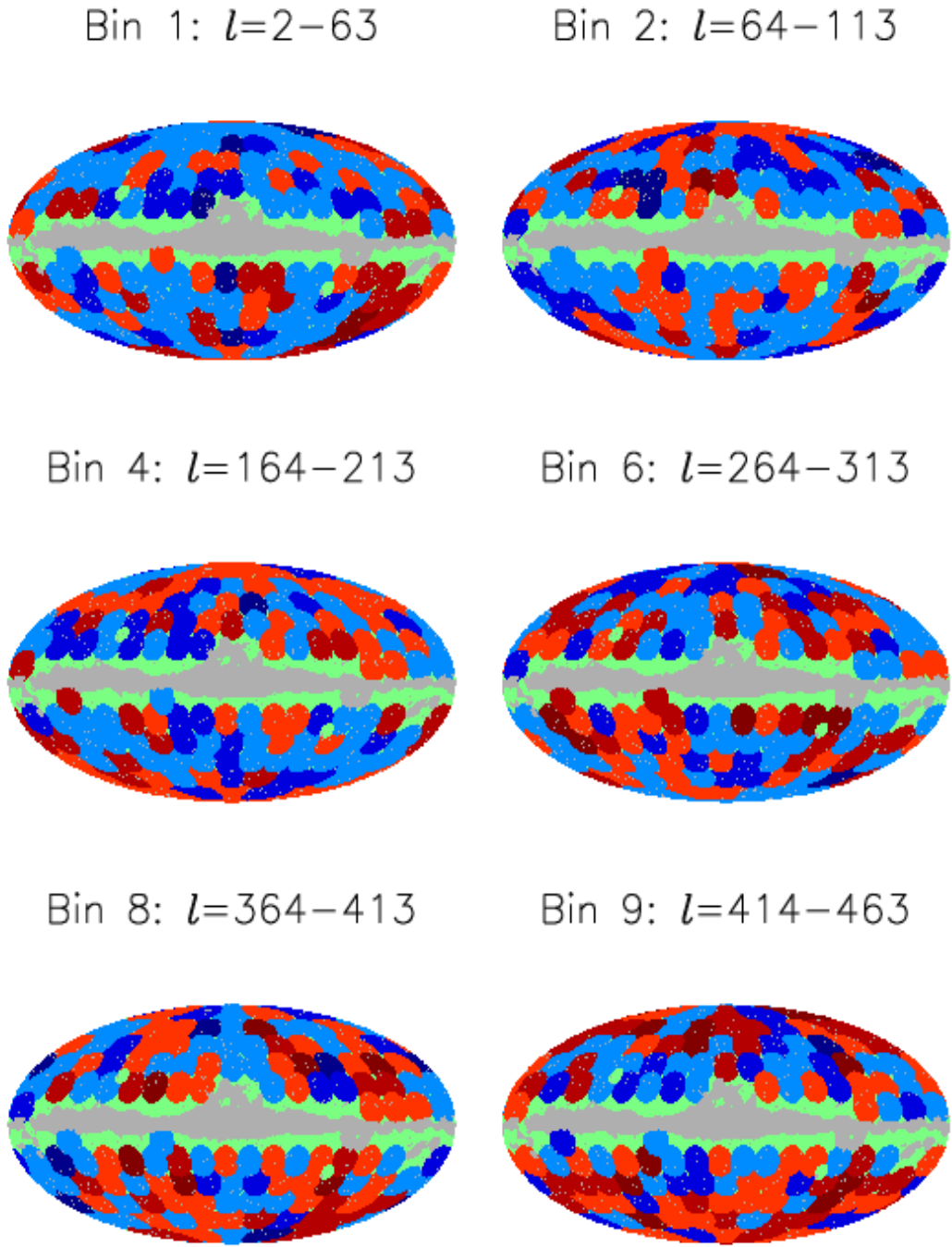,width=15cm,height=18cm}
\caption{The local power in the bin $\ell=2-63$ (upper left plot), $\ell=64-113$ (upper right plot), $\ell=164-213$ (middle left plot), $\ell=264-313$ (middle right plot) , $\ell=364-413$ (lower left plot) and the bin $\ell=414-463$ (lower right plot) estimated on $9.5^\circ$ discs. The discs with power above the average set by simulations have yellow (below 1 $\sigma$), red (between 1 and 2 $\sigma$) and dark red (above 2 $\sigma$) colour. The discs with power below the average have blue colour: light blue (above 1 $\sigma$), blue (between 1 and 2 $\sigma$) and dark blue (below 2 $\sigma$). The two green discs show the positions of the ecliptic poles.}
\label{fig:plotdisc4_b034}
\end{center}
\end{figure}

Figure (\ref{fig:plotdisc4_b034}) attempts to summarise 
the spatial distribution of the power spectra derived from the $9.5^\circ$ 
discs. We have plotted disc results for the largest angular scales (bin 1 and 2, upper row), 
the first peak (bin 4 and 6, middle row) and the first trough 
(bin 8 and 9, lower row). The colour coding indicates where these bins are above 
(light red:below 1 $\sigma$, red:between 1 and 2 $\sigma$, dark red: 
above 2 $\sigma$) and below (light blue:above 1 $\sigma$, blue:between 
1 and 2 $\sigma$, dark blue: below 2 $\sigma$) the average estabished by 
simulations.  The bin 1 results seems to suggest that the putative north-south 
asymmetry is actually associated more with the direction of the
ecliptic poles. The exact direction of maximum 
asymmetry will be investigated further in section (\ref{sect:hemis}). 

Bin 4 corresponds to the dent in the full 
sky spectrum as estimated by the \emph{WMAP} team.
There is some evidence that low amplitudes are concentrated close to
the ecliptic poles. It also appears that, for scales corresponding to 
the first trough in the power spectrum (bins 8 and 9), amplitudes are
relatively high close to the southern ecliptic pole. These features will be discussed 
further using the joint likelihood estimates in the next section. 
Apart from this, we cannot find other deviations of isotropy in 
the distribution of power looking at this figure. Studying the 
darkest colours (the $2\sigma$ outliers), we do not find that 
they appear in some particular position on the sky. Also, for 
164 discs we would expect about 7-8 discs at $2\sigma$. None of 
the maps have more $2\sigma$ outliers. We will test the isotropy 
more systematically in the next subsection. Finally, note that 
the disc-disc correlation matrix for one of the discs with an 
abnormally low value and one of the discs with an abnormally 
high value are shown in figure (\ref{fig:corcut}). We see that 
there is nothing anomalous with the correlations for these discs.

\subsection{Results of the analysis combining several discs}

\begin{figure}
\begin{center}
\leavevmode
\psfig {file=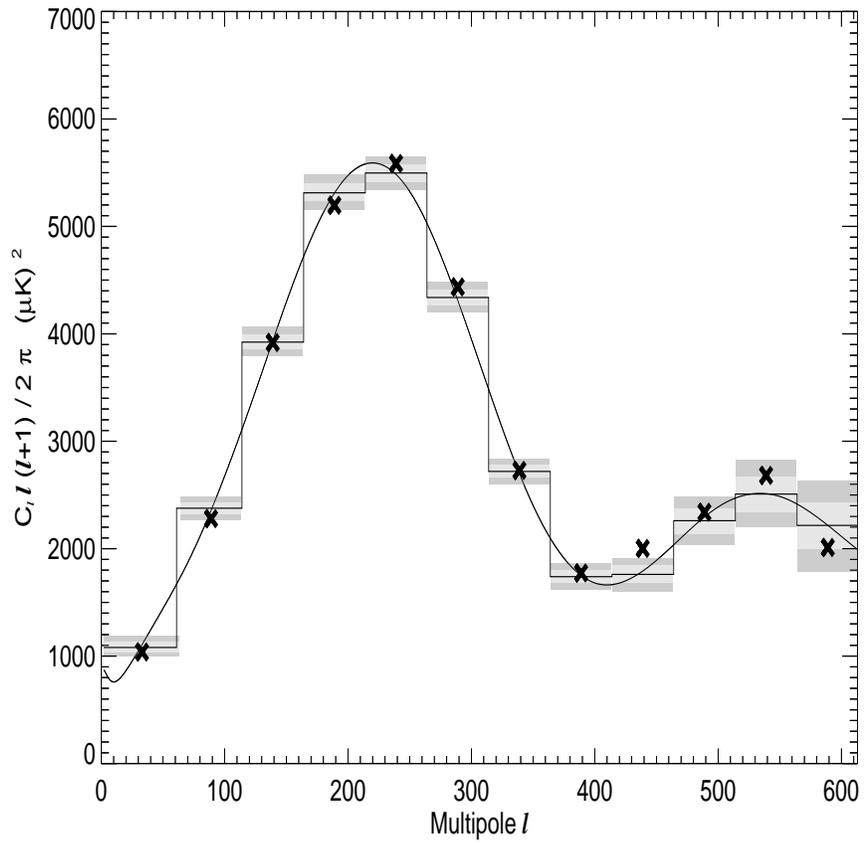,width=12cm,height=12cm}
\caption{The joint likelihood estimation of 130 discs of radius $9.5^\circ$. The histogram shows the mean and the shaded areas the 1 and 2 sigma levels from 1536 simulated maps with the same input spectrum. The crosses show the result of the same estimation procedure on the \emph{WMAP} data. }
\label{fig:10degjoint}
\end{center}
\end{figure}

We now proceed to a joint likelihood analysis of the
locally estimated spectra as described in HGH and outlined here in equation
(\ref{eq:jointlik}). 
Figure (\ref{fig:10degjoint}) shows the results 
from a likelihood estimation including all 130 discs. 
The gray bands indicate the 1 and 2 sigma levels derived from
simulations based on the \emph{WMAP} best fit spectrum, 
and the crosses show our result for \emph{WMAP}. 
Relative to the simulations, bin 4 (centred at $\ell=189$) 
is too low at the 1-sigma level, consistent with the \emph{WMAP} team
results (see figure \ref{fig:fullspectrum}).
Bin 9 (centred at $\ell=439$) is a strong outlier,
but aside from this our full sky estimate at this multipole resolution is in good agreement
with the best fit model.

\begin{figure}
\begin{center}
\leavevmode
\psfig {file=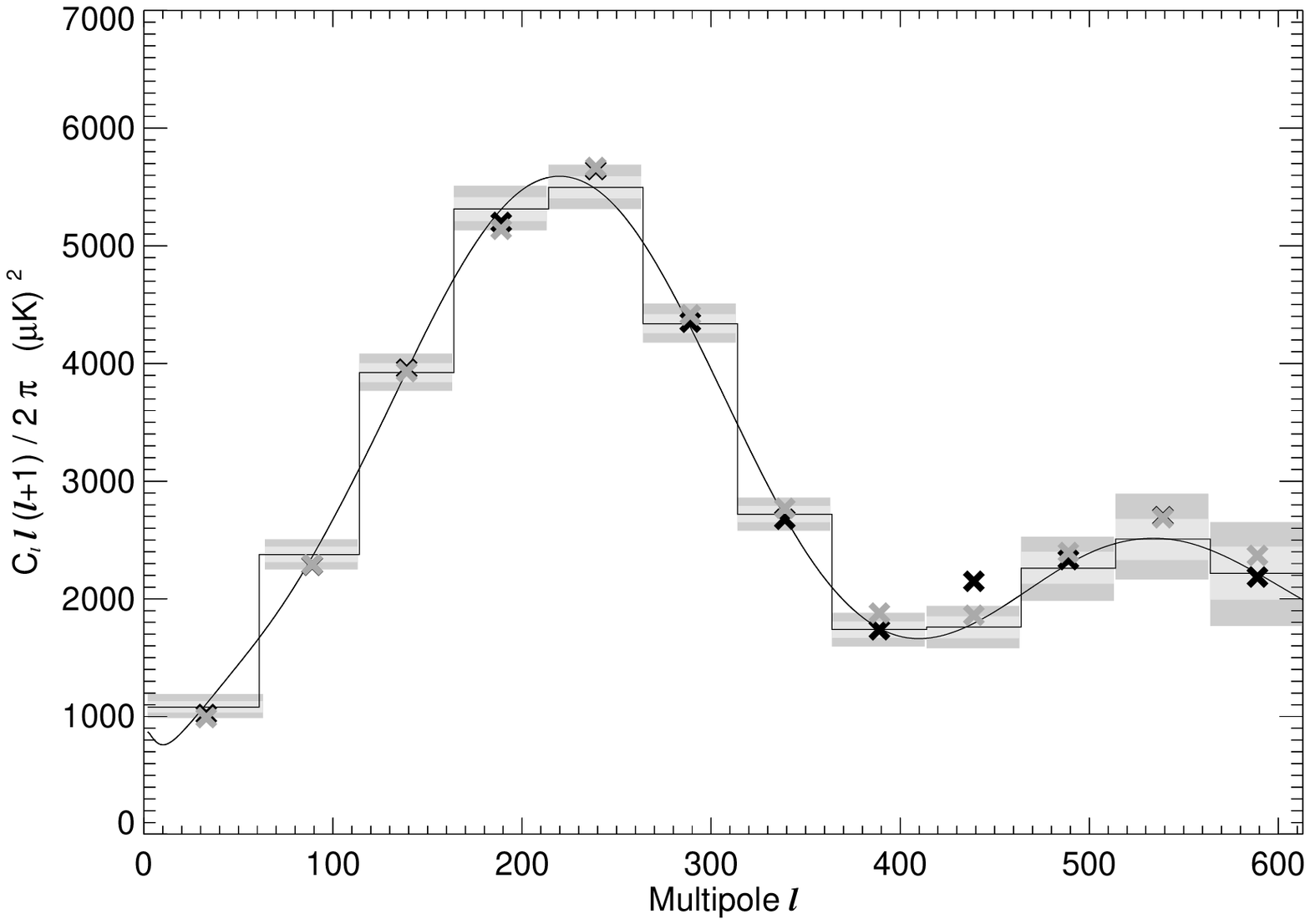,width=8cm,height=8cm}
\psfig {file=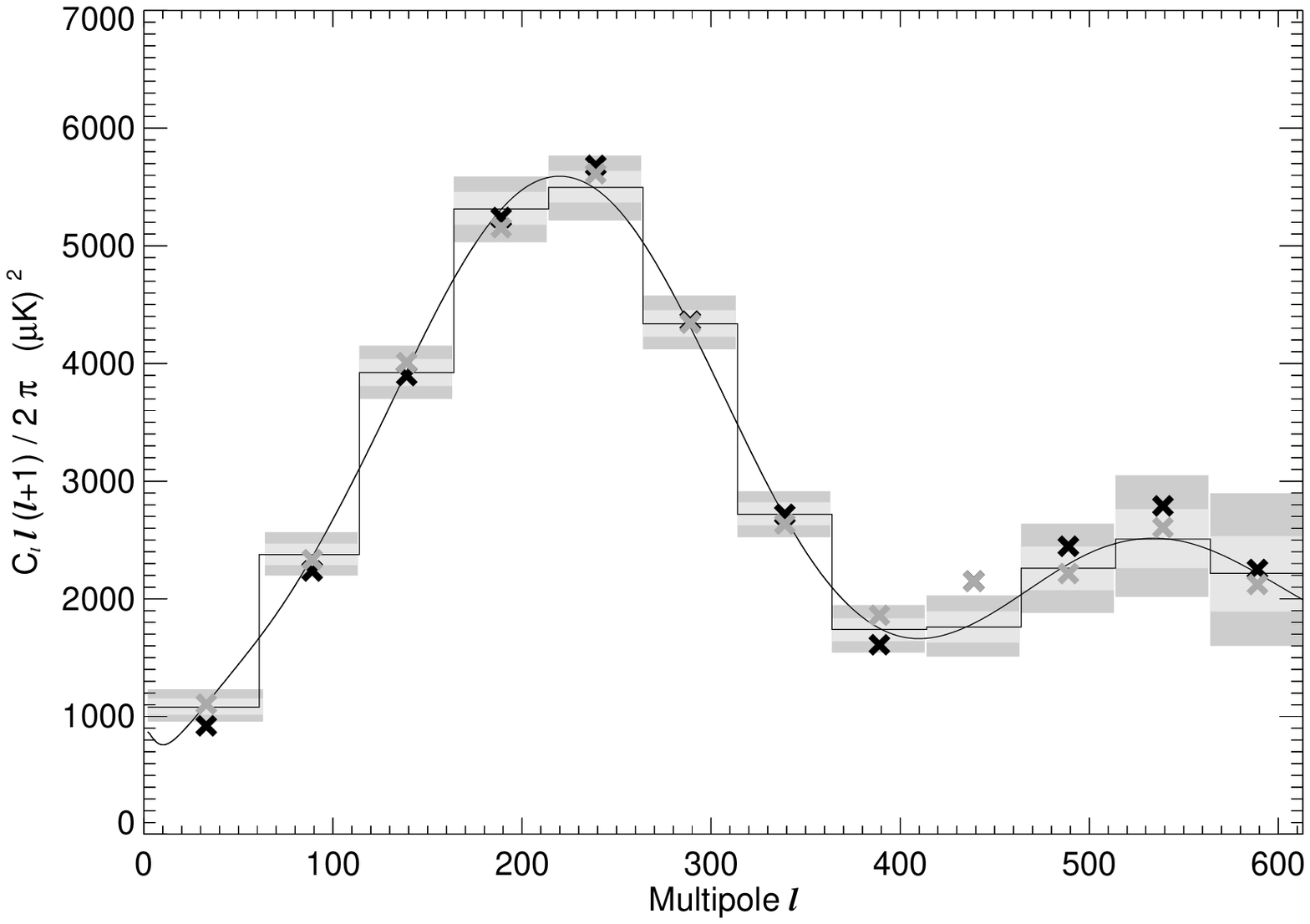,width=8cm,height=8cm}
\psfig {file=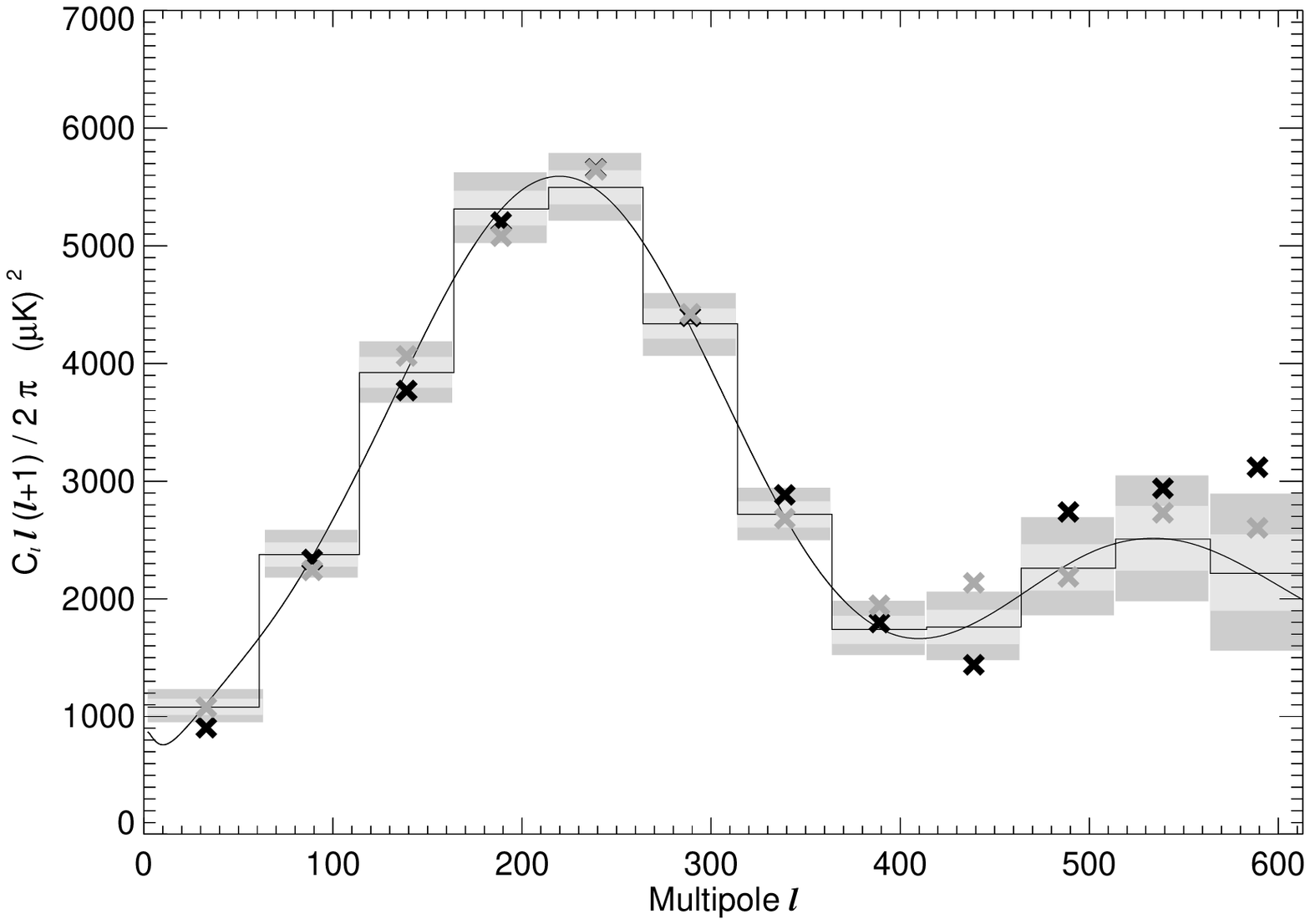,width=8cm,height=8cm}
\psfig {file=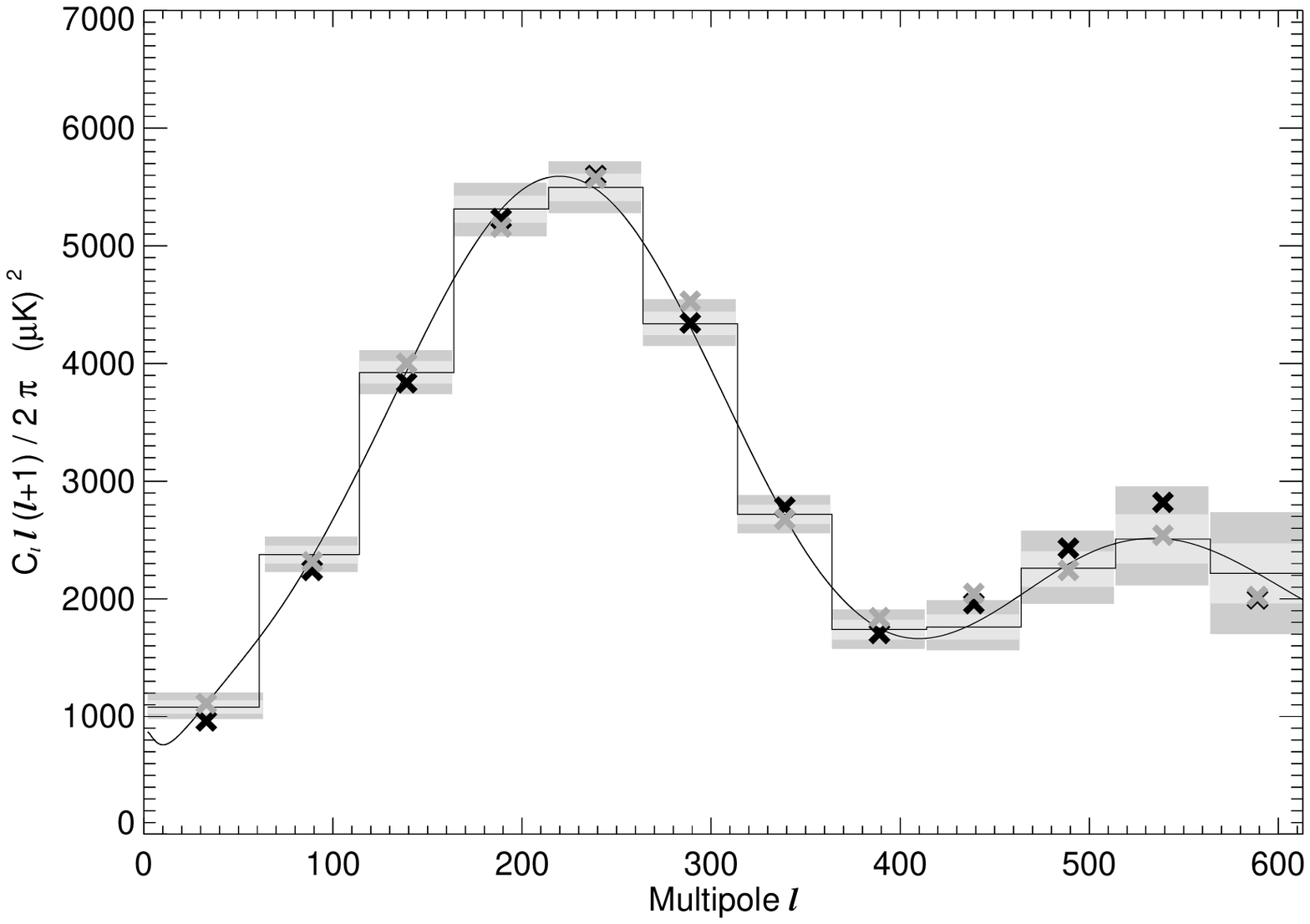,width=8cm,height=8cm}
\caption{Results of the power spectrum analysis in different parts of
the sky using a joint analysis of $9.5^\circ$ discs in the Galactic
reference frame (same as figure (\ref{fig:10degjoint}) on a limited
number of discs). The shaded zones indicate the 1 and 2 sigma errorbars calculated
in each plot for one of the spectra shown. The errorbars of the other
spectrum is very similar and is not shown. {\bf Upper left plot:}
black crosses:polar regions (discs 0-44, 66-110), gray crosses:
equatorial regions (discs 29-65, 95-129). {\bf Upper right plot:}
black crosses: polar north (discs 0-44), gray crosses: polar south
(discs 66-110). {\bf Lower left plot:} black crosses: equatorial region north(discs 29-65), gray crosses: equatorial region south (discs 95-129). {\bf Lower right plot:} black crosses: norther hemisphere (discs 0-65), gray crosses: southern hemisphere (discs 66-129).}
\label{fig:multi_galref}
\end{center}
\end{figure}

\begin{figure}
\begin{center}
\leavevmode
\psfig {file=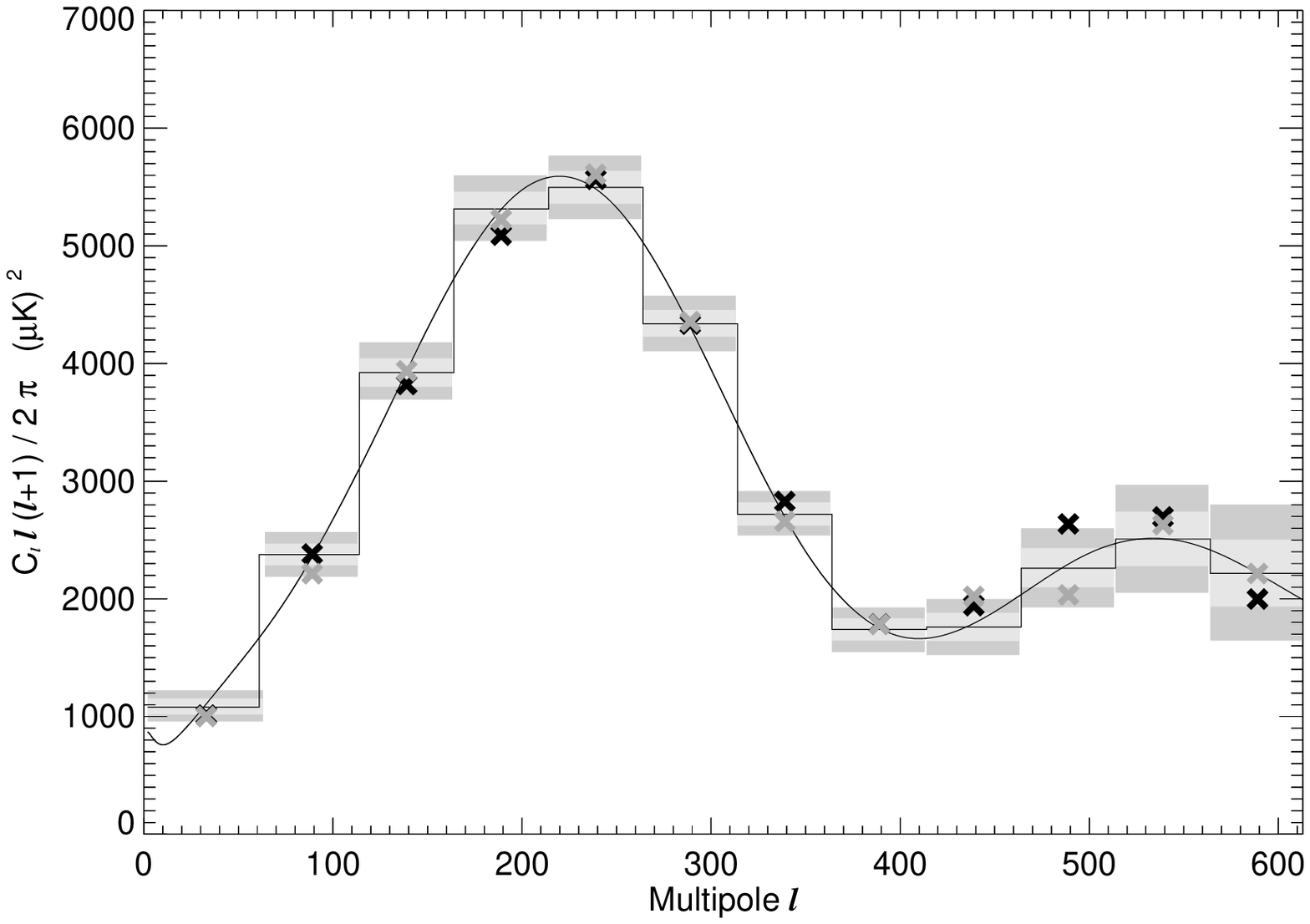,width=8cm,height=8cm}
\psfig {file=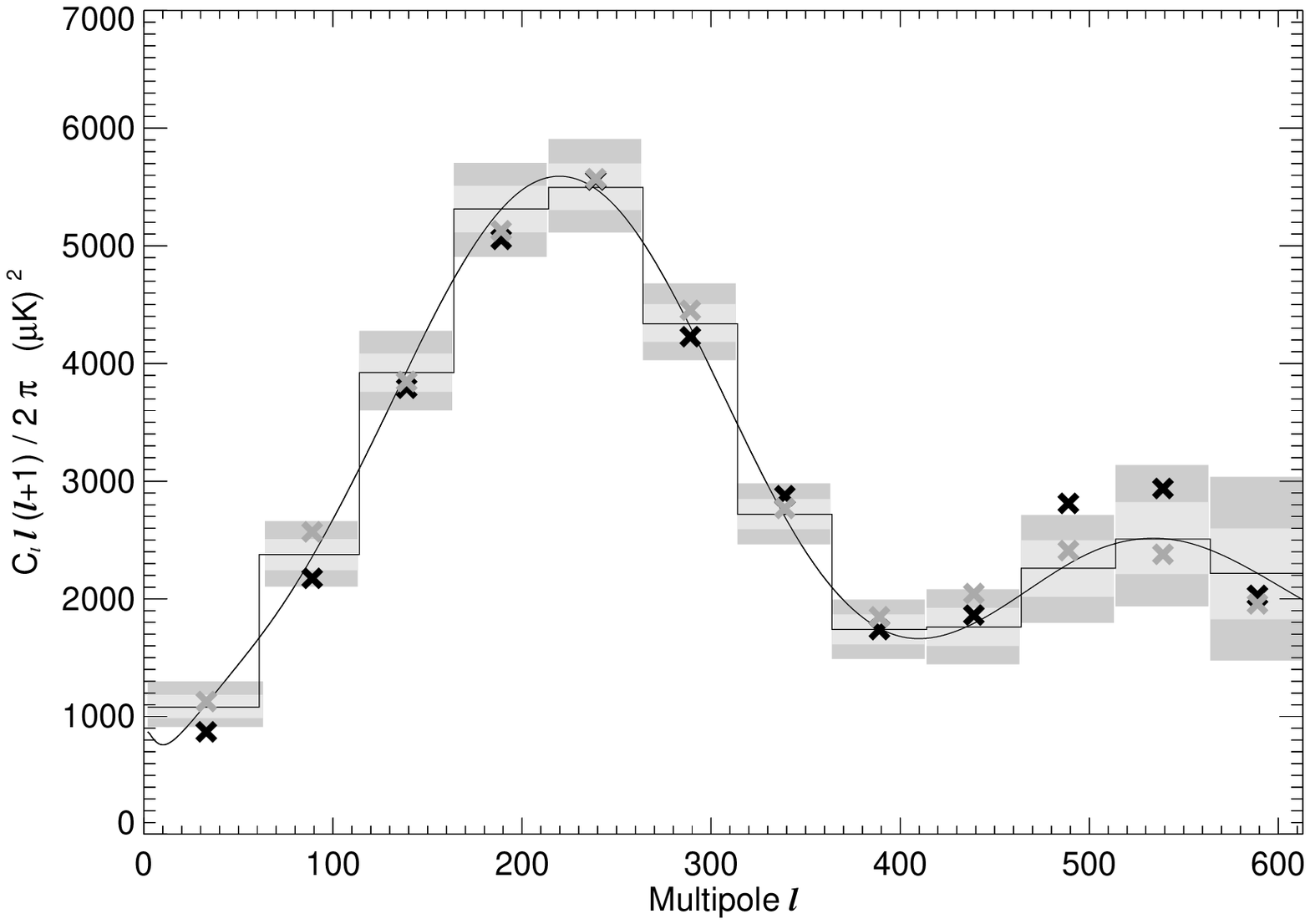,width=8cm,height=8cm}
\psfig {file=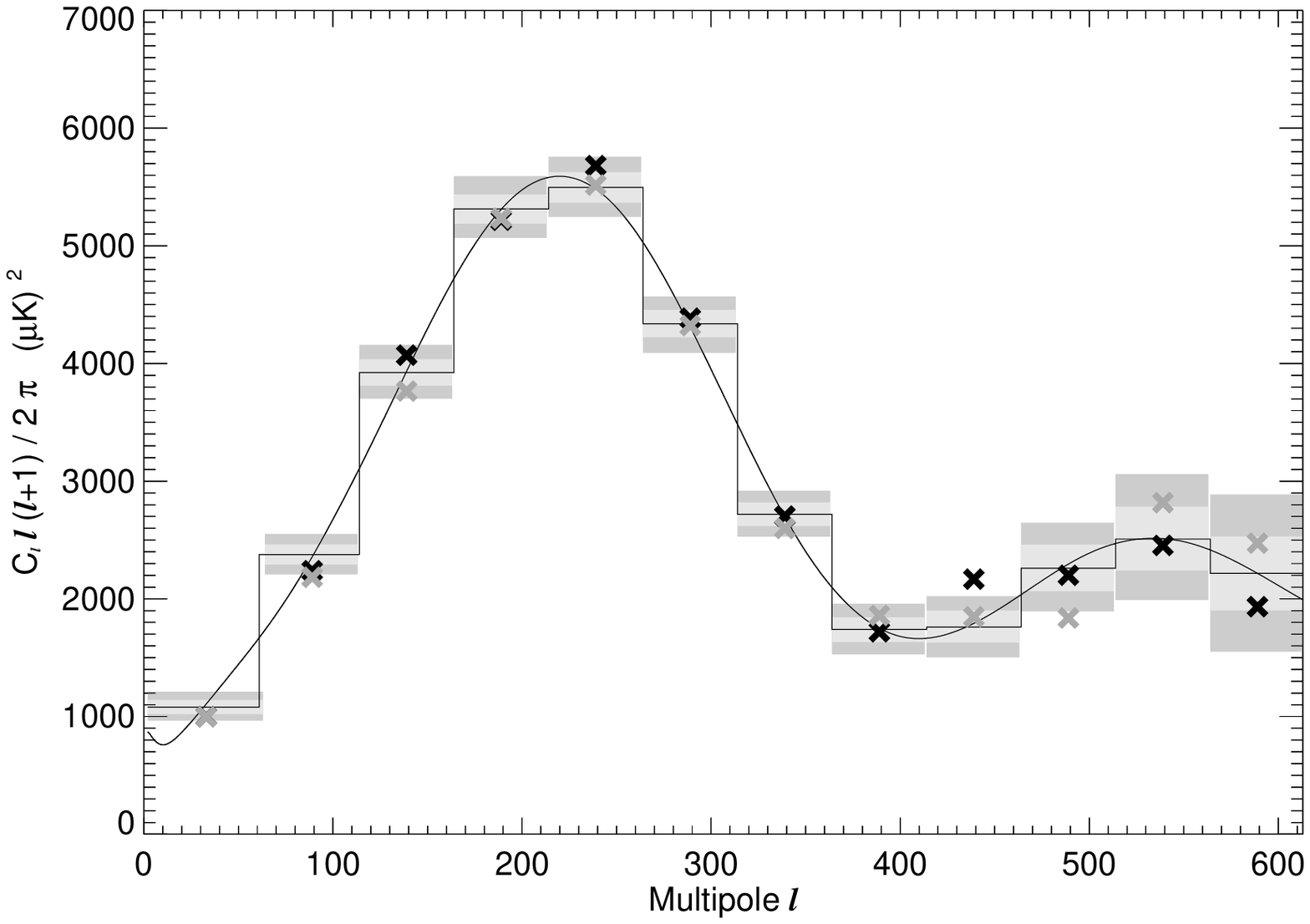,width=8cm,height=8cm}
\psfig {file=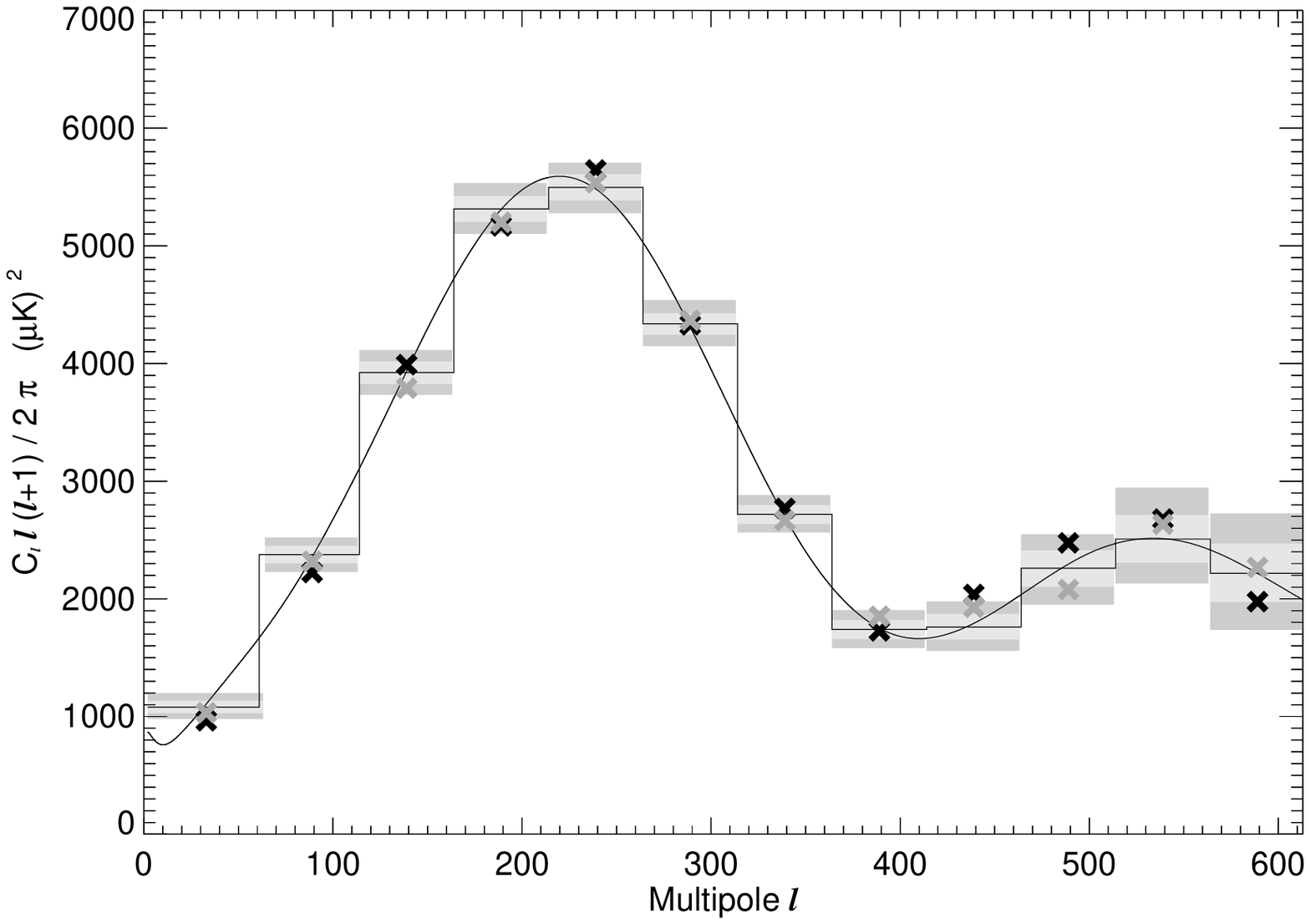,width=8cm,height=8cm}
\caption{Results of the power spectrum analysis in different parts of
the sky using a joint analysis of $9.5^\circ$ discs in the
ecliptic reference frame (same as figure (\ref{fig:10degjoint}) on a
limited number of discs). The shaded zones indicate the 1 and 2 sigma errorbars calculated in each plot for one of the spectra shown. The errorbars of the other spectrum is very similar and is not shown (this is not the case for the ecliptic polar regions versus ecliptic plane, but still the qualitative results are independent of the set of errorbars chosen). {\bf Upper left plot:} black crosses:polar regions (distance to the poles less than $50^\circ$), gray crosses: equatorial regions (distance from the poles larger than $50^\circ$). {\bf Upper right plot:} black crosses: polar north (distance to the north pole closer than $50^\circ$), gray crosses: polar south. {\bf Lower left plot:} black crosses: equatorial region north (distance from the poles larger than $50^\circ$ but inside the northern hemisphere), gray crosses: equatorial region south. {\bf Lower right plot:} black crosses: northern hemisphere, gray crosses: southern hemisphere.}
\label{fig:multi_eclref}
\end{center}
\end{figure}

Nevertheless, in the previous section we found tentative evidence
for some asymmetry in the large-scale power distribution. In
order to confirm this result, and as a test of
residual Galactic foregrounds or systematic effects, 
we also performed a joint likelihood analysis of
several partitions of discs selected over larger areas of the sky 
to probe the variation of the power spectrum in the Galactic and 
ecliptic reference frames. In the Galactic reference frame we have 
elected to test the effect on the spectrum when excluding discs $45-65$ 
and $111-129$ (see fig. \ref{fig:discnumbers}) which are the two 
rows closest to the Galactic plane. The surviving discs 
on the northern and southern hemispheres were analysed both separately 
and together. The spectrum for the three rows of discs
closest to the Galactic plane were also computed, for the northern and southern 
hemisphere separately and together. The results are collected 
in figure (\ref{fig:multi_galref}). A corresponding analysis was then
performed in the ecliptic reference frame, excluding those discs with
nominal centres further than $50^\circ$ from the ecliptic poles. The results are 
shown in figure (\ref{fig:multi_eclref}).

The most striking feature in these plots, is the north-south asymmetry 
visible both in the Galactic and ecliptic frame of reference on
large angular scales.
Moreover one can see that the low value of bin 4 (centred at 
$\ell=189$) seems to be strongest (below the $1\sigma$ band) 
in the ecliptic polar regions (both north and south) whereas 
it is weaker (inside of $1\sigma$) in the equatorial 
region (both north and south). This seems to confirm our observation 
in figure (\ref{fig:plotdisc4_b034}) and a similar observation 
made by the \emph{WMAP} team \cite{hinshaw}. However, comparing the ratio 
of this bin value from the ecliptic polar and ecliptic plane regions
to simulations, we find that 
it is not a significant difference in power.
In fact most of the simulations had a higher ratio between 
the ecliptic polar regions and ecliptic plane for this bin. At 
the multipole resolution obtained by $9.5^\circ$ discs we 
find bin 4 to be consistent with the best fit model and isotropically 
distributed on the sky.

Bin 9 (centred at $\ell=439$) appears to be of 
particularly high amplitude both in Galactic polar regions, where 
one would expect foregrounds to be less dominant, 
and in the southern ecliptic polar regions.
There are no obvious systematic effects or foregrounds which could
cause higher power at this particular scale over such diverse regions 
of the sky. One plausible solution, however, is connected to the fact
that the white noise model adopted is an approximation, i.e. if 
the estimated noise level is slightly too low, this would explain 
the outlier in bin 9 and also the fact that the multipole bins 
above $\ell>400$ are systematically above the best fit model. 
Furthermore, if this is the case then it is necessary for the
\emph{WMAP} team to provide a better model for the
noise on these angular scales for future work.
Alternately, a similar enhancement in power could result if 
the unresolved foreground due to point sources has not 
been removed correctly. For source removal, we have adopted the 
procedure described in \cite{hinshaw} which is intended for full 
sky measurements. If the distribution of unresolved sources is not 
completely uniform, applying this procedure to the local disc spectra 
could introduce errors at high multipoles. However, this is unlikely
in practise as the point source distribution is essentiall poissonian
in nature.

The conclusion to be drawn from the small disc analysis is that there is little
evidence for departures from isotropy on small angular scales. However, we found a possible 
north-south asymmetry on the largest scales, which we will return to later in section (\ref{sect:hemis}).

\section{Results on medium sized discs}

\label{sect:medium}

In this section we aim to test the isotropy of the power distribution on the
sky with greater spectral resolution by using discs of radius $19^\circ$. 
In particular, we take advantage of the higher resolution 
in multipole space to place more bins close to the first Doppler 
peak in order to test whether its location in $\ell$-space exhibits any
spatial dependence as traced by the disc positions.

\subsection{The analysis}
\begin{figure}
\begin{center}
\leavevmode
\psfig {file=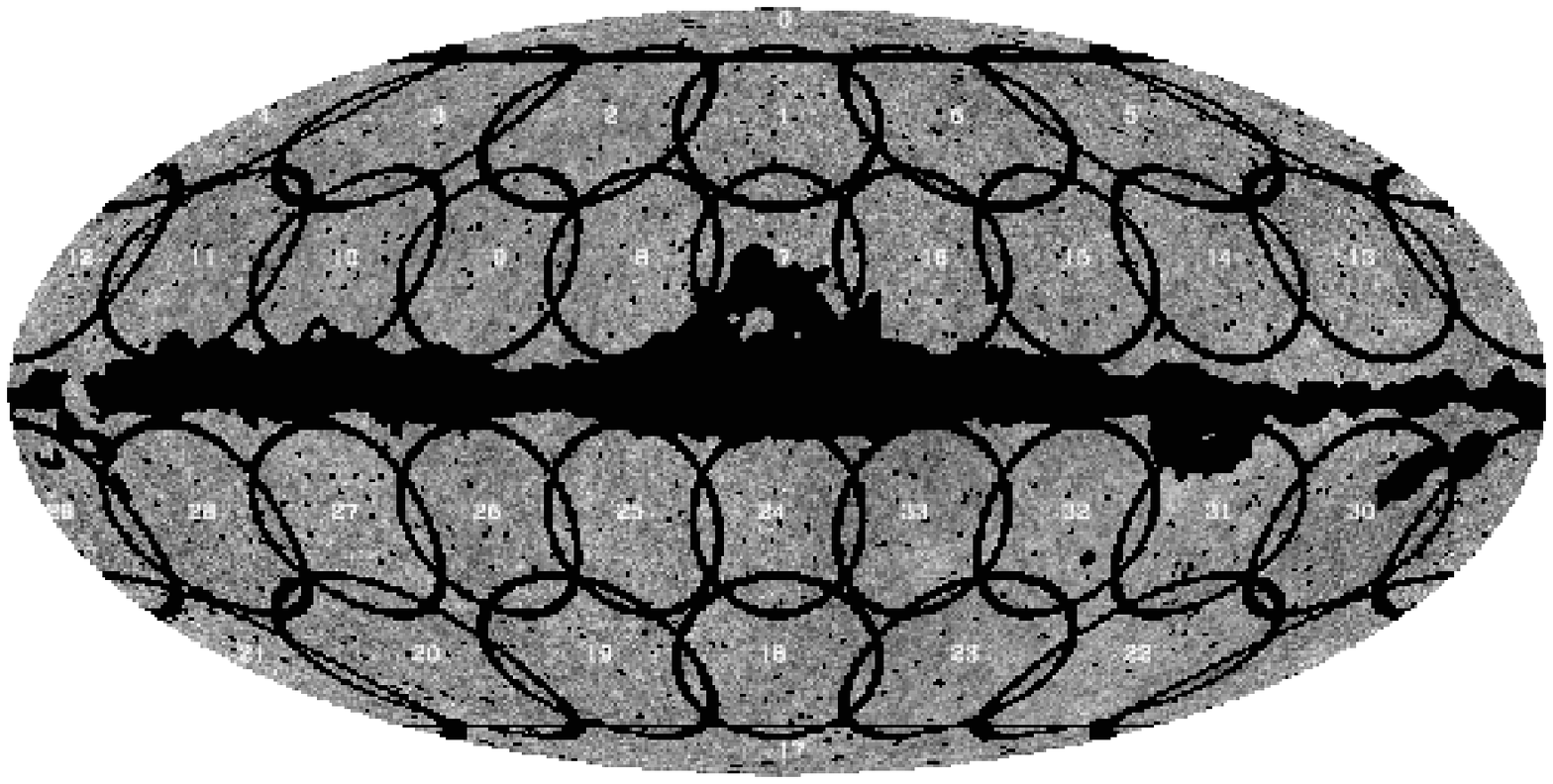,width=10cm,height=7cm}
\caption{The position and numbering of the $19^\circ$ discs on which the local power spectra have been calculated.}
\label{fig:discnumbers2}
\end{center}
\end{figure}

Using the same criteria as for the small discs, we have positioned 
34 discs of radius $19^\circ$
on the sphere as shown in figure (\ref{fig:discnumbers2}). 
The distance between the discs and 
Galactic poles is $35^\circ$ for the first ring of discs and
$65^\circ$ for the second ring. Again, the distance between discs in 
the $\phi$ direction is the
same for all discs at a given Galactic latitude. The number of input 
samples of the pseudo spectrum was chosen to be $N_\textrm{sam}=133$ 
and the number of estimated bins $N_\textrm{bin}=16$. We have positioned
smaller and more bins around the first peak in order to constrain its
location in $\ell$-space more accurately.  The bins are defined in table 
(\ref{tab:mediumbins}). We do not show the disc-disc and bin-bin 
correlation matrices since these are very similar those for
the small discs. Even for the small bins around the first peak, the 
correlations are limited to a small anti-correlation between the neighbouring bins.

\begin{center}
\begin{table}
\caption{The power spectrum bins for the $19^\circ$ disc estimates.\label{tab:mediumbins}}

%\begin{ruledtabular}
\begin{tabular}{|l|c|c|c|c|c|}
\hline
bin & 1 & 2 & 3 & 4 & 5 \\
\hline
range & $\ell=2-63$&$\ell=64-113$&$\ell=114-163$&$\ell=164-188$&$\ell=189-213$\\
centre & $\ell=33$&$\ell=89$&$\ell=139$&$\ell=176$&$\ell=201$\\
\hline
bin & 6 & 7 & 8 & 9 & 10 \\
\hline
range & $\ell=214-238$&$\ell=239-263$&$\ell=264-288$&$\ell=289-313$&$\ell=314-363$\\
centre & $\ell=226$&$\ell=251$&$\ell=276$&$\ell=301$&$\ell=339$\\
\hline
\hline
bin & 11 & 12 & 13 & 14 & 15 \\
\hline
range & $\ell=364-413$&$\ell=414-463$&$\ell=464-513$&$\ell=514-563$&$\ell=564-613$\\
centre & $\ell=389$&$\ell=439$&$\ell=489$&$\ell=539$&$\ell=589$\\
\end{tabular}
%\end{ruledtabular}
\end{table}
\end{center}

\subsection{Results on the \emph{WMAP} data}

\begin{figure}
\begin{center}
\leavevmode
\psfig {file=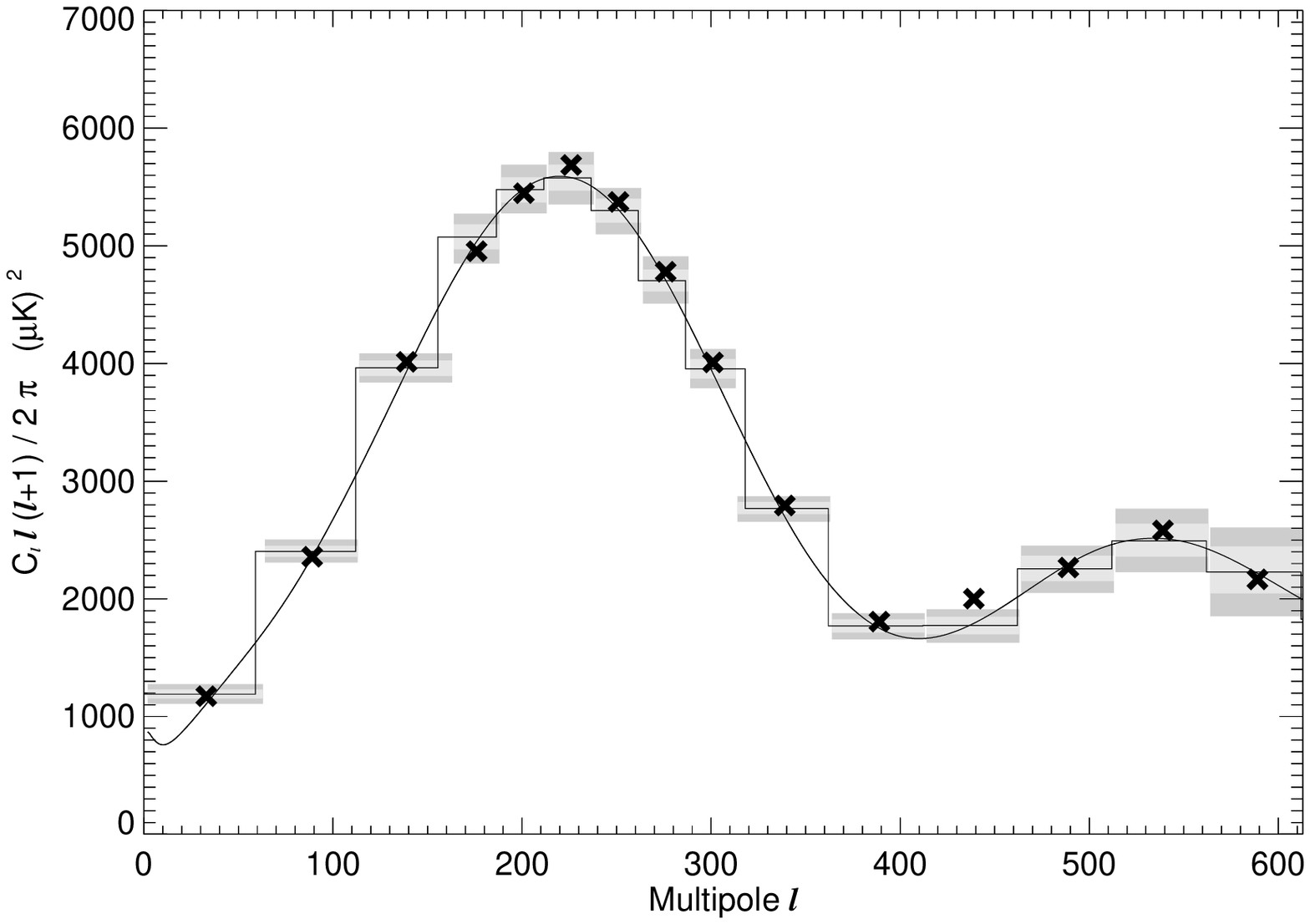,width=12cm,height=12cm}
\caption{The full sky \emph{WMAP} power spectrum from a joint likelihood estimation of 34 discs of radius $19^\circ$. The histogram shows the binned  \emph{WMAP} best fit running index spectrum, the shaded areas the 1 and 2 sigma levels from 512 simulated maps with the same input spectrum. The crosses show the result of the joint disc estimation procedure on the \emph{WMAP} data. }
\label{fig:20degjoint}
\end{center}
\end{figure}

Figure (\ref{fig:20degjoint}) shows the result of the joint 
likelihood estimation of the power spectrum by combining all 34 discs. We find 
that the result is consistent with the small disc result (figure 
\ref{fig:10degjoint}). Specifically,  there is a dip at the  $1\sigma$ level at
bin 4 centred at $\ell=176$ and the spectrum is slightly too high 
after the first peak. There is also a strong outlier at bin 12 
centred on $\ell=439$, 
but otherwise the inferred full sky spectrum is consistent with the \emph{WMAP}
best-fit model at this higher multipole resolution. 
Nevertheless, we proceed to assess the significance of the spatial
distribution of the bin values and the position of outliers
as a test of isotropy.

\begin{figure}
\begin{center}
\leavevmode
\psfig {file=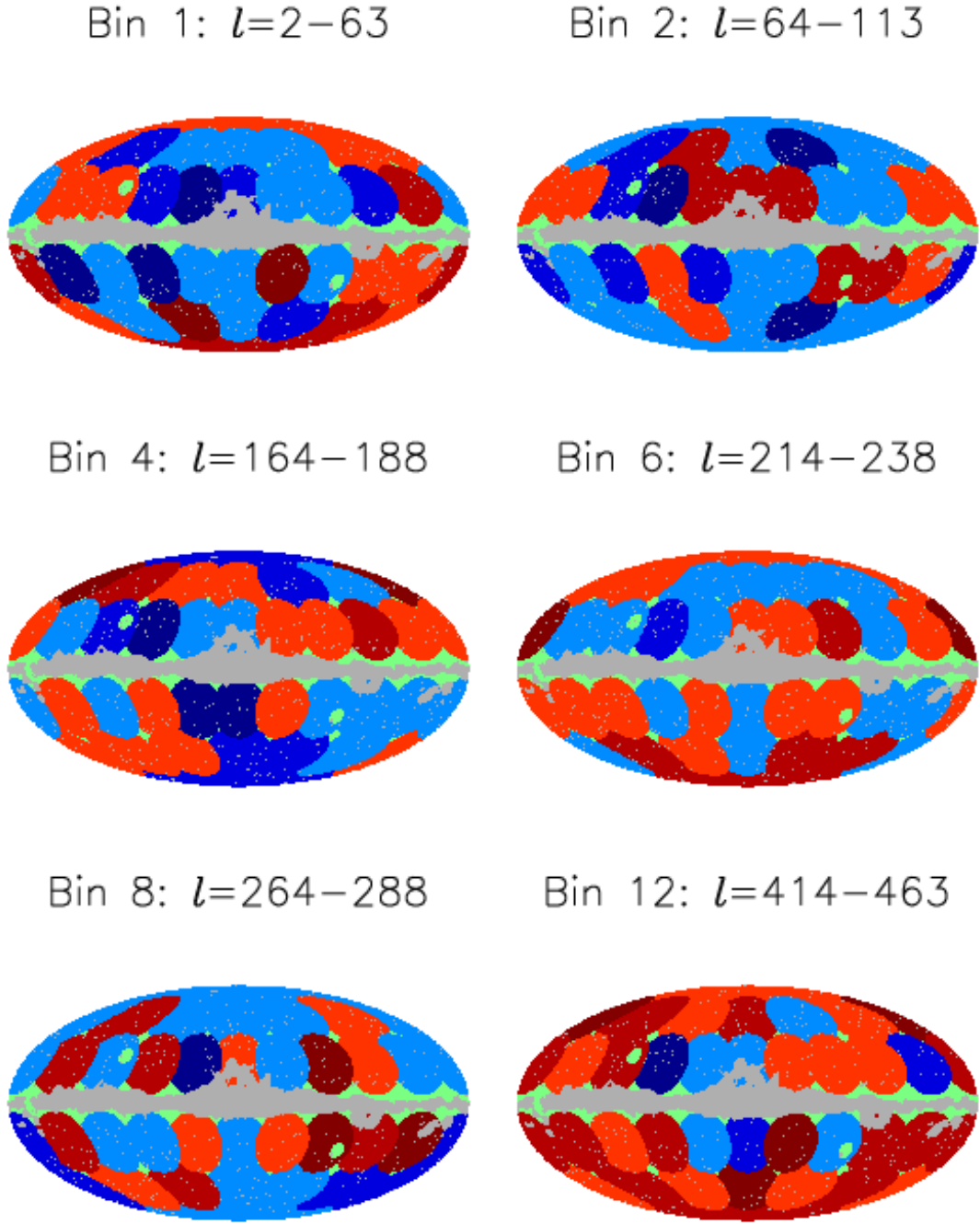,width=15cm,height=18cm}
\caption{The local power in the bin $\ell=2-63$ (upper left plot), $\ell=64-113$ (upper right plot), $\ell=164-188$ (middle left plot), $\ell=214-238$ (middle right plot) , $\ell=264-288$ (lower left plot) and the bin $\ell=414-463$ (lower right plot) estimated on $19^\circ$ discs. The discs with power above the average set by simulations are red and the discs with power below the average are blue.The intensity of the colour indicates whether the power is within $1\sigma$ (light), between $1$ and $2\sigma$ (medium) and outside of $2\sigma$ (dark). The two green discs show the positions of the ecliptic poles.}
\label{fig:plotdisc4_20deg_b03}
\end{center}
\end{figure}

In figure (\ref{fig:plotdisc4_20deg_b03}) we show the positional 
dependence of the first two bins, the 'dip' at bin 4 ($\ell=164-188$), 
bin 6 ($\ell=214-238$) and bin 8 ($\ell=264-288$) which are at or just
follow the first peak, and bin 12 ($\ell=414-463$) in the first
trough. It is clear by comparison with 
figure (\ref{fig:plotdisc4_b034}) that the distribution of these 
bins is largely consistent with the small disc results. Note however that 
bin 4 has three negative $2\sigma$ outliers close to the Galactic plane. 
This is also the case for bin 8 which has four $2\sigma$ outliers close to 
the Galactic plane. In our simulations, we find three $2\sigma$ outliers 
in the Galactic region for bin 4 in $10\%$ of the maps, so this feature
should not be regarded as highly significant. 
However, four $2\sigma$ outliers in this region for bin 8 is found 
only in $2\%$ of the maps.

\begin{figure}
\begin{center}
\leavevmode
\psfig {file=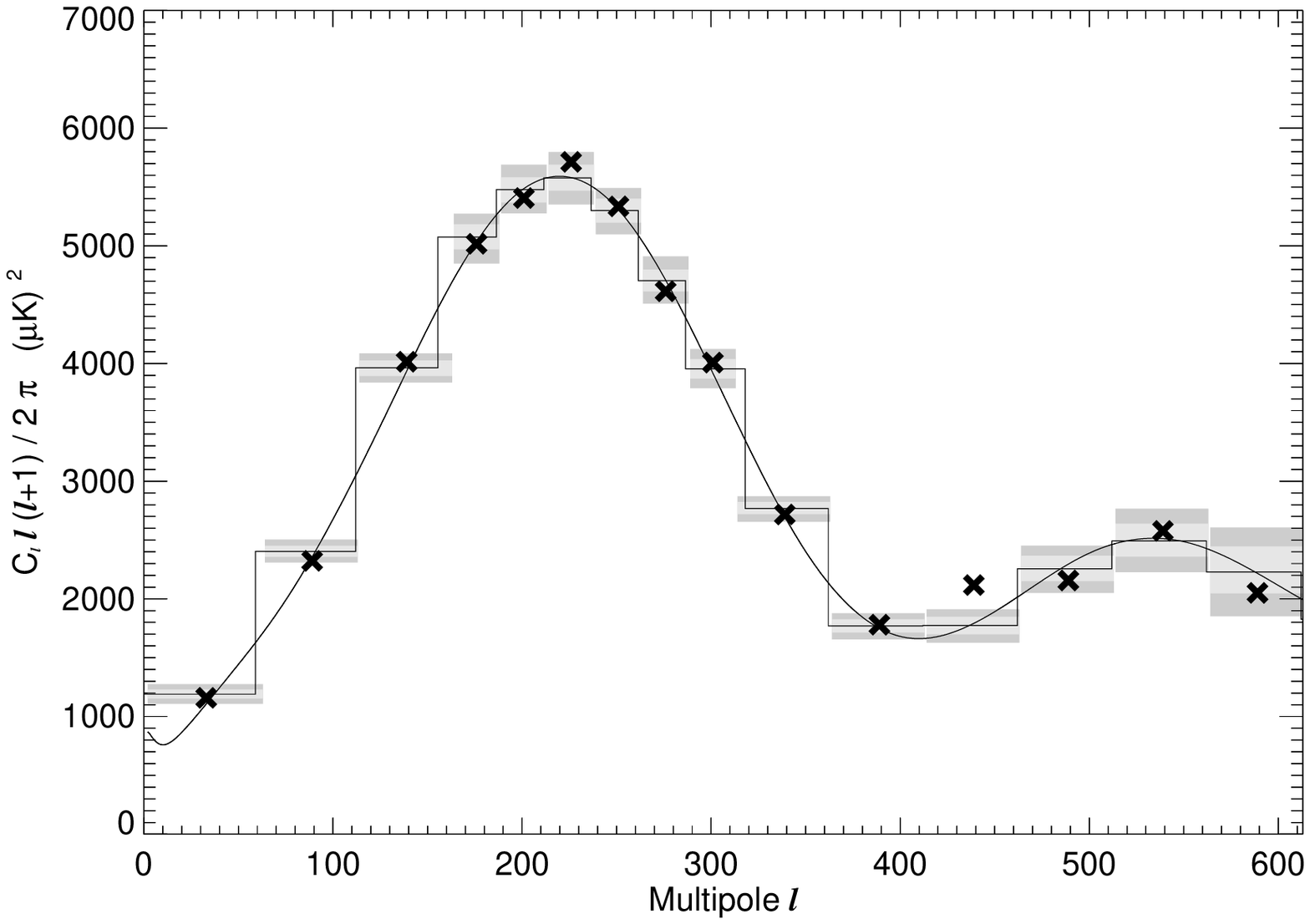,width=12cm,height=12cm}
\caption{The full sky \emph{WMAP} power spectrum from a joint likelihood
 estimation of 19 discs of radius $19^\circ$. The histogram shows the binned  \emph{WMAP} best fit running index spectrum, the shaded areas the 1 and 2 sigma levels from 512 simulated maps with the same input spectrum. The crosses show the result of the joint disc estimation procedure on the \emph{WMAP} data. }
\label{fig:20degjoint_exl}
\end{center}
\end{figure}

Assuming that these deviations might have other than a cosmological
origin, it is difficult to speculate what their origin might be.
Certainly, errors in Galactic foreground modelling could be a plausible
source of the outliers, either due to under- or over-prediction
of the template-based contribution in these regions, resulting in either over-
or under-subtraction of power. Whatever the cause
of the features, we examine their impact on the power spectrum estimation
by excluding the discs associated with the $2\sigma$ outliers in bins
4 and 8, and repeating the joint estimate of the power spectrum using
the remaining $19^\circ$ discs. 
In figure (\ref{fig:20degjoint_exl}) we see the resulting power spectrum. 
Interestingly, the dent before the first Doppler peak has now disappeared
and a slightly higher peak amplitude is preferred.

\begin{figure}
\begin{center}
\leavevmode
\psfig {file=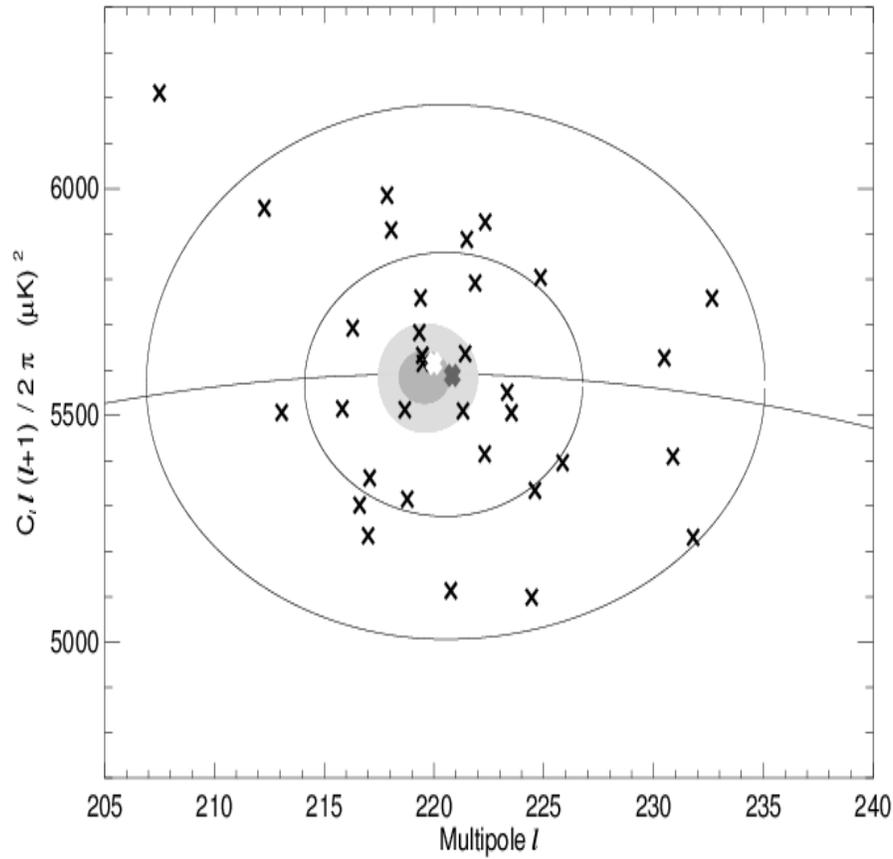,width=12cm,height=12cm}
\caption{The position and amplitude of the first peak. The solid line shows the best fit \emph{WMAP} power spectrum. The shaded zones show the result of the full sky power spectrum estimation using the joint likelihood estimation procedure of 34 discs with radius $19^\circ$. The zones indicate the $1$ and $2\sigma$ spread of the peak position over 512 simulations, and the dark bold cross shows the result of the same procedure applied to the \emph{WMAP} data. The white bold cross shows the peak position when some discs showing indications of Galactic foreground contamination are excluded. The crosses show the peak positions estimated on individual discs using the \emph{WMAP} data, and the two circles show the corresponding $1$ and $2\sigma$ spread over 34 discs in 512 simulations. Note that individual discs may have larger $1$ and $2\sigma$ contours and are therefore not automatically outside of their individual contours even if they are outside of the circles in the plot.}
\label{fig:peak1}
\end{center}
\end{figure}

\begin{figure}
\begin{center}
\leavevmode
\psfig {file=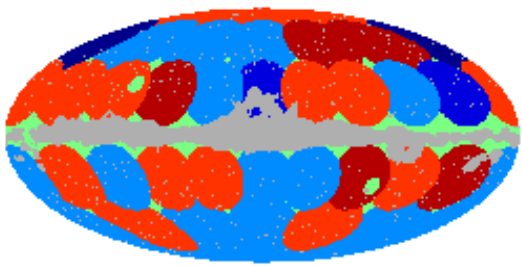,width=7.5cm,height=12cm}
\caption{The peak position (upper plot) and peak amplitude (lower
plot) estimated on $19^\circ$ discs. Red colour indicates that for the given disc, the estimate on \emph{WMAP} was above the mean of simulations and blue colour shows that the \emph{WMAP} estimate was below the mean. The significance is indicated by the intensity of the colour: light means within $1\sigma$, medium means between $1$ and $2\sigma$ and dark means outside of $2\sigma$. The two green dots show the position of the ecliptic poles.}
\label{fig:discplot_peak}
\end{center}
\end{figure}

Finally we exploit the higher multipole space resolution to estimate 
the position and amplitude of the first peak in different directions.
Following \cite{wmap_peak}, we approximate the first peak as a Gaussian. 
Using the estimated power spectra we applied a maximum likelihood 
procedure to find the best fit peak position and amplitude. Figure 
(\ref{fig:peak1}) shows the result of the peak position estimates on 
512 simulated maps. The shaded zones indicate the $1$ and $2\sigma$ 
spread of the peak position. The circles show the mean $1$ and $2\sigma$ 
contours for the individual disc estimates. Note that any specific
disc will have significance levels different from the mean levels shown by 
the circles. The crosses show the peak position on the individual 
$19^\circ$ discs using the \emph{WMAP} data, and the bold dark cross 
shows the joint result of the 34 discs of the \emph{WMAP} data. We 
find the estimated peak position, amplitude and the distribution of 
the position and amplitude in individual discs consistent with the 
best fit model. The white bold cross shows the position of the first 
peak when excluding those discs with possible residual Galactic 
contamination. In this case, the agreement with the 
model spectrum is improved. In figure (\ref{fig:discplot_peak}) we 
show the discs where the peak position and amplitude are above (red) or 
below (blue) the mean, and where they are within the $1\sigma$ level (bright 
colour), between $1$ and $2\sigma$ (medium) and outside of  $2\sigma$ (dark 
colour). We find no particular positional dependence of the first peak position and amplitude. 

In conclusion, the power on small scales appears to be
isotropically distributed, although there is some evidence 
that contamination of several discs close to the Galactic
plane may be associated with the dent in the power spectrum
close to the first peak.
When these discs are excluded from the analysis, the peak amplitude
is somewhat higher.

\section{Results of the spectra estimated on hemispheres}

\label{sect:hemis}

In previous sections we have presented evidence for a possible
asymmetry in the large angular scale power present in the \emph{WMAP}
data. In order to pursue this effect further, and to allow
greater spectral resolution in the low order multipole bins,
we have computed the power spectrum amplitudes determined for the
northern and southern hemispheres as defined in a particular
coordinate system.
Using the small and medium sized discs, 
the multipole resolution was limited and the asymmetry could 
only be studied using a relatively large multipole binning range. By estimating the 
power spectrum on hemispheres the positional dependence of smaller 
multipole intervals can be investigated in order to check if the asymmetry 
is related to specific angular scales. In addition, we will study the two strong 
outliers at $\ell=21$ and $\ell=39$ found in the results of the global analysis
obtained by the \emph{WMAP} team.

\subsection{The analysis}

In order to check the positional dependence of the lowest multipoles 
we have computed the power spectra on hemispheres defined 
with respect to those coordinate frames for which the north pole pierces the centre 
of the discs in
fig. (\ref{fig:discnumbers}), plus a further set of 34 discs
which were previously rejected in the small disc analysis since they
were sufficiently
compromised by the Kp2 mask. Clearly, all 164 disc positions
can now be used since they are merely used to define reference directions.
For each hemisphere, bandpower estimates were computed for the first 64 multipoles 
in bins of width 3, defined by
$\ell_b\epsilon\{\ell_\mathrm{min},\ell_\mathrm{max}\}$ where
$\ell_\mathrm{min}=3b-1$ and $\ell_\mathrm{max}=3b+1$. 
The spectra were then compared to a Monte-Carlo ensemble of hemisphere 
spectra estimated on simulated maps. 

\subsection{Results on the northern and southern hemispheres}

\begin{figure}
\begin{center}
\leavevmode
\psfig {file=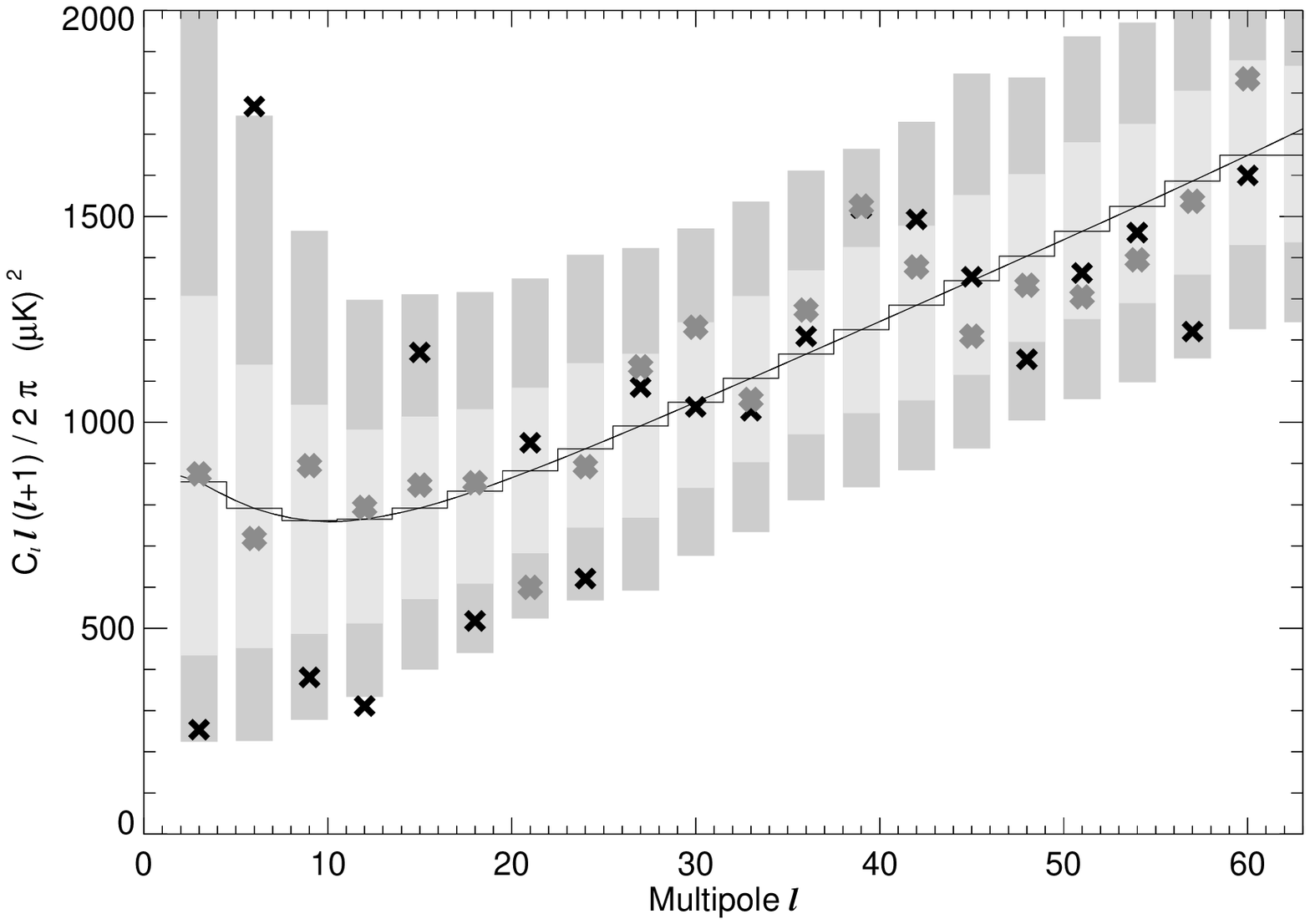,width=7cm,height=7cm}
\psfig {file=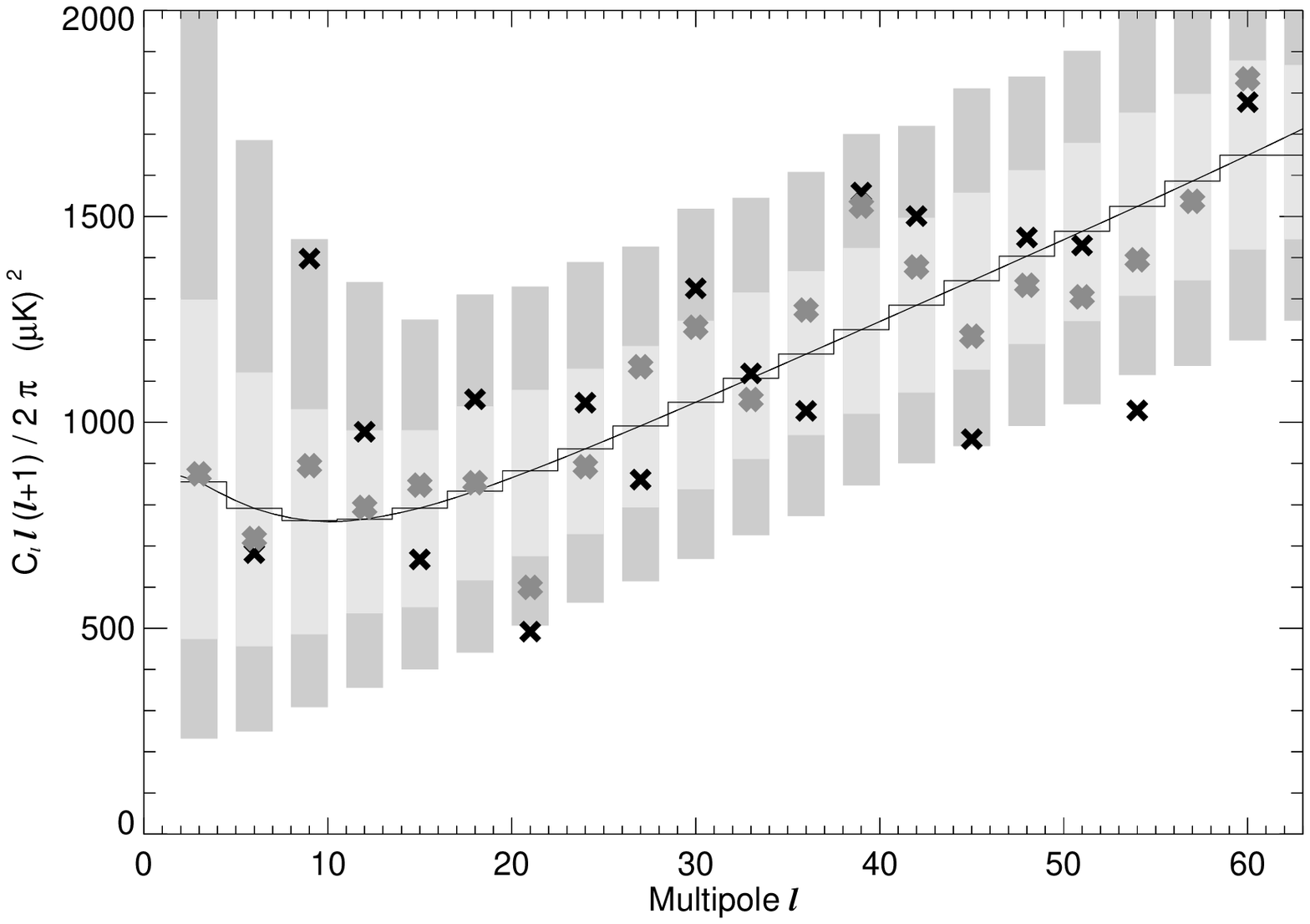,width=7cm,height=7cm}
\psfig {file=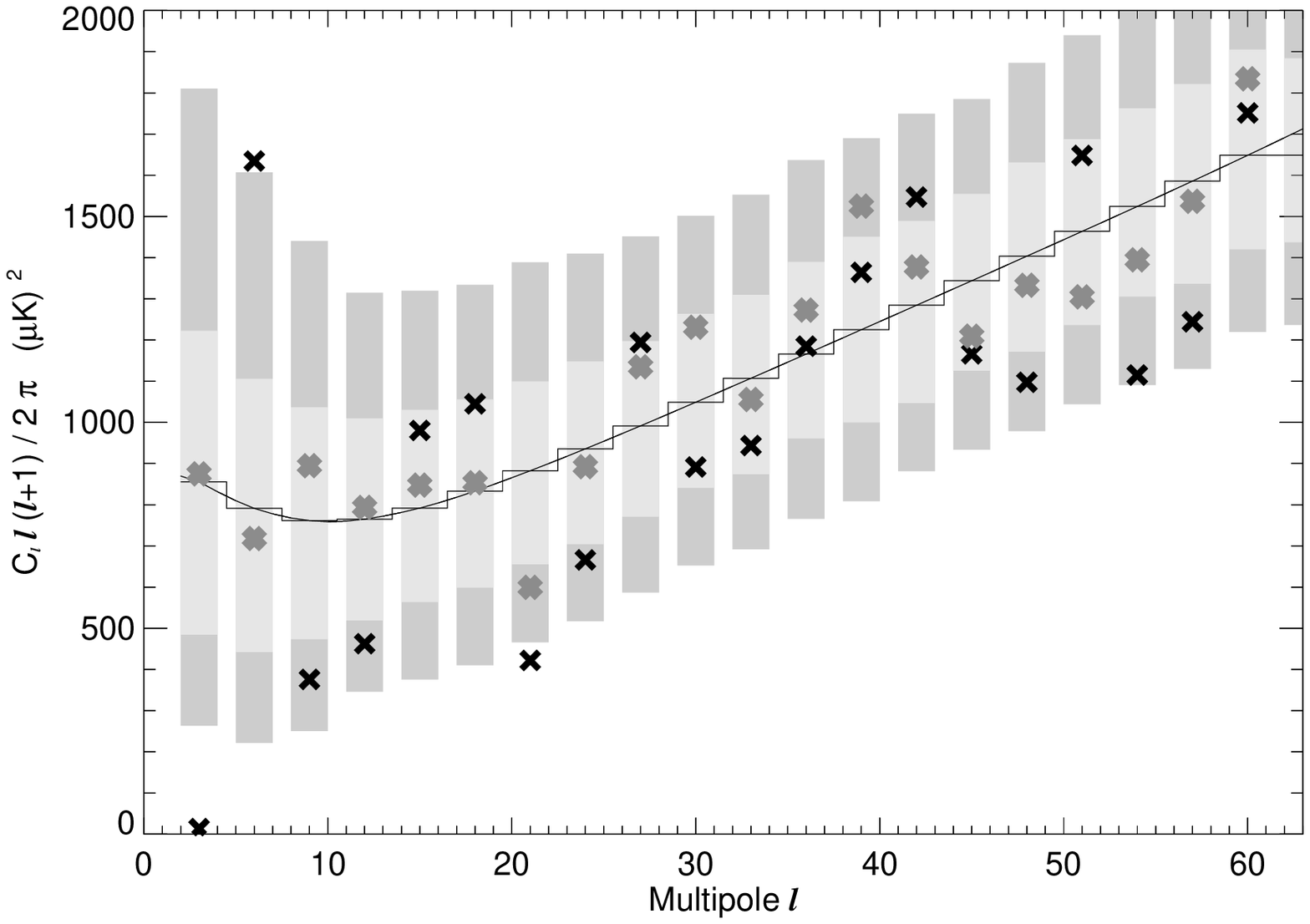,width=7cm,height=7cm}
\psfig {file=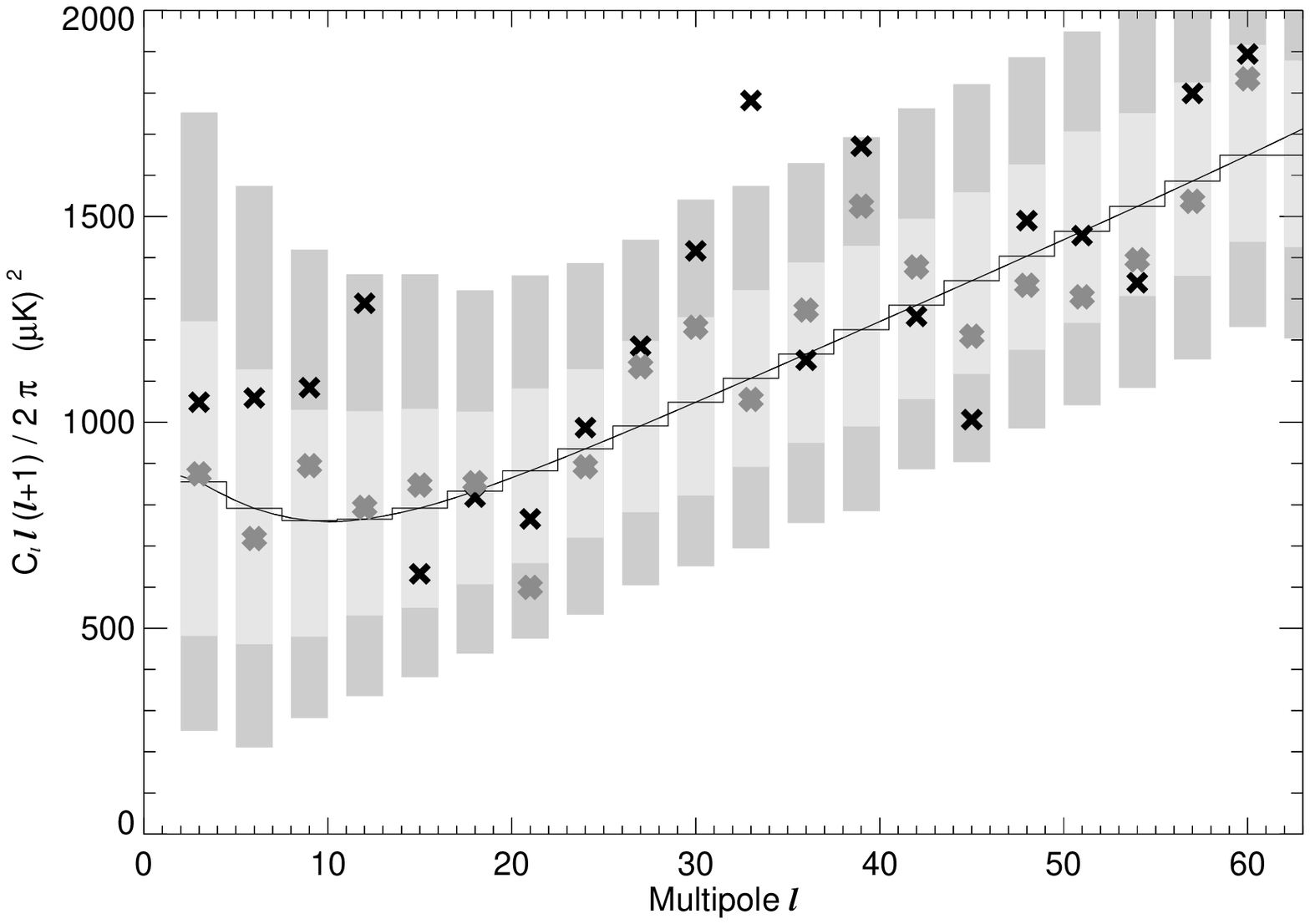,width=7cm,height=7cm}
\psfig {file=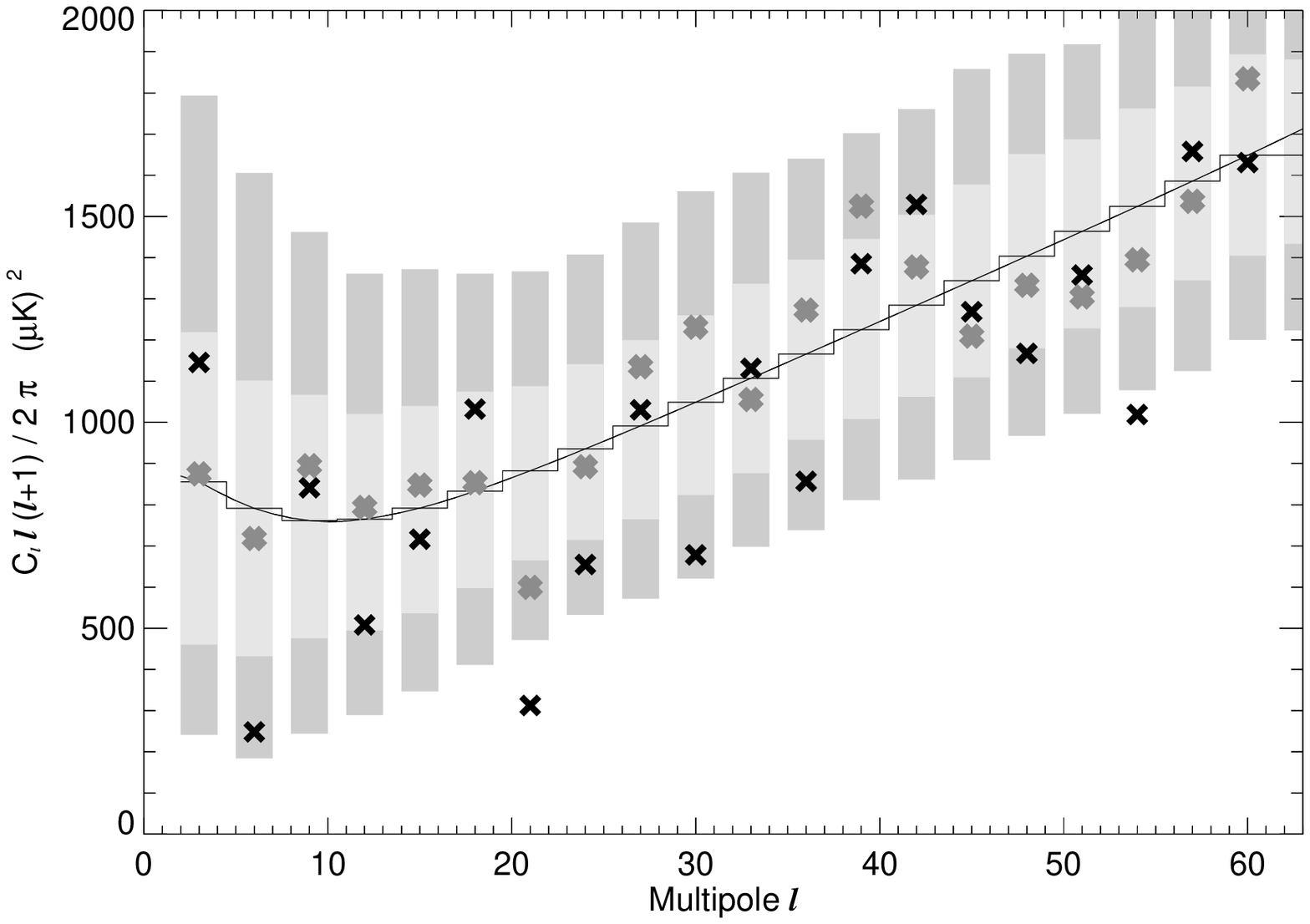,width=7cm,height=7cm}
\psfig {file=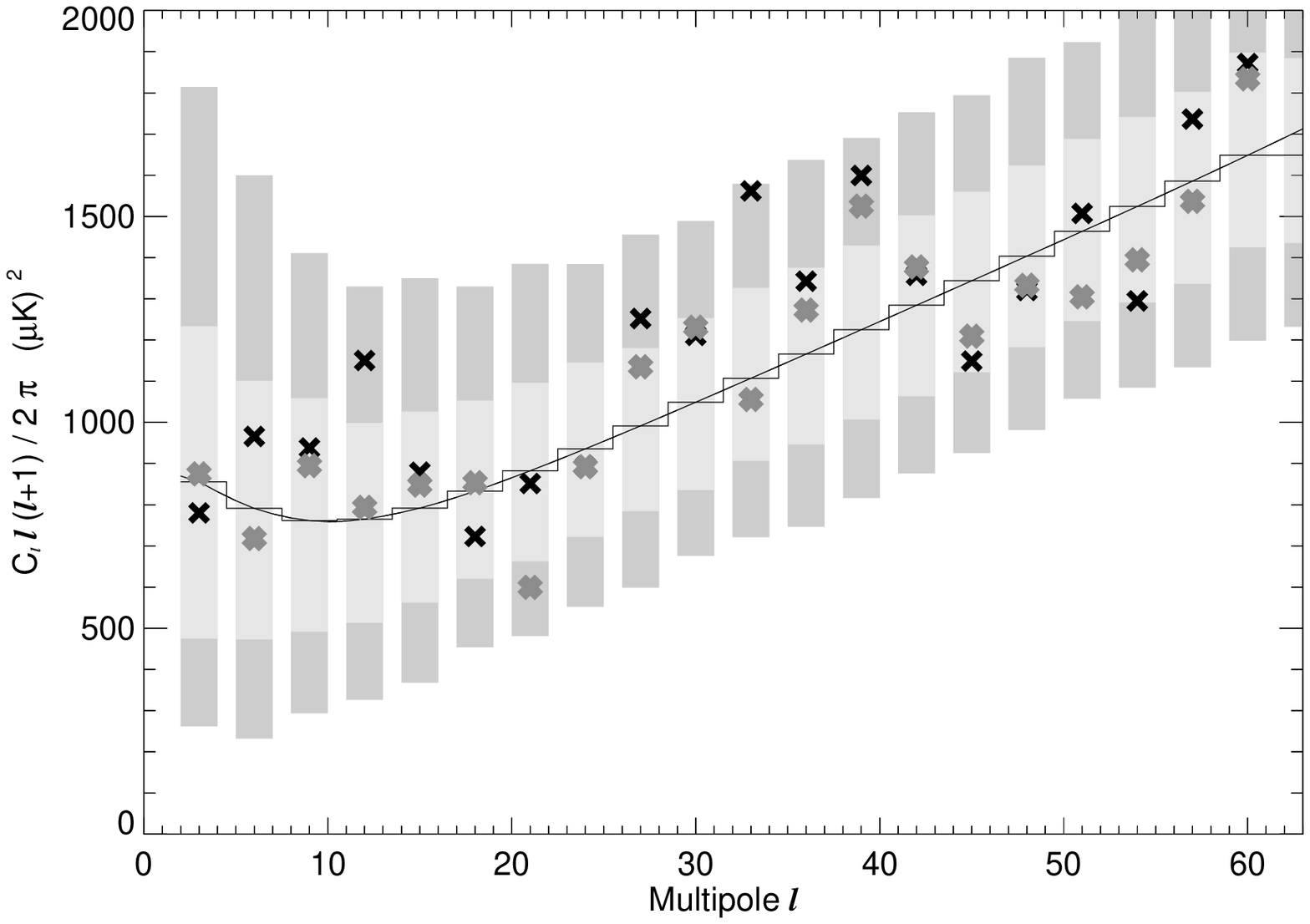,width=7cm,height=7cm}
\caption{Results of power spectrum estimation on hemispheres: the left
column shows the northern spectra, the right column shows the southern
spectra. The first row is taken in the Galactic reference frame, the second row in the ecliptic reference frame and the last row in the reference frame of maximum asymmetry, where the north pole points at $(80^\circ,57^\circ)$. The solid line shows the best fit \emph{WMAP} spectrum, the histogram is the same spectrum binned, the shaded zones show the one and two sigma error bars from simulations of the given hemisphere, grey crosses show the \emph{WMAP} estimate on the full sky (with the Kp2 mask) and the black crosses show our estimate on the given hemisphere. Please note that the shaded zones show the errorbars for a hemisphere, the errorbars for the \emph{WMAP} estimates (grey crosses) are smaller by a factor of roughly $\sqrt{2}$.}
\label{fig:hemispec}
\end{center}
\end{figure}

In figure (\ref{fig:hemispec}) we have plotted the spectra estimated
for the northern and southern hemispheres in the 
Galactic and ecliptic reference frames. We
have also included the two hemispheres for which the large scale
asymmetry is found to be largest (which will be defined in the last part of this
section). In each plot, the histogram shows the binned best fit \emph{WMAP}
spectrum (the full spectrum shown as a solid line) used in the
simulations (2048 for each hemisphere), the shaded zones show the 1
and 2 sigma error bars from the simulations on the given hemisphere,
the gray thick crosses show the \emph{WMAP} global estimate of the given
bin and the black crosses show our estimate on the given hemisphere
of the \emph{WMAP} map. Note that the errorbars shown as shaded zones are 
valid only for the hemisphere results. The full sky \emph{WMAP} results 
have error bars smaller by a factor of roughly $\sqrt{2}$. 

On the largest angular scales, note that bin 1 ($\ell=2-4$) 
which includes the quadrupole 
is low in the northern Galactic hemisphere, exceptionally so in the ecliptic reference
frame, yet takes values close to the best fit average prediction in the 
corresponding southern hemispheres.
Moreover, in the reference frame of maximum asymmetry the amplitude no
longer seems to be anomalous. 

Next we consider those bins centred at $\ell=21$ and $\ell=39$
which were determined to be outliers in the analysis of \cite{hinshaw}.
It is striking that the negative outlier at $\ell=21$ is particularly strong for
the northern hemispheres of the ecliptic and maximum asymmetry
reference frames, but absent in the corresponding southern hemispheres. In the
case of the $\ell=39$ bin, the opposite trend is observed with the outlier 
being associated with only the southern ecliptic hemisphere. 
We suspect that these features may be related to
the asymmetry suggested by the small disc results since both
features are consistent with the observed trend of high power in the southern hemisphere 
and low power in the northern hemisphere (relative to the canonical
best-fit model). 

A systematic effect could preferentially produce artifacts in the
ecliptic reference frame (as a consequence of the scan pattern of the
\emph{WMAP} satellite), but it is difficult to envisage how an
asymmetric signal could manifest itself over a range of angular scales.
Indeed, the suppression of power in the northern hemisphere
would itself require a convenient alignment of structure generated by
the parasitic signal with the underlying CMB anisotropy. An
astrophysical origin might also be aligned with the ecliptic reference
frame, but such signals would be predominantly confined to the
ecliptic plane region.

\begin{figure}
\begin{center}
\leavevmode
\psfig {file=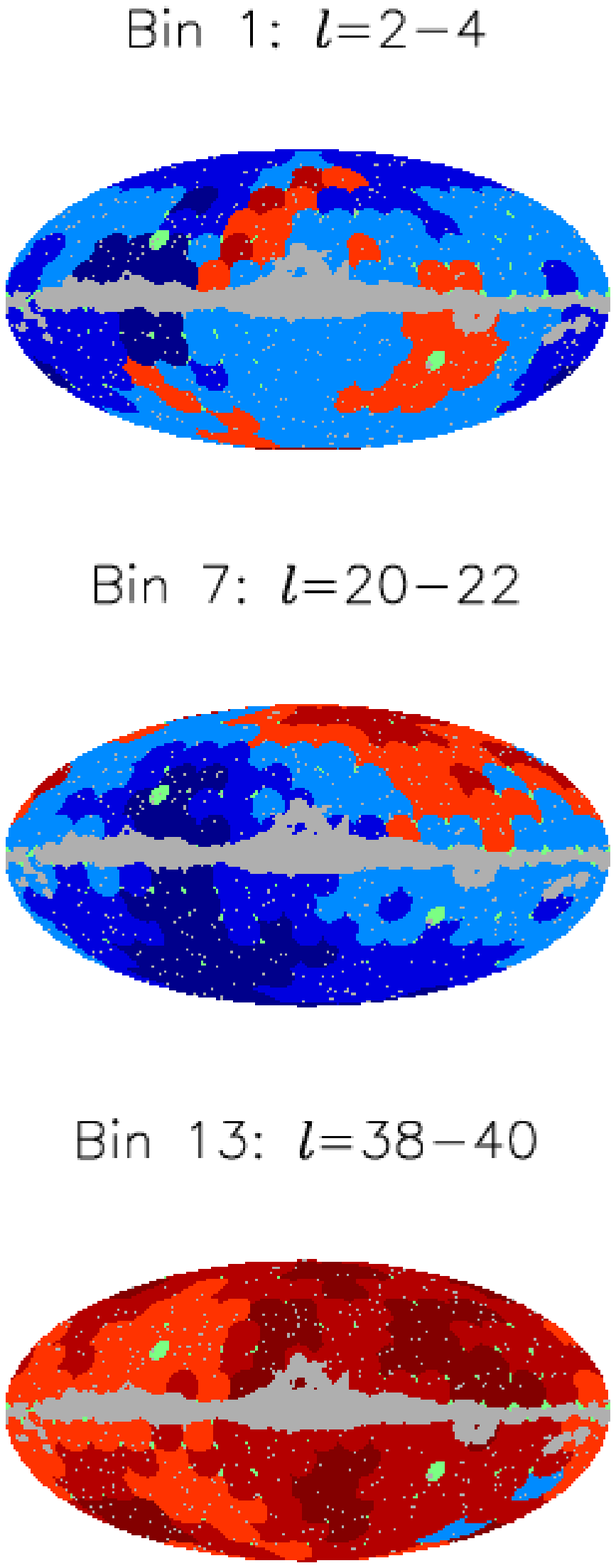,width=7.5cm,height=18cm}
\caption{ The local power in the bin $\ell=2-4$ (upper plot), $\ell=20-22$ (middle plot) and $\ell=38-40$ (lower plot)  estimated on hemispheres centred on the positions indicated by the discs. The hemispheres with power above the average set by simulations are indicated by red discs and the hemispheres with power below the average with blue discs. The significance is indicated by the intensity of the colour: light means within $1\sigma$, medium means between $1$ and $2\sigma$ and dark means outside of $2\sigma$. The two green discs show the positions of the ecliptic poles.}
\label{fig:indbin}
\end{center}
\end{figure}

Figure (\ref{fig:indbin}) shows the positional dependence of 
the three bins described above. The discs delineate the location of 
the centres of the hemispheres (ie. where the north or south pole
direction vectors pierce the sphere), red indicate a bin value above the average, blue below. 
The values within $1\sigma$, between $1$ and $2\sigma$ and outside of 
$2\sigma$ are indicated by light, medium and dark colours respectively. 
From these figures one can easily see how the distribution of bin 7 ($\ell=21$) 
and 13 ($\ell=39$) gives rise to differences in the northern and southern 
ecliptic hemispheres as well as the differences between the polar and 
equatorial ecliptic regions. We observe that although bin 1 is lower 
than average for almost all hemispheres, the distribution is still 
asymmetric as there are particularly low values around the northern 
ecliptic hemisphere. The opposite is the case for bin 13 which is 
generally high, but particularly around the southern ecliptic hemisphere. 
For bin 7 the bin values are distributed both above and below the average 
value. 

\begin{figure}
\begin{center}
\leavevmode
\psfig {file=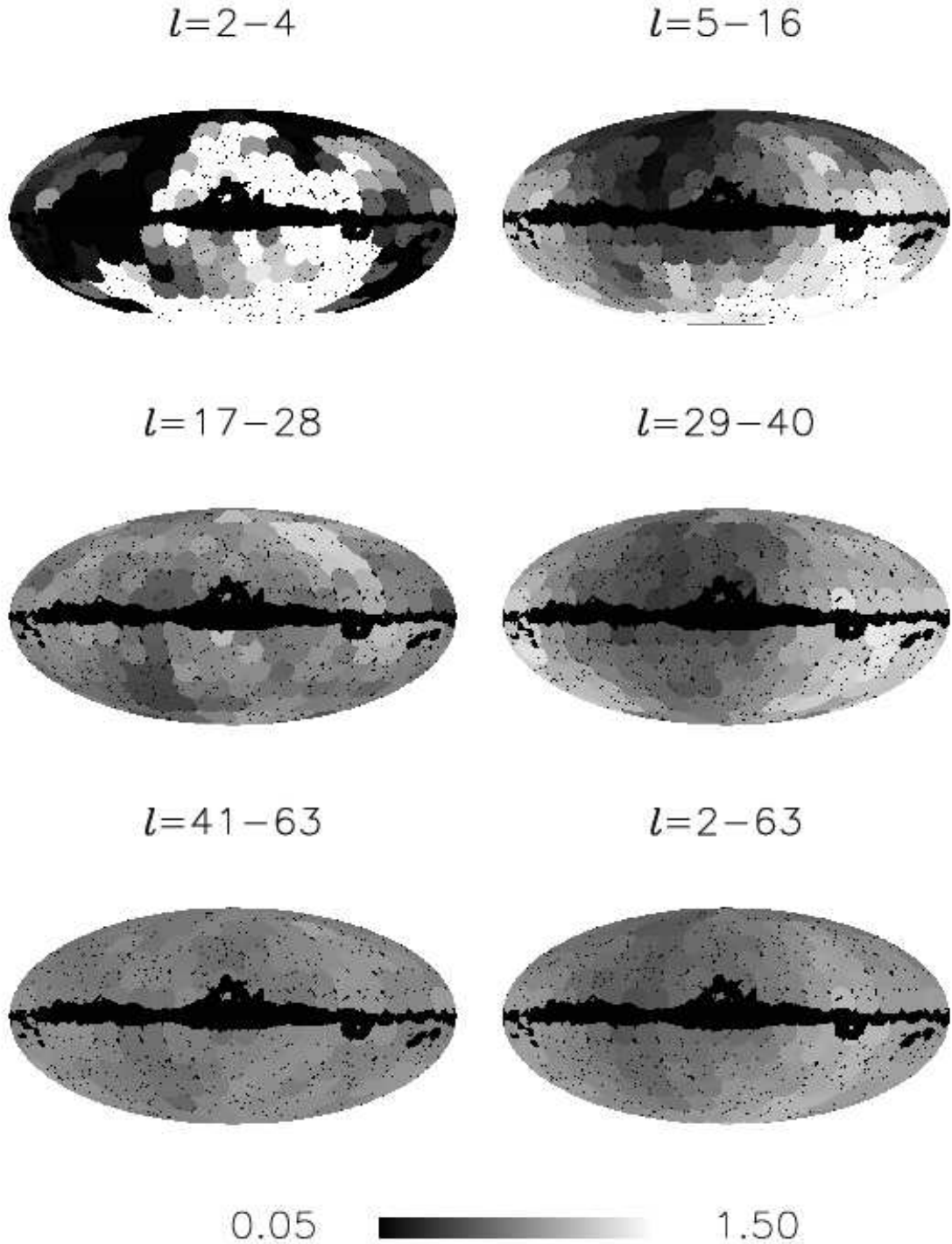,width=15cm,height=19cm}
\caption{The discs show the centres of the hemispheres where the power spectrum has been estimated. The colours show the ratio of the power spectrum bin $\ell=2-4$ (upper left), $\ell=5-16$ (upper right), $\ell=17-28$ (middle left), $\ell=29-40$ (middle left), $\ell=41-63$ (lower left) and $\ell=2-63$ (lower right) between the opposite hemispheres.}
\label{fig:allbins}
\end{center}
\end{figure}

Finally, we utilise the hemisphere estimates to check the positional 
dependence of the large scale power over larger multipole intervals
which include the three interesting ranges above. 
Using the above defined 3-bins, we constructed new larger bins. 
Initially, we cross-check the bin $\ell=2-63$ for which we detected a north-south 
asymmetry using the smaller $9.5^{\circ}$ and $19^{\circ}$ discs.
Figure (\ref{fig:allbins})
indicates the ratio of the bandpower amplitude 
between the hemisphere centred at the point where the disc is plotted
and the value in the opposite hemisphere.
A similar asymmetric structure is present in this map as found for the 
small and medium discs (figure \ref{fig:plotdisc4_b034} and
\ref{fig:plotdisc4_20deg_b03}). 
By considering the bins, $\ell=2-4$, $\ell=5-16$, 
$\ell=17-28$, $\ell=29-40$ and $\ell=41-63$
it appears that 
the asymmetry seen for $\ell=2-63$ seems to be dominated by the 
$\ell=5-16$ and $\ell=29-40$ multipole ranges. For these two ranges 
the asymmetry pattern is almost identical. For multipoles beyond $\ell=40$ 
the asymmetry seems to be gone. There is also a multipole range $\ell=17-28$ 
which does not have exactly the asymmetry pattern of the other
multipoles. 
Nevertheless, we conclude that separate multipole ranges between $\ell=5-40$ show similar asymmetric patterns.

\subsection{The significance of the maximum asymmetry}

In order to avoid the introduction of bias in our 
interpretation of the significance of the asymmetry results,
in particular due to the consideration 
of hemispheres defined only with respect to `special' coordinate systems,
we have adopted the following approach.

\begin{enumerate}
\item The power spectrum for each hemisphere centred on the 164
possible orientations on the sky was computed for 2048 simulated maps.
\item The results were binned over the desired interval in $\ell$ and 
the ratio of bandpower $r$ computed between a specific (northern)
hemisphere and its opposite (southern) hemisphere.
\item For each simulation, the axis with the maximum ratio,
$r_\mathrm{max}$, was determined (NB. this may or may not be the same as the axis of maximum 
asymmetry for the \emph{WMAP} data itself). 
\end{enumerate}

Table \ref{tab:maxratop} summarises the number of simulations with a higher maximum 
ratio $r_\mathrm{max}$ than the maximum ratio in the \emph{WMAP} data
for different multipole ranges. Table \ref{tab:maxratpos} shows the position of the pole 
of maximum asymmetry for the same $\ell$-intervals. 
We note that the position of 
maximum asymmetry for those bin ranges starting with $\ell=2$ are all at the 
Galactic north pole. Later, we will speculate that the lowest
multipole band, $\ell=2-4$, may contain residual Galactic contamination,
although it is not clear how this would lead to the north Galactic
pole being the preferred asymmetry direction.
For the remaining multipole combinations, there are intervals
where only 0-5 percent of the simulated maps show a similar asymmetry
to the data, and nearly all of these share the same axis of asymmetry,
specifically that for which the north pole of the reference frame
pierces $(80^\circ,57^\circ)$ in galactic coordinates\footnote{
Here, $\theta$ and $\phi$ are measured in the HEALPix convention thus
corresponding to \emph{co}-latitude and longitude.}.

\begin{figure}
\begin{center}
\leavevmode
\psfig {file=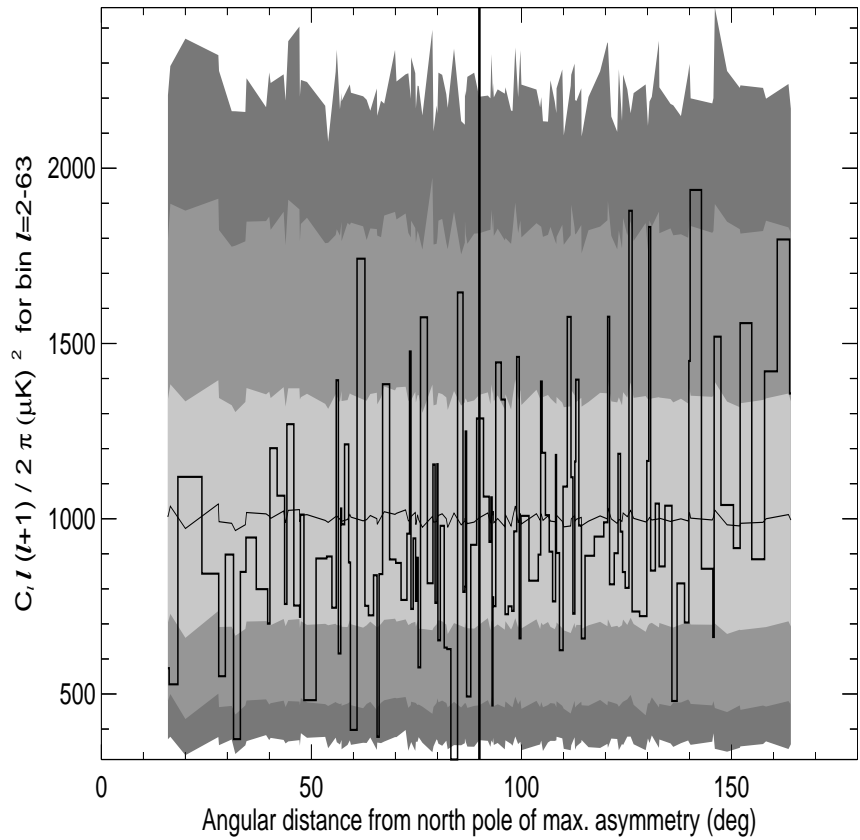,width=12cm,height=12cm}
\caption{The disc estimates of the bin $\ell=2-63$. The discs are ordered such that the values to the left are close to the north pole of the axis $(80^\circ,57^\circ)$ of maximum asymmetry and the values to the right are close to the corresponding south pole. The angular distance from this north pole is shown. The shaded zones indicate the 1, 2 and 3 sigma spread of the estimated bin for the given disc as found from Monte Carlo simulations.}
\label{fig:plotdiscs2max}
\end{center}
\end{figure}

Figure (\ref{fig:plotdiscs2max}) serves to emphasize that this
asymmetry cannot be associated with one particular region on the 
sky, but is an extended feature seen by the locally estimated spectra in 
most directions. In fact, this is simply Fig.\ref{fig:plotdiscs2bin0} --
showing the local large power estimates on small discs --
replotted in our preferred reference frame. We see that within $60^\circ$
of this north pole, the bin $\ell=2-63$ is particularly low 
(only 4 discs above the average, but several discs below the lower 
$1$ and $2\sigma$ levels)  whereas around this south pole, the bin 
values are particularly high (very few discs below the lower $1\sigma$ 
level but several above the upper $1$ and $2\sigma$ level). 

\begin{figure}
\begin{center}
\leavevmode
\psfig {file=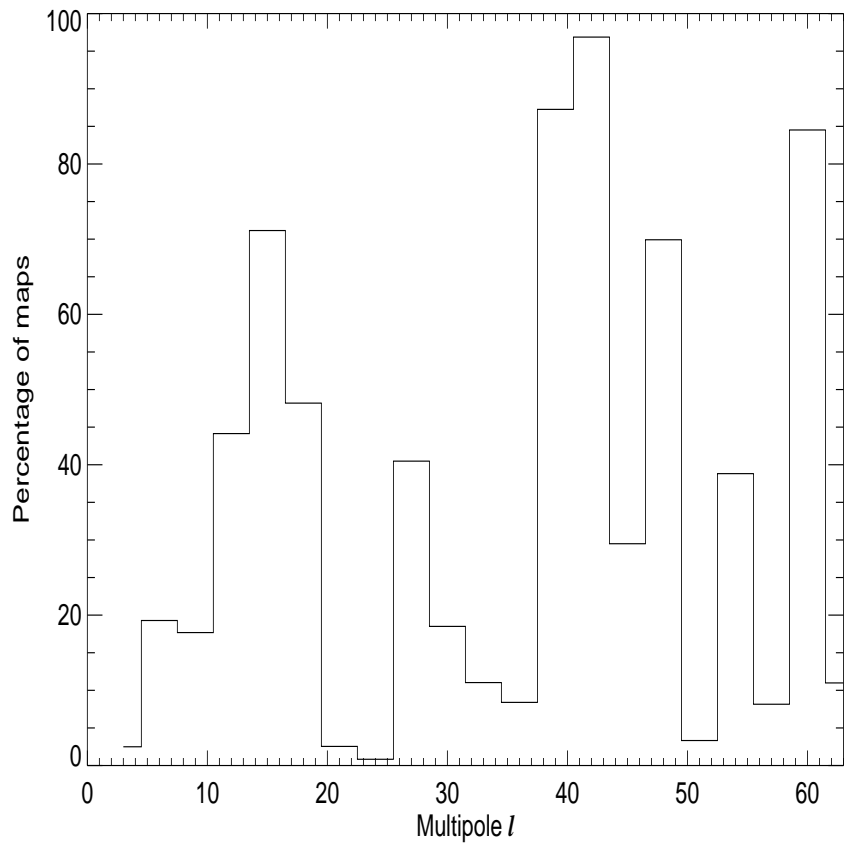,width=12cm,height=12cm}
\caption{The percentage of maps with a higher asymmetry than in the \emph{WMAP} data for a single multipole bin.}
\label{fig:binforbin}
\end{center}
\end{figure}

\begin{figure}
\begin{center}
\leavevmode
\psfig {file=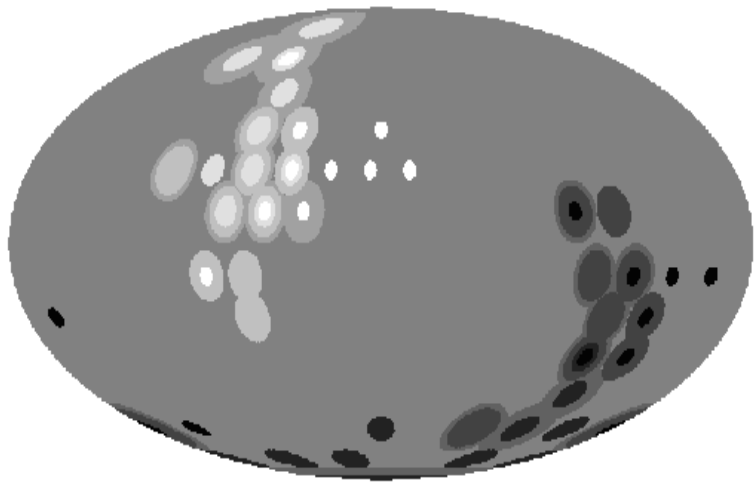,width=10cm,height=7cm}
\caption{The discs show the positions of the hemispheres with the 10 highest (black discs) and 10 lowest (white discs) bin values. The power spectrum bins considered were $\ell=2-40$ (large discs), $\ell=8-40$ (second largest discs), $\ell=5-16$ (second smallest discs) and $\ell=29-40$ (smallest discs).}
\label{fig:maxasspos}
\end{center}
\end{figure}

Figure (\ref{fig:binforbin}) shows the percentage of simulated maps with 
a higher maximum asymmetry for each single bin. Note that there is a dip 
for the bins close to $\ell=21$ and a peak close to $\ell=39$ where there 
are particular features in the full sky power spectrum as detailed above. 
At the first feature, the asymmetry is particularly large, whereas for 
the second feature the spatial distribution is particularly uniform. In 
figure (\ref{fig:maxasspos}) we show the position of the hemispheres with 
the 10 highest(black discs) and lowest (white discs) powers for the bins 
$\ell=2-40$ (large discs), $\ell=8-40$ (second largest discs), $\ell=5-16$ 
(second smallest discs) and $\ell=29-40$ (smallest discs). We see that the 
maximum/minimum positions are in good agreement for all these multipole
ranges. We also see than the 10 maximum/minimum positions are all relatively 
close to each other. We conclude that the asymmetry, apart from being
similarly orientated for several multipole ranges, is also highly significant for
separate ranges as compared to 2048 simulations and is not associated with one
particular spot on the sky.

\subsection{The asymmetry in other channels and with other sky cuts}

In order to constrain possible foreground residual contributions
to our analysis, we have analysed the data with more aggressive 
Galactic cuts -- 
the \emph{WMAP} $Kp0$ mask and a $30^\circ\times2$ ($30^\circ$ north of equator and 
$30^\circ$ south of equator) Galactic cut --
and also considered data at a lower frequency --
the combined Q1+Q2 channels evaluated with the $Kp0$ cut. 
To save CPU time, we have only computed pseudo power spectrum
estimates, with simple corrections for the sky coverage, beam response
and noise contribution, rather than employing the full maximum likelihood
machinery\footnote{We first verified that this approach when applied to 
the co-added W+V channels rendered results consistent with the 
maximum likelihood power spectrum estimation}. 
Table \ref{tab:pseudoassym_gal} and
\ref{tab:pseudoassym_max} summarise the results. We have considered the
Galactic reference frame and the reference frame of maximum asymmetry
and found that the fraction of simulations (from a total of 512 realisations)
with a higher asymmetry ratio than for the \emph{WMAP}data over specific multipole ranges. 
We see that when excluding the lowest 
multipoles, the asymmetry persists for any sky cut and channel in either
frame of reference. 
Furthermore, when considering the $Kp0$ cut, the strength of the asymmetry for the 
W+V and Q-band maps is identical. If Galactic foregrounds had played 
an important role, one would expect some clear differences between these two frequency channels.
The fact that the asymmetry is present at a significant
level in several frequency channels and with different sky cuts
excludes a simple explanation based on systematic effects and foregrounds.

\begin{figure}
\begin{center}
\leavevmode
\psfig {file=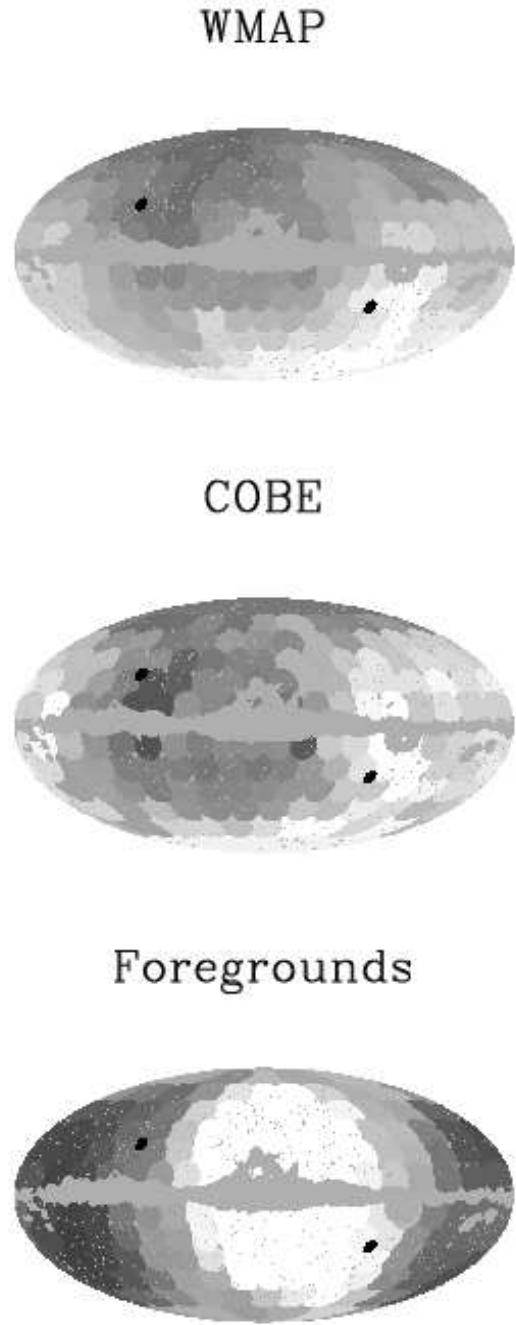,width=7.5cm,height=18cm}
\caption{The discs show the centres of hemisphere on which the power spectrum has been estimated. The colour in each disc represents the ratio between the power in the given hemisphere and the opposite hemisphere in the multipole range $\ell=5-20$. The upper plot is for the \emph{WMAP} W+V channel, the middle plot for the COBE data and the lower plot for the foreground template used for \emph{WMAP}.}
\label{fig:wmap_vs_cobe}
\end{center}
\end{figure}

As a useful cross-check which should help to mitigate against
systematic effects as the cause of the observed structure,
the same hemisphere exercise has been performed for the co-added 53+90 GHz
\emph{COBE}-DMR map, considering multipoles in the range $\ell=5-20$ where the signal is
dominant. 
A comparison of the results with the \emph{WMAP} data is shown in figure (\ref{fig:wmap_vs_cobe}). 
We see immediately the same asymmetric structure in both data sets even if the 
\emph{COBE}-DMR map is somewhat noisier, the significance of the
asymmetry being at $\sim 1\sigma$ level. 
The fact that the \emph{COBE}-DMR data is susceptible to different
parasitic signals compared to \emph{WMAP} argues against an
explanation for the results in terms of a systematic effect. 

Finally, in the same figure, we have plotted 
the positional dependence of the power ratio for the effective
total foreground template contribution to the \emph{WMAP} data.
There is no clear dependence between the structures 
seen here and the direction of maximum asymmetry, which again makes the 
explanation of the asymmetry in terms of Galactic foregrounds less
probable. 
Over the range $\ell=2-4$ the distribution of power in the foregrounds
template shows the same morphology as for the interval $\ell=5-20$
seen in the plot. If we then compare the distribution of power 
for the multipoles $\ell=2-4$ determined from the \emph{WMAP} data
(figure \ref{fig:allbins}) with the distribution of the foregrounds template, 
there is a striking similarity. This strengthens the hypothesis that this 
bin is contaminated by foregrounds as we noted above looking at the 
direction of maximum asymmetry. Thus, Galactic contamination in this bin 
could contribute to the low value of the quadrupole, but cannot easily 
explain the asymmetry seen in the other multipole ranges. 

We conclude that the remarkable lack of large scale structure in the 
area around the ecliptic north pole (more precisely around the point 
$(80^\circ,57^\circ$))  
over the multipole range from $\ell=5$ to $\ell=40$ cannot easily be explained 
by either Galactic foregrounds or systematic effects.

\begin{center}
\begin{table}
\caption{The asymmetry ratio as computed in a coordinate frame selected such that the
  asymmetry is maximised. 
  The numbers indicate the percentage of simulations with a higher maximum
  ratio than that found in the \emph{WMAP} data for the given
  multipole range. The upper horizontal line shows the first multipole and 
  the left vertical line shows the last multipole in the multipole range.
  Note that in this case the data value is compared against values
  derived from the simulations which may or may not have the same
  preferred axis as the data.}
\label{tab:maxratop}
%\begin{ruledtabular}
\begin{tabular}{|l|l|l|l|l|l|l|l|l|l|}
bin & $\ell=2$&$\ell=5$&$\ell=8$&$\ell=14$&$\ell=20$&$\ell=26$&$\ell=32$&$\ell=38$&$\ell=44$\\
\hline
$\ell=19$ & 0.2 & 7.6  & 20.3 & 23.3 &   &  &  & &  \\ 
$\ell=25$ & 0.1 & 12.5  & 24.5 & 74.9 & 24.9 &  &  & &  \\ 
$\ell=31$ & 0.1 & 5.5  & 16.8 & 24.8 & 7.0 & 15.4 &  & &  \\
$\ell=37$ & 0.2 & 0.7  & 4.7 & 2.2 & 0.5 & 1.8 & 1.7 & &  \\
$\ell=40$ & 0.3 & 0.7  & 4.7 & 9.0 & 0.9 & 13.7 & 35.1 & 90.9 &  \\
$\ell=43$ & 0.5 & 2.4  & 10.5 & 23.9 & 6.2 & 23.4 & 41.9 & 52.4 &  \\
$\ell=49$ & 0.5 & 3.4  & 5.6 & 14.5 & 2.5 & 8.5 & 20.7 & 36.7 & 90.3 \\
$\ell=55$ & 1.9 & 1.4  & 5.9 & 5.9 & 2.8 & 2.3 & 4.7 & 14.0 & 16.8 \\
$\ell=61$ & 0.3 & 1.1  & 5.5 & 13.4 & 4.6 & 23.6 & 40.7 & 77.7 & 75.2 \\
\hline
\end{tabular}
%\end{ruledtabular}
\end{table}
\end{center}

\begin{center}
\begin{table}
\caption{Position of the axis of maximum asymmetry for the multipole ranges given in table (\ref{tab:maxratop}). The upper horizontal line shows the first multipole and the left vertical line shows the last multipole in the multipole range. The given position is the centre of the hemisphere with most power of the pair of opposite hemispheres. The angels $(\theta,\phi)$ are given in degrees in Galactic coordinates, using co-latitude as in the Healpix software package, i.e. $\theta$ goes from $0$ to $180^\circ$ going from the north to the south pole}
\label{tab:maxratpos}
%\begin{ruledtabular}
\begin{tabular}{|l|l|l|l|l|l|l|l|l|l|}
bin & $\ell=2$&$\ell=5$&$\ell=8$&$\ell=14$&$\ell=20$&$\ell=26$&$\ell=32$&$\ell=38$&$\ell=44$\\ 
\hline
$\ell=19$ &(0,0)&(14,72)&(28,180)&(100,331.6)&  &  &  & &  \\ 
$\ell=25$ &(0,0)&(14,72)&(54,180)&(54,23)&(80,57)&  &  & &  \\ 
$\ell=31$ &(0,0)&(80,57)&(80,76)&(80,38)&(80,57)&(100,28)&  & &  \\
$\ell=37$ &(0,0)&(80,57)&(80,57)&(80,38)&(80,57)&(80,38)&(67,6)& &  \\
$\ell=40$ &(0,0)&(80,57)&(80,57)&(80,57)&(80,57)&(80,57)&(67,346)&(126,203)&  \\
$\ell=43$ &(0,0)&(80,57)&(80,57)&(80,57)&(80,57)&(100,85)&(152,72)&(113,86)&  \\
$\ell=49$ &(0,0)&(80,57)&(100,85)&(100,85)&(100,85)&(100,85)&(152,72)&(113,86)&(126,113)  \\
$\ell=55$ &(0,0)&(80,57)&(80,57)&(139,47)&(152,72)&(139,47)&(152,72)&(152,72)&(126,46)  \\
$\ell=61$ &(0,0)&(80,57)&(80,57)&(80,57)&(80,57)&(80,57)&(152,72)&(152,72)&(54,0)  \\
\hline
\end{tabular}
%\end{ruledtabular}
\end{table}
\end{center}

\begin{center}
\begin{table}
\caption{The percentage of simulations with a higher ratio of the pseudo spectra between the northern and southern Galactic hemisphere for the given multipole, frequency channel and sky cut. }
\label{tab:pseudoassym_gal}
%\begin{ruledtabular}
\begin{tabular}{|l|c|c|c|c|}
channel: & W+V Kp2 & W+V Kp0 & W+V $30^\circ\times2$ & Q Kp0 \\
\hline
$\ell=2-19$& 0.7 & 1.4 & 3.1 & 1.4\\
$\ell=2-40$& 0.9 & 1.4 & 3.5 & 1.4\\
$\ell=2-61$& 1.0 & 1.4 & 3.9 & 1.4\\
$\ell=5-40$& 0.0 & 0.4 & 2.1 & 0.4\\
$\ell=8-19$& 0.0 & 0.0 &  0.4 & 0.0\\
$\ell=8-40$& 0.0 & 0.0 &  0.4 & 0.0\\
$\ell=8-61$& 0.0 & 0.0 &  0.4 & 0.0\\
$\ell=20-40$&0.0 & 0.4 &  4.5 & 0.4\\
$\ell=20-61$&0.0 & 0.2 &  6.4 & 0.2\\
\hline
\end{tabular}
%\end{ruledtabular}
\end{table}
\end{center}

\begin{center}
\begin{table}
\caption{The percentage of simulations with a higher ratio of the pseudo spectra between the maximally asymmetric hemispheres (with the axis piercing $(80^\circ,57^\circ)$ for the given multipole, frequency channel and sky cut.}
\label{tab:pseudoassym_max}
%\begin{ruledtabular}
\begin{tabular}{|l|c|c|c|c|}
channel: & W+V Kp2 & W+V Kp0 & W+V $30^\circ\times2$ & Q Kp0 \\
\hline
$\ell=2-19$& 0.6 & 1.2 & 1.9 & 1.4\\
$\ell=2-40$& 0.7 & 1.4 & 0.0 & 1.4\\
$\ell=2-61$& 0.8 & 1.4 & 3.1 & 1.4\\
$\ell=5-40$& 0.0 & 0.0 & 0.0 & 0.0\\
$\ell=8-19$& 0.0 & 0.0 &  0.0 & 0.0\\
$\ell=8-40$& 0.0 & 0.0 & 0.0 & 0.0\\
$\ell=8-61$& 0.0 & 0.0 &  0.0 & 0.0\\
$\ell=20-39$&0.0 & 0.0 &  3.3 & 0.0\\
$\ell=20-61$&0.0 & 0.0 &  5.1 & 0.0\\
\hline
\end{tabular}
%\end{ruledtabular}
\end{table}
\end{center}

\clearpage

\section{Conclusions}

\label{sect:concl}

In this paper, we have applied a method of local power spectrum
estimation to small regions of the CMB sky as measured by the 
\emph{WMAP} satellite in order to search for evidence
of spatial dependence in the computed spectra beyond that due to
sampling variance alone. In this way, we are able to confront one of
the central tenets of modern cosmology, namely that of cosmological
isotropy. As an important consequence of this analysis, we are also
able to mitigate against foreground or systematic artefacts in the data.
Specifically, we have estimated the angular power spectrum in different spatial 
directions within discs of radius $9.5^\circ$ and $19^\circ$ and 
on hemispheres. Since by increasing the disc size we increase the 
resolution of the power spectrum estimation in multipole space but
lower the sensitivity to 
the spatial dependence of the estimate, the combination of these
scales allows us to achieve high fidelity in both real and $\ell$
space.

On small angular scales and specifically for two power spectrum bins,
$\ell=164-188$ and $\ell=264-288$, close to the first Doppler peak, 
we found that the highest deviations, relative to the best fit model
preferred by the \emph{WMAP} team, was seen in the power spectrum
estimates for those medium sized discs close to the Galactic plane. 
This was also the case for an interval in $\ell$-space spanning the
`dent' in the power spectrum as estimated by the global 
analysis of the \emph{WMAP} team. By excluding those discs 
close to the Galactic plane with a local spectrum deviating by
more than $2\sigma$ from the best fit model, we find a resulting 
power spectrum with no dent and a slightly higher Doppler peak.
We consider this to be a tentative indication of possible 
residual Galactic foreground contamination. 

When studying the distribution of large scale power using the small and
medium sized discs, we found that the northern Galactic and ecliptic 
hemispheres seemed to show a remarkable lack of large scale power. 
In order to further investigate this apparent lack of isotropy,
we estimated the power spectrum for the first 60 multipoles 
on hemispheres centred on various directions on the sky. 
The multipole range $\ell=2-4$ was found to be strongly 
asymmetric with the Galactic north-south axis as the axis of maximum 
asymmetry. Furthermore we found that the spatial dependence of the
variations in amplitude for this multipole range was similar to the
morphology seen in the integrated foreground template contribution,
suggesting that the low order multipoles may remain compromised by
residual Galactic emission (this was also noted by \cite{oc,eriksen3,schwarz}).
The other multipoles between $\ell=5-40$ however, are asymmetric with the axis pointing in 
the direction $(80^\circ,57^\circ)$ in Galactic co-latitude and
longitude, close to the ecliptic axis. 
In the northern hemisphere of the reference frame 
of maximum asymmetry, almost the full multipole range between 
$\ell=5-40$ is below the average amplitude. The strong negative outlier at 
$\ell=21$ found in the \emph{WMAP} global estimate of the power 
spectrum, is associated specifically with this hemisphere.
Conversely, in the southern hemisphere of maximum asymmetry, almost all multipoles 
in the range $\ell=5-40$ are above the average amplitude, and 
a positive outlier at $\ell=39$, as seen in the \emph{WMAP} global spectrum, 
is found.
The asymmetry does not appear to be associated with a particular
anomalous region on the sky, but extends over a large area,
as evidenced by the small and medium disc results.

We have checked the possibility that incorrectly removed foregrounds 
could cause the observed large scale asymmetry by testing the 
dependency on Galactic cut and frequency channel. 
The observed asymmetry is remarkably stable with respect to both frequency 
and sky coverage, thus arguing against this possibility.
Moreover, the results seem unlikely to be compromised by systematics 
since we find supporting evidence (albeit at reduced statistical
significance) for similar features in the \emph{COBE}-DMR maps, 
which are susceptible to \emph{different} parasitic signals.
Similar asymmetric structures have been determined by a number
of other groups \cite{park,eriksen1,vielva,eriksen2,copi,curvat}
using several alternative analysis techniques. Besides finding 
asymmetry, non-Gaussian features have also been
detected in the northern and/or southern hemispheres.
Since some of the non-Gaussian estimators used are power spectrum
dependent, the detection of non-Gaussianity on opposite hemispheres 
may be related to the uneven distribution of large scale power studied in this paper. 
Given the large number of detections with different methods on 
different sky cuts and frequency channels,
it seems inescapable that the \emph{WMAP} data does indeed 
contain unexpected properties on large scales. 
In the absence of compelling evidence for a Galactic or systematic 
origin for the asymmetry, the intriguing possibility is raised that the 
cosmological principle of isotropy is violated and 
that fundamentally new physics on large scales in the universe
is required.
Further clarification of this scenario awaits 
further observations from \emph{WMAP}, and ultimately the forthcoming Planck 
satellite mission.

\clearpage

\clearpage

\clearpage

\clearpage

\clearpage

\clearpage

\clearpage

\clearpage

\clearpage

\clearpage

\clearpage

\clearpage

\clearpage

\clearpage

\clearpage

\clearpage

\clearpage

\clearpage

\clearpage

\clearpage

\clearpage

\clearpage

\clearpage

\clearpage

\clearpage

\clearpage

\section*{Acknowledgements}

We are grateful to Gary Hinshaw, Hans Kristian Eriksen and Domenico Marinucci for useful discussions. FKH acknowledges financial support from the CMBNET Research Training Network. We acknowledge use of the HEALPix \cite{healpix} software and analysis package for deriving the results in this paper. This research
used resources of the National Energy Research Scientific Computing
Center, which is supported by the Office of Science of the U.S. Department of Energy under Contract No. DE-AC03-76SF00098. We acknowledge the use of the Legacy Archive for Microwave Background Data Analysis (LAMBDA). Support for LAMBDA is provided by the NASA Office of Space Science.

\begin{appendix}

\section{Monte Carlo precalculations}
\label{app:mcpre}
For a general Gabor window $G_i$ as a function of pixel number $i$, one can write the pseudo spherical harmonic coefficients as
\begin{equation}
\tilde a_{\ell m}=\sum_{\ell' m'}a_{\ell'm'}h(\ell,\ell',m,m')
\end{equation}
where
\begin{equation}
h(\ell,\ell',m,m')=\sum_iG_iY_{\ell m}^iY_{\ell'm'}^i
\end{equation}
To use the likelihood ansatz on the pseudo spectrum coefficients, we need to sample the observed spectrum at $N^{in}$ multipoles $\ell_i$. How dense the sampling is, depends on the size of the window \ref{paper1}. Also, we want to estimate the underlying full sky spectrum in $N^{bin}$ flat bins $b$.\\

For calculating the likelihood, one needs the theoretical expectation value $C_i\equiv\VEV{\tilde C_{\ell_i}}$ and the correlation matrix $C_{ij}\equiv\VEV{\tilde C_{\ell_i}\tilde C_{\ell_j}}-\VEV{\tilde C_{\ell_i}}\VEV{\tilde C_{\ell_j}}$. These can be expressed, using the above formulae as
\begin{equation}
C_i=\sum_b C_b \underbrace{\frac{1}{2\ell_i+1}\sum_{\ell'\epsilon b}\frac{B_\ell'}{\ell'(\ell'+1)}\sum_{mm'}h^2(\ell_i,\ell',m,m')}_{K(i,b)}
\end{equation}
and
\begin{equation}
C_{ij}=\sum_{bb'}C_bC_{b'}\underbrace{\frac{2}{(2\ell_i+1)(2\ell_j+1)}\sum_{\ell'\epsilon b}\sum_{L'\epsilon b'}\sum_{mM}\sum_{m'M'}h(\ell_i,\ell',m,m')h(\ell_j,\ell',M,m')h(\ell_i,L',m,M')h(\ell_j,L',m',M')}_{K'(i,j,b,b')}.
\end{equation}
For a general Gabor window, the factors $K(i,b)$ and $K(i,j,b,b')$ are very heavy to calculate using analytic formulae. We propose to calculate these factors using a Monte Carlo approach.

The Monte Carlo approach to the calculation of these factors can be done as follows;
\begin{itemize}

\item Generate a set of independent random Gaussian variables $A_{\ell m}$ with variance
\begin{equation}
<A_{\ell m}A_{\ell'm'}>=\delta_{\ell\ell'}\delta_{mm'}\frac{B_\ell}{\ell(\ell+1)}.
\end{equation}

\item For each bin $b$, calculate a set of skies $T_i^b$ as a function of pixel number $i$ defined as
\begin{equation}
T^b_i=\sum_{\ell'\epsilon b}\sum_{m'}A_{\ell'm'}Y_{\ell'm'}^i.
\end{equation}

\item For each sky, calculate the pseudo spectrum $\tilde a_{\ell m}^b$,
\begin{equation}
\tilde a_{\ell m}^b=\sum_i T_i^bG_iY_{\ell m}^i
\end{equation}

\item Form the pseudo cross-bin spectra $C^{bb'}_i$ defined as
\begin{equation}
C^{bb'}_i=\sum_m\frac{\tilde a_{\ell_i m}^b\tilde a_{\ell_i m}^{b'}}{2\ell_i+1}
\end{equation}
and the covariance matrices $C^{bb'}_{ij}$
\begin{equation}
C^{bb'}_{ij}=(C^{bb'}_i-<C^{bb'}_i>)(C^{bb'}_j-<C^{bb'}_j>)
\end{equation}

\end{itemize}

Using the formulae above, one sees that $\VEV{C^{bb}_i}=K(i,b)$ and $2\VEV{C^{bb'}_{ij}}=K'(b,b',i,j)$ for $b\ne b'$ and $\VEV{C^{bb}_{ij}}=K'(b,b,i,j)$. Using about 1000 simulations, we get an accuracy of the $C_\ell$ estimate that is within one percent of the estimate found using the analytical expression.

When noise is present, the correlation matrix $C_{ij}$ can be written as
\begin{equation}
C_{ij}=C^S_{ij}+C^X_{ij}+C^N_{ij},
\end{equation}
where $C^S_{ij}$ is the signal correlation matrix calculated above, $C^N_{ij}=\VEV{\tilde C^N_{\ell_i}\tilde C^N_{\ell_j}}-\VEV{\tilde C^N_{\ell_i}}\VEV{\tilde C^N_{\ell_j}}$ which can be calculated analytically for white noise and by Monte Carlo of noise realisations for correlated noise. Finally there is a mixed signal-noise term which can be written as
\begin{equation}
C^X_{ij}=\sum_b C_b\underbrace{\frac{4}{(2\ell_i+1)(2\ell_j+1)}\sum_{\ell'\epsilon b}\sum_{mM}\sum_{m'}h(\ell_i,\ell',m,m')h(\ell_j,\ell',M,m')h'(\ell_i,\ell_j,m,M)}_{K''(i,j,b)},
\end{equation}
where $h'(\ell,\ell',m,m')=\sum_i\sigma^2_iG_i^2Y_{\ell m}^iY_{\ell'm'}^i$ in the case of white noise for a given noise pixel variance $\sigma_i^2$. This factor is also heavy to calculate for a general window and for general noise. It can however be calculated using a similar Monte-Carlo approach:

\begin{itemize}

\item Generate a set of independent random Gaussian variables $A_{\ell m}$ with variance
\begin{equation}
<A_{\ell m}A_{\ell'm'}>=\delta_{\ell\ell'}\delta_{mm'}\frac{B_\ell}{\ell(\ell+1)}.
\end{equation}

\item For each bin $b$, calculate a set of skies $T_i^b$ as a function of pixel number $i$ defined as
\begin{equation}
T^b_i=\sum_{\ell'\epsilon b}\sum_{m'}A_{\ell'm'}Y_{\ell'm'}^i.
\end{equation}

\item Generate a noise realisation $T^N_i$.

\item For each sky, calculate the signal pseudo spectrum $\tilde a_{\ell m}^b$,
\begin{equation}
\tilde a_{\ell m}^b=\sum_i T_i^bG_iY_{\ell m}^i
\end{equation}
and the noise pseudo spectrum
\begin{equation}
\tilde a_{\ell m}^N=\sum_i T_i^NG_iY_{\ell m}^i
\end{equation}

\item Form the signal-noise pseudo cross-bin spectra $C^{NSb}_i$ defined as
\begin{equation}
C^{NSb}_i=\sum_m\frac{\tilde a_{\ell_i m}^b\tilde a_{\ell_i m}^N}{2\ell_i+1}
\end{equation}
and the covariance matrices $C^{NSb}_{ij}$
\begin{equation}
C^{NSb}_{ij}=C^{NSb}_iC^{NSb}_j
\end{equation}

\end{itemize}

Using the formulae above, one can show that $4\VEV{C^{NSb}_{ij}}=K''(i,j,b)$. The noise part converges slower than the signal part, and for that reason we suggest to calculate 5 noise realizations for each signal realization so that a total 5000 noise realizations are used.\\

Finally, we note that the estimates of the lowest multipoles are biased \cite{bjk}. This comes from the fact that we are estimating the variance of the likelihood. This normally introduces a bias of the form $1/\sqrt{\sigma^2+1}$ which in our case is $2\ell/(2\ell+1)$. Correcting with this factor, the bias vanishes.

\end{appendix}

\end{document}